\documentclass[aps, superscriptaddress, twocolumn, amsmath, amssymb,  longbibliography, 10pt]{revtex4-1} 

\usepackage{epsfig}
\usepackage{subfigure}
\usepackage{graphicx}
\usepackage{amsfonts}
\usepackage[figuresright]{rotating}
\usepackage{amssymb}
\usepackage{amsmath}
\usepackage{psfrag}
\usepackage{esint} 
\usepackage{bm} 
\usepackage{hyperref}
\hypersetup{colorlinks,allcolors=blue}
\usepackage[version=4]{mhchem}
\usepackage{multirow}
\usepackage{siunitx}
\usepackage{xcolor}
\usepackage{tikz}
\usepackage{lipsum}
\usepackage{bbm}
\usepackage{soul}

\newcommand{\black}[1]{{\color{black} #1}}

\usepackage{physics,xcolor}

\begin{document}

\title{Designing phase sensitive probes of monopole superconducting order}

\author{Grayson R. Frazier}
\affiliation{Department of Physics and Astronomy, Johns Hopkins University, Baltimore, Maryland 21218, USA}
\author{Junjia Zhang}
\affiliation{Department of Physics and Astronomy, Johns Hopkins University, Baltimore, Maryland 21218, USA}
\affiliation{Department of Physics, Princeton University, Princeton, New Jersey 08544, USA}
\author{Junyi Zhang}
\affiliation{Department of Physics and Astronomy, Johns Hopkins University, Baltimore, Maryland 21218, USA}
\author{Xinyu Sun}
\affiliation{Department of Physics and Astronomy, Johns Hopkins University, Baltimore, Maryland 21218, USA}
\author{Yi Li}
\affiliation{Department of Physics and Astronomy, Johns Hopkins University, Baltimore, Maryland 21218, USA}

\date {November 22, 2024}

\begin{abstract}
    Distinct from familiar $s$-, $p$-, or $d$-wave pairings, the monopole superconducting order represents a novel class of pairing order arising from nontrivial monopole charge of the Cooper pair. 
    In the weak-coupling regime, this order can emerge when pairing occurs between Fermi surfaces with different Chern numbers in, for example, doped Weyl semimetal systems. 
    However, the phase of monopole pairing order is not well-defined over an entire Fermi surface, making it challenging to design experiments sensitive to both its symmetry and topology. 
    To address this, we propose a scheme based on symmetry and topological principles to identify this elusive pairing order through a set of phase-sensitive Josephson experiments. 
    By examining the discrepancy between global and local angular momentum of the pairing order, we can unveil the monopole charge of the pairing order,
    including for models with pair monopole charge $|q_p| = 1$, $2$, and $3$. 
    We demonstrate the proposed probe of monopole pairing order through analytic and numerical studies of Josephson coupling in models of monopole superconductor junctions.
    This work opens a promising avenue to uncover the unique topological properties of monopole pairing orders and to distinguish them from known pairing orders based on spherical harmonic symmetry. 
\end{abstract}

\maketitle

\section{Introduction}

Monopole superconducting order~\cite{Li2018} is an exotic family of three-dimensional topological superconducting orders.
This pairing order arises from the nontrivial geometric phase of Cooper pairs~\cite{Murakami2003a} formed between two Fermi surfaces with different Chern numbers, which can occur in weak-coupling superconducting materials supporting Weyl Fermi surfaces~\cite{Armitage2018,Baidya2020, Jeon2021,Xiao2022}.
For example, when superconductivity is developed in a three-dimensional inversion symmetric magnetic Weyl system, intrinsically or by proximity, monopole harmonic pairing exists as the zero center-of-mass momentum pairing between two inversion-related Fermi surfaces enclosing Weyl points of opposite chiralities~\cite{Li2018, Li2024,Bobrow2022}. 
Unlike many other known topological superconductors, where only Bogoliubov-de Gennes (BdG) quasiparticle states cannot be adiabatically connected to a Bose-Einstein condensate~\cite{Read2000, Volovik2003}, the pairing order of a monopole superconductor also fails to be globally well-defined over the entire Fermi surface due to topological obstruction in the $\mathrm{U}(1)$ pairing phase. 
The topological obstruction in the pairing order also fundamentally changes the symmetry representation: monopole pairing order is no longer describable by familiar symmetry in terms of $s$-, $p$-, and $d$-waves based on spherical harmonics but is rather characterized by the so-called monopole harmonic functions~\cite{Wu1976,Haldane1983,Li2018, Li2024}. 
Moreover, the existence of pairing nodes in monopole superconducting order is independent of specific pairing mechanisms and is instead dictated by nonzero monopole charge of Cooper pairs, which results in unconventional pairing with nontrivial winding of pairing phase over an entire Fermi surface~\cite{Murakami2003a, Li2018}. 

It has been known that, in the presence of a magnetic monopole in real space, the orbital angular momentum of a charged particle is no longer conserved. However, the total angular momentum, which includes the angular momentum of the electromagnetic field, remains conserved in both classical and quantum scatterings of charged particles~\cite{Goddard1978,Atiyah1985}. 
In a monopole superconductor, Cooper pairs scatter in the presence of a momentum space ``magnetic" monopole.
The total angular momentum of the pairing order, which we refer to as the ``global angular momentum," 
includes the contribution from the pair Berry flux and is a conserved physical observable that characterizes the global symmetry of the pairing order. 
However, even though the global angular momentum is conserved, the pairing order is not well-defined globally, \textit{i.e.}   
for any choice of gauge, there exists at least a singular point at the Fermi surface at which the pairing phase cannot be well-defined. 
Only away from the singular point is the pairing order locally well-defined and able to be described in terms of spherical harmonic functions, which determine the ``local angular momentum" of the pairing order. 
The shift, or say, the discrepancy, between the conserved global angular momentum and local angular momentum is a unique signature of the monopole charge of the pairing order.

To detect superconducting pairing symmetries, phase-sensitive experiments have long played an essential role~\cite{Josephson1962, Geshkenbein1987, Sigrist1992, Wollman1993, Tsuei1994, Harlingen1995, Tsuei2000}. 
For example, the $d_{x^2-y^2}$ pairing symmetry of a high-temperature cuprate superconductor has been revealed by a corner Josephson junction.
The relative sign difference of the pairing order along the $a$ and $b$ axes of a cuprate superconductor leads to a destructive Fraunhofer interference pattern at zero magnetic flux~\cite{Harlingen1995}.
For $p$-wave topological superconductors,  Majorana modes localized at boundaries have been proposed to contribute to single particle tunneling and the $4\pi$-periodic Josephson current phase relation~\cite{Kitaev2001, Kwon2004, Fu2008, Chung2009, Sau2010, Alicea2010, Grosfeld2011a, Alicea2012}. 
For monopole pairing order, the pairing phase is not well-defined over an entire Fermi surface, making it challenging to design experiments sensitive to both its symmetry and topology.

We propose a phase-sensitive approach to address the outstanding question of experimentally identifying the monopole superconducting order. 
Guided by symmetry and topological principles, we develop the design of a set of Josephson junctions that can extract the shift between the global and local angular momentum of the monopole pairing order, thereby determining the pair monopole charge. 
Specifically, we investigate two classes of Josephson junctions and study the Josephson currents in these junctions by combining linear response and numerical tight-binding calculations. 
The first class consists of junctions aligned along the common high-symmetry rotational axis between a monopole superconductor and a superconductor with known spherical harmonic pairing symmetry. 
This junction probes the global angular momentum of the monopole pairing order, as only when the angular momentum components at two sides of the junction are identical does the system exhibit nonzero first-order Josephson coupling. 
The second class consists of junctions between identical monopole superconductors, probing the local pairing symmetry at the momenta conserved at the junction interface. 
Our investigation provides guiding principles for designing phase-sensitive experiments to probe the exotic monopole pairing order and distinguish it from all known superconducting orders with spherical harmonic symmetry.

This paper is organized as follows. 
In Sec.~\ref{sec:JC}, we discuss the first-order Josephson coupling between two uniform superconductors based on linear response theory, initially considering spin-orbit interactions at the interface and then extending to spin-orbit coupling in the bulk, emphasizing the role of the symmetry of superconducting orders in terms of a scattering form factor in this microscopic formulation.  
In Sec.~\ref{sec:JJ_MSC}, we focus on Josephson junctions involving monopole superconductors, first deriving the form factor discussed in Sec.~\ref{sec:JC} based on linear response theory, and then analyzing the symmetry and topological principles which are independent of microscopic tunneling processes.
We propose designing Josephson junctions aligned along or perpendicular to the high-symmetry rotational axis of the monopole superconductor to reveal the shift between the global and local angular momentum of the pairing order and extract the pair monopole charge. 
In Sec.~\ref{sec:JG}, we design Josephson junctions to extract the global angular momentum of the monopole superconducting order, presenting results from both linear response theory and numerical results of a tight-binding model. 
Sec.~\ref{sec:JL} explores the design of Josephson junctions to probe the local angular momentum component by considering junctions between identical monopole superconductors. Finally, Sec.~\ref{sec:distinguish_from_chiral_SCs} addresses the presence of spin-orbit interactions at the junction interface, proposing additional junction designs to distinguish monopole superconducting order from chiral spherical harmonic pairing orders.

\section{First-order Josephson coupling in the presence of spin-orbit coupling}
\label{sec:JC}

In this section, we employ linear response theory to derive the first-order Josephson coupling~\cite{Ambegaokar1963, Sigrist1991} in a superconductor-insulator-superconductor (SIS) junction in the presence of spin-orbit coupling in the bulk of the superconductors. 
Initially, we examine the scenario where only spin-orbit interactions occur at the junction interface, assuming the absence of spin-orbit coupling in the bulk. 
We emphasize the role of a form factor, which determines the symmetry selection rules for the first-order Josephson current. 
Then, we additionally consider spin-orbit coupling in the bulk of superconductors. 
In the weak-coupling regime, by projecting to states near Fermi surfaces, we demonstrate that the bulk spin-orbit coupling can be treated as an effective interface spin-orbit interaction. 
Using this formalism, we can obtain the first-order Josephson current in the presence of spin-orbit interactions in the bulk superconductors or at the interface.

We restrict our consideration to the Josephson effects between two uniform superconductors separated by a thin insulating barrier, as described by the following Hamiltonian, 
\begin{equation}
    H_\mathrm{JJ} = H_{\mathrm{BdG},L} + H_{\mathrm{BdG},R} + H_{\mathrm{link}}.
    \label{BdG_JJ}
\end{equation}
Here, we use subscripts $L$ and $R$ to denote the two sides of a junction, irrespective of its spatial orientation. 
The Bogoliubov-de-Gennes (BdG) Hamiltonians $H_{\mathrm{BdG},\alpha=L,R}$ of the superconductors on opposite sides of the junction generally take the form 
\begin{equation}
    \begin{aligned}
        &H_{\mathrm{BdG},\alpha} = 
        \\
        &{\sum_{\mathbf{k}}}'\Psi^\dagger_{\alpha}  (\mathbf{k}) 
        \left(
        \begin{array}{cc}
            \mathcal{H}_{\mathrm{kin}, \alpha}(\mathbf{k})  & \Delta_{\alpha}(\mathbf{k})e^{i\phi_{\alpha}}
            \\
            \Delta^\dagger_{\alpha}(\mathbf{k})e^{-i\phi_{\alpha}}
            & -\mathcal{H}^\mathrm{T}_{\mathrm{kin},\alpha}(-\mathbf{k})
        \end{array}
        \right)  
        \Psi_{\alpha}  (\mathbf{k}),
        \label{BdG_hamiltonian_general}
    \end{aligned}
\end{equation}
where ${\sum_{\mathbf{k}}}'$ denotes the sum over half the Brillouin zone and $\Psi_{\alpha}(\mathbf{k})=(c_{\alpha, \mathbf{k}, \uparrow}, c_{\alpha, \mathbf{k},\downarrow}, c_{\alpha, -\mathbf{k}, \uparrow}^\dagger, c_{\alpha, -\mathbf{k}, \downarrow}^\dagger)^{\mathrm{T}}$ denotes the annihilation operator of Nambu spinor. 
Here, $c_{\alpha= L(R), \mathbf{k}, \sigma}$ annihilates a single-particle state with momentum $\mathbf{k}$
and spin $\sigma =\uparrow,\downarrow$ at the left (right) side of the junction. 
$\mathcal{H}_{\mathrm{kin},\alpha}$ is the kinetic Hamiltonian kernel defined in the particle-hole channel, and $\Delta_\alpha e^{i\phi_{\alpha}}$ is the superconducting pairing matrix defined in the particle-particle channel with the overall $\mathrm{U}(1)$ pairing phase $e^{i\phi_{\alpha}}$ separated out here. 
$H_{\mathrm{link}}$ describes the single particle tunnelings across the insulating junction link and takes the form of the Bardeen-Josephson Hamiltonian~\cite{Bardeen1961, Cohen1962, Josephson1962} as follows,
\begin{equation}
    H_{\mathrm{link}} 
    = \sum_{\mathbf{k},\mathbf{k}'} 
    \sum_{\sigma, \sigma'=\uparrow,\downarrow} 
    T_{\sigma, \sigma'}(\mathbf{k}, \mathbf{k}') c^\dagger_{R, \mathbf{k}, \sigma} c_{L, \mathbf{k}', \sigma'} + \mathrm{h.c.}
    \label{tunneling_hamiltonian.spin_basis}
\end{equation}
The tunneling amplitudes $T_{\sigma, \sigma'}(\mathbf{k}, \mathbf{k}')$ form a $2 \times 2$ matrix $T(\mathbf{k}, \mathbf{k}')$ which can be decomposed into spin-independent and spin-dependent tunneling amplitudes as $T(\mathbf{k}, \mathbf{k}') = T_0(\mathbf{k}, \mathbf{k}') \sigma_0 + {\mathbf{T}(\mathbf{k}, \mathbf{k}') \cdot \boldsymbol{\sigma}}$, where  $\sigma_0$ and  $\sigma_{i=x,y,z}$ are the identity and Pauli matrices in spin space. 
$\mathbf{T}(\mathbf{k}, \mathbf{k}')$, the spin-dependent tunneling amplitude, results from the spin-orbit interactions at the junction interface. 
The results are also generalizable to systems with additional internal degrees of freedom, like valley or orbital.

We now derive the first-order Josephson current in this SIS junction at zero bias, where the dissipationless DC Josephson current running in thermal equilibrium is the only contributor to the tunneling current through the junction. 
Using linear response theory, 
to first order in the perturbation $H_\mathrm{link}$ in the expansion of the thermal weight $e^{- \beta H_{\mathrm{link}}}$ with $\beta=1/(k_B T)$, 
the DC Josephson current is given by
\begin{align}
    I_J (\phi) 
    = 
    \mathrm{lim}_{\delta \rightarrow 0^+}
     &\frac{2e}{\hbar} 
     \mathrm{Im} \Big\{
    \frac{1}{\beta}\sum_{ip_n} \sum_{\mathbf{k}, \mathbf{k}'}
     \mathrm{Tr}[
    \mathcal{F}_L(\mathbf{k}, ip_n - i \delta)
    \nonumber \\
    & T^\mathrm{T}(-\mathbf{k}, -\mathbf{k}')
    \mathcal{F}^{*}_R(\mathbf{k}, ip_n)
    T(\mathbf{k}, \mathbf{k}')
    ] 
    \Big\},
    \label{Josephson_current}
\end{align}
where 
$\phi = \phi_R - \phi_L$ is the overall U($1$) phase difference of superconducting pairing across the junction, the trace is taken over spin space, and $p_n=(2n+1)\pi/\beta $ with $n\in \mathbb{Z}$ are fermion Matsubara frequencies. 
$\mathcal{F}_{\alpha}(\mathbf{k}, ip_n)$ is the anomalous Green's function of the unperturbed BdG Hamiltonian $H_{\mathrm{BdG},\alpha}$ in Eq.~\eqref{BdG_hamiltonian_general}. 
Generally, the total first-order Josephson current can be expressed as 
\begin{equation}
    I_J(\phi) = 
    I_c \sin(\phi),
\end{equation}
where $I_c$ 
is the Josephson critical current at zero bias.

When the band Hamiltonian $\mathcal{H}_{\mathrm{kin}, \alpha}$ 
has no spin dependence
and when the pairing is unitary, $\Delta^\dagger \Delta \propto \sigma_0$, 
the anomalous Green's function $\mathcal{F}_\alpha(\mathbf{k}, ip_n)$ can be simplified to $- e^{i \phi_\alpha} \Delta_\alpha(\mathbf{k})/(p_n^2 + E^2_{\alpha, \mathbf{k}})$.  
Here, $E_{\alpha, \mathbf{k}} = \sqrt{\xi^2_{\alpha, \mathbf{k}}+ \mathrm{Tr}(\Delta_\alpha^\dagger(\mathbf{k})\Delta_\alpha(\mathbf{k}))/2}$ is the dispersion of the unperturbed BdG quasiparticle on the $\alpha$ side of the junction with $\xi_{\alpha, \mathbf{k}}$ being the spin-degenerate band dispersion. 
The form of $\mathcal{F}(\mathbf{k}, ip_n)$ for more general pairings, including non-unitary spin triplet pairing, is shown in Appendix~\ref{appendix.subec:josephson_coupling}.

The trace part of Eq.~\eqref{Josephson_current} can be regarded as the following product of three distinct factors, 
\begin{align}
    \frac{1}{\beta} \sum_{ip_n} \mathrm{Tr}[\cdots]
    =e^{-i\phi}
    w\left(E_{L, \mathbf{k}'}, E_{R, \mathbf{k}}; \beta \right) 
    \mathfrak{F}(\mathbf{k}, \mathbf{k}').
    \label{correlation_fnc.simplified}
\end{align}
Here, the first term  $e^{-i\phi}$ gives rise to $2\pi$-periodicity with respect to the pairing phase difference $\phi$ in the Josephson current $I_J(\phi)$. 
This reflects the first-order process corresponding to a single Cooper pair tunneling through the junction barrier. 
The second term $w(E_{L;\mathbf{k}'}, E_{R,\mathbf{k}}; \beta)
\equiv \frac{1}{\beta} \sum_{ip_n}
(p_n^2 + E^2_{L, \mathbf{k}'})^{-1}
(p_n^2 + E^2_{R, \mathbf{k}})^{-1}$ 
includes the temperature dependence and contribution from
the unperturbed BdG quasiparticle energy dispersions, $E_{\alpha, \mathbf{k}}$.  
After frequency summation, it becomes 
\begin{eqnarray}
        &&w(E_{L;\mathbf{k}'}, E_{R,\mathbf{k}}; \beta)\nonumber \\
        & = &
        \frac{1}{4E_{L,\mathbf{k}'} E_{R,\mathbf{k}}} \mathrm{lim}_{\omega \rightarrow 0^+}[
        (
        n_F(E_{L,\mathbf{k}'})- n_F(E_{R,\mathbf{k}})
        ) \times
        \nonumber \\
        && 
        (
        \frac{1}{E_{L, \mathbf{k}'} - E_{R, \mathbf{k}} - i\omega}
        +
        \frac{1}{E_{L, \mathbf{k}'} - E_{R, \mathbf{k}} + i\omega}
        )
        \nonumber \\
        &&+
        (1-n_F(E_{L, \mathbf{k}'}) - n_F(E_{R, \mathbf{k}})) \times \nonumber \\
        && 
        (
        \frac{1}{E_{L, \mathbf{k}'} + E_{R, \mathbf{k}} + i\omega}
        +
        \frac{1}{E_{L, \mathbf{k}'} + E_{R, \mathbf{k}} - i\omega}
        )
        ],
    \label{f_fnc.full.maintext}
\end{eqnarray}
where $n_F$ is the Fermi-Dirac distribution.

The last and the most essential term is the form factor $\mathfrak{F}(\mathbf{k}, \mathbf{k}')$,  defined as 
\begin{equation}
    \mathfrak{F}(\mathbf{k}, \mathbf{k}')
    \equiv
    \mathrm{Tr}
    \Big[
    {\Delta_{L}(\mathbf{k}')}
    {T^\mathrm{T}(-\mathbf{k},-\mathbf{k}')}
    {\Delta_{R}^*(\mathbf{k})}
    {T(\mathbf{k},\mathbf{k}')}
    \Big].
    \label{form_factor}
\end{equation} 
Though the magnitude of the form factor depends on the microscopic details of tunneling at the junction interface via $T(\mathbf{k},\mathbf{k}')$, whether it vanishes or not is sensitive to the superconducting orders at the two sides of the junction, $\Delta_{L}$ and  $\Delta_{R}$, as well as the symmetry of the junction geometry. 
Therefore, the form factor $\mathfrak{F}(\mathbf{k}, \mathbf{k}')$ gives rise to symmetry selection rules which determine whether the first-order Josephson current vanishes.  

To demonstrate the symmetry selection rule provided by the form factor, consider a Josephson junction between a spin singlet and a spin triplet superconductor without spin-orbit coupling in the bulk superconductors but with spin-dependent tunneling at the junction interface. 
The pairing matrices of superconductors at two sides of the junction generally take the form $\Delta_{L}(\mathbf{k}) = d_{L, 0}(\mathbf{k}) i \sigma_y$ and $\Delta_{R}(\mathbf{k}) = (\mathbf{d}_{R}(\mathbf{k})\cdot \boldsymbol{\sigma})i\sigma_y$, with $d_{L, 0}(\mathbf{k})$ the spatial part of the singlet pairing order and $\mathbf{d}_{R}(\mathbf{k})$ the $d$-vector of the triplet pairing order~\cite{Balian1963, Leggett1975}.
Employing trace identities of Pauli matrices, we simplify the form factor for this singlet-triplet junction, denoted as $\mathfrak{F}_{\mathrm{sing}-\mathrm{trip}}$, to the following form,
\begin{eqnarray}
       \mathfrak{F}_{\mathrm{sing}-\mathrm{trip}}
       (\mathbf{k}, \mathbf{k}')
        &=&
        2d_{L,0}(\mathbf{k}') \{
        T_0(-\mathbf{k}, -\mathbf{k}')
        \mathbf{d}^*_{R}(\mathbf{k}) \cdot \mathbf{T}(\mathbf{k}, \mathbf{k}') \nonumber \\
        && -i \mathbf{d}^*_{R}(\mathbf{k}) \cdot 
        [
        \mathbf{T}(\mathbf{k}, \mathbf{k}')
        \times
        \mathbf{T}(-\mathbf{k}, -\mathbf{k}')
        ] \nonumber \\
        && -T_0(\mathbf{k}, \mathbf{k}')
        \mathbf{d}^*_{R}(\mathbf{k})\cdot
        \mathbf{T}(-\mathbf{k}, -\mathbf{k}')  \}.
    \label{form_factor.triplet_singlet}
\end{eqnarray}
Consequently, in the absence of spin-orbit interaction at junction interface, $\mathbf{T}(\mathbf{k}, \mathbf{k}')=0$, the form factor vanishes and forbids first-order Josephson coupling. 
This reflects the symmetry principle that when pairing orders are in different total spin channels, the first-order Josephson coupling vanishes in the system where spin and orbital angular momentum are separately conserved.

Further consider Rashba-type spin-orbit interaction at the junction interface, as will be used later in Sec.~\ref{sec:distinguish_from_chiral_SCs},  which has the form $\mathbf{T}(\mathbf{k}, \mathbf{k}') = T_\mathrm{SO} (\hat{\mathbf{n}} \times \hat{\mathbf{k}})
\delta_{\mathbf{k}, \mathbf{k}'}$,  with $\hat{\mathbf{n}}$ being the junction orientation. 
The form factor reduces to
\begin{equation}
    \begin{aligned}
        \mathfrak{F}_{\mathrm{sing}-\mathrm{trip}}
        (\mathbf{k}, \mathbf{k}')
        =
        4
        T_0
        T_\mathrm{SO}
        d_{L,0}(\mathbf{k}')
        [
        \hat{\mathbf{n}} \cdot (\hat{\mathbf{k}} \times \mathbf{d}_R^* (\mathbf{k}) )
        ]
        \delta_{\mathbf{k}, \mathbf{k}'},   
    \end{aligned}
    \label{single_triplet_form_factor}
\end{equation}
where we take $T_0(\mathbf{k}, \mathbf{k}')=T_0$, a constant, for simplicity. 
Together with Eq.~\eqref{Josephson_current}, the first-order Josephson current takes the form, $I_J = (8e/\hbar) T_0 T_{\mathrm{SO}} \mathrm{Im} [e^{-i\phi} \sum_{\mathbf{k}} w(E_{L, \mathbf{k}}, E_{R, \mathbf{k}}; \beta) d_{L,0}(\mathbf{k}) \hat{\mathbf{n}} \cdot (\hat{\mathbf{k}} \times \mathbf{d}_R^*(\mathbf{k}))]$, which agrees with the result obtained by Geshkenbein and Larkin~\cite{Geshkenbein1986}. 
Then, when  $d_{L,0}(\mathbf{k}') = \Delta_{L,0}$ for an $s$-wave superconductor, $\mathbf{d}_R(\mathbf{k}) = \Delta_{R,0}(0, 0, (k_x + ik_y)/k_F)$ for a chiral $p$-wave triplet superconductor with total spin $s_z=0$, and the junction is oriented along $\hat{\mathbf{n}}=\hat{z}\parallel \mathbf{d}_R$, the form factor in Eq.~\eqref{single_triplet_form_factor} vanishes, forbidding first-order Josephson coupling even though there is nonvanishing Rashba spin-orbit interaction.
This result can be understood from the following spin symmetry selection rules.
In the spin-independent tunneling channel, as the singlet and triplet superconductors have different total spins, their first-order Josephson coupling is zero. 
In the spin-dependent tunneling channel, the junction geometry determines the form of Rashba spin-orbit interaction as consisting of only in-plane spins which change $s_z$, the total spin-$z$ component of the Cooper pair, by $1$. 
However, $s_z$ of the pairing orders at two sides of the junction are identical, $s_z = 0$. 
Hence, the first-order Josephson coupling in this example vanishes.

In this work, to take into account spin-orbit couplings in bulk topological superconductors,
we introduce a convenient formalism to treat the spin-orbit coupling in terms of effective spin-dependent tunneling at the junction barrier. 
We consider the tunneling of band eigenstates through the junction, where $H_\mathrm{link}$ in Eq.~\eqref{tunneling_hamiltonian.spin_basis} now takes the form
\begin{equation}
    H_\mathrm{link} = \sum_{\mathbf{k}, s; \mathbf{k}', s'}
    T_{s, s'}^{(b)}(\mathbf{k}, \mathbf{k}') \psi^\dagger_{R, \mathbf{k}, s} \psi_{L, \mathbf{k}', s'} + \mathrm{h.c.}
    \label{link.band_basis}
\end{equation}
Here, $\psi_{\alpha, \mathbf{k}, s}$ with $\alpha =L(R)$ annihilates a band eigenstate with band index $s=\pm$ and momentum $\mathbf{k}$ at the left (right) of the junction.  
For a generic two-band Hamiltonian,  $\psi_{\alpha, \mathbf{k}, s}$ is related to the annihilation operators in spin basis, $c_{\alpha, \mathbf{k}, \sigma}$, by a unitary transformation, 
$\psi_{\alpha, \mathbf{k}, s} = \sum_{\sigma=\uparrow, \downarrow} 
[U^\dagger_{\alpha}(\mathbf{k})]_{s, \sigma} 
c_{\alpha, \mathbf{k}, \sigma}$, where $U_{\alpha} (\mathbf{k})$ encodes the spin-orbit coupling in band Hamiltonian $\mathcal{H}_{\mathrm{kin}, \alpha}(\mathbf{k})$. 
It is convenient to define the complex four-vector $(u_{\alpha, 0}(\mathbf{k}), \mathbf{u}_\alpha (\mathbf{k}))$, such that
$U_{\alpha} (\mathbf{k}) = u_{\alpha,0} (\mathbf{k}) \sigma_0 + \mathbf{u}_{\alpha}(\mathbf{k}) \cdot \boldsymbol{\sigma}$, with $u_{\alpha,i=0, \cdots, 3} (\mathbf{k})$ being complex functions. 
Correspondingly, the tunneling matrix defined in the band representation, denoted by the superscript ``$b$'', is given by $T^{(b)} (\mathbf{k}, \mathbf{k}') =  U^\dagger_R(\mathbf{k}) T(\mathbf{k}, \mathbf{k}') U_L(\mathbf{k}')$ with $T(\mathbf{k}, \mathbf{k}')$ defined below Eq.~\eqref{tunneling_hamiltonian.spin_basis} in the spin basis. 
Therefore, even in the absence of bare spin-orbit interaction at the junction interface, $\mathbf{T}(\mathbf{k}, \mathbf{k}') = 0$, the effective tunneling in the band eigenbasis takes the following form, 
\begin{equation}
    T^{(b)}(\mathbf{k}, \mathbf{k}') = T^{(b)}_0(\mathbf{k}, \mathbf{k}') + \mathbf{T}^{(b)}(\mathbf{k}, \mathbf{k}')\cdot \boldsymbol{\sigma},
\end{equation}
where $T^{(b)}_0(\mathbf{k}, \mathbf{k}') = T_0(\mathbf{k}, \mathbf{k}') 
(u^*_{R,0} (\mathbf{k})
u_{L,0} (\mathbf{k}')
+
\mathbf{u}^*_{R}(\mathbf{k}) \cdot \mathbf{u}_L(\mathbf{k}')
)$
and
$\mathbf{T}^{(b)}(\mathbf{k}, \mathbf{k}') = T_0(\mathbf{k}, \mathbf{k}') 
(
u^*_{R,0}(\mathbf{k}) \mathbf{u}_L(\mathbf{k}')
+
\mathbf{u}^*_{R}(\mathbf{k}) {u}_{L,0}(\mathbf{k}')
+
i \mathbf{u}^*_{R}(\mathbf{k}) \times \mathbf{u}_L(\mathbf{k}')
)$. 
For most parts of the manuscript, we consider the case of zero bare interface spin-orbit interaction for simplicity.
We discuss the results when the bare interface spin-orbit coupling $\mathbf{T}(\mathbf{k}, \mathbf{k}') \neq 0$ in Sec.~\ref{sec:distinguish_from_chiral_SCs}.

Furthermore, in a weak-coupling superconductor, we can project its pairing order to low-energy states near the Fermi surfaces participating in Cooper pairing. 
For pairing order already expressed in the band representation $\Delta_\alpha^{(b)}(\mathbf{k}) =  U_\alpha^\dagger(\mathbf{k}) \Delta_\alpha(\mathbf{k})  U^*_\alpha(-\mathbf{k})$, 
we capture the key ingredients of the low-energy pairing by focusing on the Fermi surface projected pairing order in this representation, defining $\Delta^{(bp)}_\alpha(\mathbf{k}) \equiv P_+^{(b)} \Delta_\alpha^{(b)}(\mathbf{k})P_+^{(b)}$, where $P_+^{(b)} = \mathrm{diag}(1,0)$ is the projection operator in band representation. 
Here, without loss of generality, we have ordered the band basis so that the states at the Fermi surface of interest correspond to the first eigenvalues in the particle and hole parts (see Appendix~\ref{appendix:effective_pairing_channels}). 
The superscript ``$bp$'' denotes the band diagonal basis with bands projected to the Fermi surfaces of interest. 
The anomalous Green's function of the corresponding BdG Hamiltonian in this band projected basis takes a simple form, $\mathcal{F}^{(bp)}_\alpha(\mathbf{k}, ip_n) \approx - e^{i \phi_\alpha} \Delta_\alpha^{(bp)}(\mathbf{k})/(p_n^2 + E^{(bp)2}_{\alpha, \mathbf{k}})$. 
This form is analogous to the simplified form of $\mathcal{F}_\alpha(\mathbf{k}, ip_n)$ in the case of spin degenerate band and unitary pairing. However, here, the energy of the BdG quasiparticle is given by $E^{(bp)}_{\alpha, \mathbf{k}} = \sqrt{\xi_{\alpha, \mathbf{k}}^{(bp) 2} + \mathrm{Tr}(\Delta^{(bp) \dagger}_{\alpha}(\mathbf{k}) \Delta^{(bp)}_{\alpha}(\mathbf{k}))}$, where $\xi^{(bp)}_{\alpha, \mathbf{k}}$ is the dispersion of the band that forms the Fermi surface. 
The expression of the form factor in Eq.~\eqref{form_factor} now reduces to $\mathfrak{F}(\mathbf{k}, \mathbf{k}') = \mathrm{Tr}[\Delta_L^{(bp)}(\mathbf{k}')T^{(b)\mathrm{T}}(-\mathbf{k}, -\mathbf{k}')\Delta_R^{(bp)*}(\mathbf{k}) T^{(b)}(\mathbf{k}, \mathbf{k}')]$, where the pairing and tunneling matrices are in the projected band basis.
We use this band projected form factor to analyze Josephson junctions of monopole superconductors, as demonstrated in the following section.

\section{Josephson coupling in junctions of monopole superconductors}
\label{sec:JJ_MSC}
Having discussed the form of the first-order Josephson current in a general SIS junction, our focus turns to junctions that include monopole superconductors. 
This discussion unfolds in two different approaches. 
We first derive the form factor that determines the Josephson current using the microscopic derivation based on linear response theory, as discussed in Sec.~\ref{sec:JC}.
We work in the weak-coupling regime, projecting the pairing order to low-energy states that comprise the Fermi surface.
Secondly, we supplement the first approach with a discussion based on the underlying symmetry and topological principles, which are essential in designing a set of Josephson junctions to probe the monopole superconducting order.

\subsection{Microscopic derivation of Josephson coupling for monopole superconductors}
\label{subsec:MSC.microscopic}

We first extend the results from linear response theory in Sec.~\ref{sec:JC} to determine the first-order Josephson current for junctions that include monopole superconductors. 
Monopole pairing order is an exotic class of topologically obstructed pairing order that arises from the nontrivial Berry phase of Cooper pairs. 
This, for example, can arise when pairing occurs between two Fermi surfaces $\mathrm{FS}_1$ and $\mathrm{FS}_2$ with different Chern numbers, $\mathcal{C}_1$ and $\mathcal{C}_2$ respectively, in doped time-reversal symmetry broken Weyl semimetals in three dimensions~\cite{Li2018}. 
In the weak-coupling regime, we describe this system using mean-field BdG Hamiltonian in Eq.~\eqref{BdG_hamiltonian_general},
with its Nambu spinor modified to $\Psi(\mathbf{k})=(c_{1;\mathbf{k}, \uparrow}, c_{1; \mathbf{k},\downarrow}, c_{2;-\mathbf{k}, \uparrow}^\dagger, c_{2;-\mathbf{k}, \downarrow}^\dagger)^{\mathrm{T}}$ to describe the inter-Fermi surface pairing $\Delta_\mathrm{inter}(\mathbf{k})$ between  parity-related $\mathrm{FS}_1$ and $\mathrm{FS}_2$, where $c_{i;\mathbf{k}, \sigma}$ denotes the annihilation operator of an electron on $\mathrm{FS}_i$ ($i=1,2$) with momentum $\mathbf{k}$ and spin $\sigma = \uparrow,\downarrow$.
Correspondingly, the BdG Hamiltonian kernel takes the following form,
\begin{equation}
    \mathcal{H}_\mathrm{BdG} (\mathbf{k})    =
    \left(
    \begin{array}{cc}
         \mathcal{H}_{\mathrm{kin},1}(\mathbf{k})         & 
         \Delta_\mathrm{inter}(\mathbf{k})
         \\
         \Delta_\mathrm{inter}^\dagger(\mathbf{k})
         & 
         -\mathcal{H}_{\mathrm{kin},2}^\mathrm{T} (-\mathbf{k})    \end{array}
    \right).
    \label{MSC.general}
\end{equation}
Here, we suppress the subscript $\alpha = L,R$ when only discussing the bulk superconductor residing at one side of the junction in this and later sections.
 
We use the following two-band models to describe the two topological Fermi surfaces with nontrivial Chern numbers $\mathcal{C}_1$ and $\mathcal{C}_2$:
\begin{subequations}
    \begin{align}
        \mathcal{H}_{\mathrm{kin},1} (\mathbf{k}=\mathbf{K}_0 + \tilde{\mathbf{k}}) &=\mathbf{h}_1(\tilde{\mathbf{k}}) \cdot \boldsymbol{\sigma} 
        -\mu_1,    
        \\
        \mathcal{H}_{\mathrm{kin},2} (\mathbf{k}=-\mathbf{K}_0 + \tilde{\mathbf{k}}) &=\mathbf{h}_2(\tilde{\mathbf{k}}) \cdot \boldsymbol{\sigma}-\mu_2,
                                    \end{align}
    \label{kinetic_ham.general}\end{subequations}
where $\tilde{\mathbf{k}}$ is the momentum relative to the corresponding Weyl nodes at $\pm \mathbf{K}_0$ and $\mathbf{h}_i(\tilde{\mathbf{k}})$ denotes the pseudo ``magnetic" field in momentum space which determines the chirality of Weyl node at $\mathbf{K}_0$ ($-\mathbf{K}_0$) for $i=1$ ($i=2$). 
There are helical eigenstates  $\chi_{i,s} (\tilde{\mathbf{k}})$ of $\mathcal{H}_{\mathrm{kin},i}$
where
$i=1,2$ refers to the valley index and $s = \pm$ denotes the band index. 
Here, we use $s=+$ to denote the band at the Fermi level. 
The Fermi surfaces have Chern numbers
$\mathcal{C}_{i} = {(1/2\pi) \oint_{\mathrm{FS}_i}\boldsymbol{\Omega}_i(\tilde{\mathbf{k}}) \cdot d\mathbf{S}_{\tilde{\mathbf{k}}}}$, where $\boldsymbol{\Omega}_i(\tilde{\mathbf{k}}) = \boldsymbol{\nabla}_{\tilde{\mathbf{k}}} \times \langle \chi_{i, +}(\tilde{\mathbf{k}})| i \boldsymbol{\nabla}_{\tilde{\mathbf{k}}}|\chi_{i,+}(\tilde{\mathbf{k}}) \rangle$ is the corresponding single particle Berry curvature.

The inter-Fermi surface pairing,
$\Delta_{\mathrm{inter}; \sigma, \sigma'} (\mathbf{k}) 
\sim \langle c_{2; -\mathbf{k},\sigma'}c_{1; \mathbf{k},\sigma} \rangle$ in the spin-$\uparrow, \downarrow$ representation, describes the zero center-of-mass momentum pairing, since FS$_1$ and FS$_2$ are related by parity. 
In the weak-coupling regime, superconducting pairing occurs at low-energy close to the Fermi surfaces;
hence, we express the pairing in the helical band representation as
$\Delta_{\mathrm{inter}; s,s'}^{(b)} (\tilde{\mathbf{k}}) \sim
\langle
\psi_{2; -\mathbf{\tilde{k}}, s'} \psi_{1; \mathbf{\tilde{k}}, s}
\rangle
$, where
$\psi_{1; \mathbf{\tilde{k}}, s} 
= \sum_{\sigma=\uparrow, \downarrow} \chi_{1;s,\sigma}^*(\tilde{\mathbf{k}}) c_{1; \tilde{\mathbf{k}}+ \mathbf{K}_0, \sigma}$ 
and
$\psi_{2; \mathbf{\tilde{k}},s} 
= \sum_{\sigma=\uparrow, \downarrow} \chi_{2;s,\sigma}^*(\tilde{\mathbf{k}}) c_{2; \tilde{\mathbf{k}}- \mathbf{K}_0, \sigma}$ 
are annihilation operators of helical band eigenstates 
of $\mathcal{H}_{\mathrm{kin}, 1}$ and $\mathcal{H}_{\mathrm{kin},2}$ 
respectively.
When the Chern numbers of two Fermi surfaces are different, the inter-Fermi surface pairing inherits the band topology of single-particle states at $\mathbf{k}$ and $-\mathbf{k}$ in a nontrivial way, and
the Cooper pair scatters in the presence of a magnetic monopole in momentum space given by the pair monopole charge, $q_p = (\mathcal{C}_1 - \mathcal{C}_2)/2$.
Consequently, this leads to a topologically obstructed superconducting pairing order that cannot be well defined over an entire Fermi surface. Instead, the pairing order is described by monopole harmonics~\cite{Wu1976, Wu1977} that all belong to the topological sector characterized by the pair monopole charge $q_p$~\cite{Li2018}.

Upon projecting to the helical Fermi surfaces in the weak-coupling regime, the effective pairing in the helical band-diagonal Nambu basis $(\psi_{1;\tilde{\mathbf{k}},+}, \psi^\dagger_{2;-\tilde{\mathbf{k}},+})^\mathrm{T}$ 
is
\begin{equation}
            \Delta^{(bp)}_{\mathrm{inter}}(\tilde{\mathbf{k}}) 
    =
    \left(
    \begin{array}{cc}
         \Delta_{\mathrm{MSC}}^{(q_p)}(\tilde{\mathbf{k}}) &0
         \\
         0 & 0
    \end{array}
    \right),
    \label{effective_pairing.band_basis}
\end{equation}
where
\begin{equation}
    {\Delta}_\mathrm{MSC}^{(q_p)}(\tilde{\mathbf{k}}) 
            = \left\langle
    \chi_{1,+}(\tilde{\mathbf{k}})
    \middle|
    \Delta_{\mathrm{inter}}(\mathbf{k})
    \middle|
    \chi_{2,+}^*(-\mathbf{\tilde{k}})
    \right\rangle.
    \label{MSC.effective_gap_function.weak_coupling}
\end{equation}
In the pseudospin space formed by the helical band eigenbasis, we define an effective $d$-vector for the monopole pairing order as
\begin{equation}
    \mathbf{d}^{(bp)}_{\mathrm{inter}}(\tilde{\mathbf{k}}) = - \frac{1}{2} \Delta_\mathrm{MSC}^{(q_p)}(\tilde{\mathbf{k}})
    (1, i, 0)^\mathrm{T},
    \label{msc_d-vector}
\end{equation}
with $\Delta_\mathrm{inter}^{(bp)}(\tilde{\mathbf{k}}) \equiv \mathbf{d}^{(bp)}_\mathrm{inter}(\tilde{\mathbf{k}})\cdot \boldsymbol{\sigma} (i \sigma_y)$.
In the following sections, we will use the effective pairing order and the symmetry principles encoded in the 
form factor, $\mathfrak{F}(\mathbf{k}, \mathbf{k}')$, to extract the symmetry and topology of the monopole superconducting order.

To note, if each Fermi surface of the doped Weyl system has additional symmetry with respect to its enclosed Weyl point, 
intra-Fermi surface pairing is also possible, corresponding to nonzero center-of-mass momentum Fulde-Ferrell-Larkin-Ovchinnikov (FFLO) pairing~\cite{Fulde1964, Larkin1964}.
As the pair monopole charge of the FFLO pairing is zero, the pairing can be fully gapped and compete with the inter-Fermi-surface pairing~\cite{Li2018}. 
However, without the additional symmetry of Fermi surfaces, the inter-Fermi-surface monopole pairing is energetically preferable to the intra-Fermi-surface FFLO pairing. 
Hence, we only consider zero center-of-mass pairing in this work.

To describe a monopole superconductor with pair monopole charge $q_p=-1$, we consider the pairing between Fermi surfaces of Chern numbers $\mathcal{C}_1=-1$ and $\mathcal{C}_2=1$. 
The Hamiltonian in Eq.~\eqref{MSC.general} takes a specific form below, with 
$\mathbf{h}^{(\nu= 1)}_1(\tilde{\mathbf{k}})
=
\hbar v_F (\tilde{k}_x, \tilde{k}_y, \tilde{k}_z)$ and $\mathbf{h}^{(\nu= -1)}_2(\tilde{\mathbf{k}})
=
\hbar v_F (\tilde{k}_x, \tilde{k}_y, -\tilde{k}_z)$
defined near the Weyl nodes at $\pm \mathbf{K}_0 = (0,0, \pm K_{0,z})$ respectively,
\begin{equation}
    \begin{aligned}&\mathcal{H}_\mathrm{BdG}^{(q_p = -1)}(\mathbf{k})
        =
        \left(
        \begin{array}{cc}
            \mathcal{H}_{\mathrm{kin}}^{(\nu= 1)}({\mathbf{k}})
                          &
             \Delta_\mathrm{inter}(\mathbf{k})
                          \\
             \Delta_\mathrm{inter}^\dagger(\mathbf{k})
                          &
                          -[\mathcal{H}_{\mathrm{kin}}^{(\nu= -1)}(-\mathbf{k})]^{\mathrm{T}}
        \end{array}
        \right).
    \end{aligned}\label{qp=-1.MSC.hamiltonian}
\end{equation}
Here, the band Hamiltonians are given by
\begin{subequations}
    \begin{align}
        \mathcal{H}_{\mathrm{kin}}^{(\nu= 1)}(
        \mathbf{k}
                )
        &=
        \mathbf{h}^{(\nu= 1)}_1(\tilde{\mathbf{k}}) \cdot \boldsymbol{\sigma} - \mu,
        \\
        \mathcal{H}_{\mathrm{kin}}^{(\nu= -1)}(
        \mathbf{k}
                )
        &=
        \mathbf{h}^{(\nu= -1)}_2(\tilde{\mathbf{k}})
        \cdot \boldsymbol{\sigma}
        -\mu,
    \end{align}
    \label{continuum_models.c=1}\end{subequations}
with the superscript $\nu$ denoting the chiralities of the Weyl nodes.  $v_F$ is the Fermi velocity, and $\mu>0$ is chosen so that we have disjoint Fermi surfaces which are related by parity via 
$\mathcal{H}^{(\nu= +1)}_1(\mathbf{k}) 
= \sigma_z \mathcal{H}^{(\nu= -1)}_2(-\mathbf{k})\sigma_z$. 
The helical eigenstate of $\mathcal{H}^{{(\nu= 1)}}_{\mathrm{kin}}$ 
in Eq.~\eqref{continuum_models.c=1} describing $\mathrm{FS}_1$ takes the form
\begin{align}
    \chi_{1,+}(\tilde{\mathbf{k}}) 
    &=
    \left( \cos \frac{\theta_{\tilde{k}}}{2} e^{i (\lambda -1)\varphi_{\tilde{k}} /2},
    \sin \frac{\theta_{\tilde{k}}}{2} e^{i (\lambda +1)\varphi_{\tilde{k}} /2}
    \right)
    \\
    \nonumber
    &=
    \sqrt{2\pi} \left( \mathcal{Y}_{\frac{1}{2}; \frac{1}{2}, - \frac{1}{2}}(\boldsymbol{\Omega}_{\tilde{k}}), 
    - \mathcal{Y}_{\frac{1}{2}; \frac{1}{2}, + \frac{1}{2}}(\boldsymbol{\Omega}_{\tilde{k}})\right),
                \end{align}
and is related to states on $\mathrm{FS}_2$ by $\chi_{2,+} (\tilde{\mathbf{k}}) = \sigma_z \chi_{1,+} (-\tilde{\mathbf{k}})$.  
Here,
$\boldsymbol{\Omega}_{\tilde{{k}}} = (\theta_{\tilde{k}}, \varphi_{\tilde{k}})$, with
$\theta_{\tilde{k}}$ and $\varphi_{\tilde{k}}$ being the polar and azimuthal angles of the spherical Fermi surfaces. $\mathcal{Y}_{q;l,l_z}(\boldsymbol{\Omega}_{\tilde{k}})$ is the monopole harmonic function~\cite{Wu1976, Wu1977} (reviewed in Appendix~\ref{appendix:table_of_MHs}),  now defined in momentum space.
The monopole harmonic function is labeled by the monopole charge $q$, the eigenvalue of angular momentum $l$, 
and the eigenvalue of $z$-component angular momentum $l_z$.  
Above, $\lambda = \pm 1$ corresponds to two different choices of gauge. The upper (lower) sign denotes the choice where the state $\chi_{1, +}(\tilde{\mathbf{k}})$ is well-defined locally near the north (south) pole of $\mathrm{FS}_1$, while $\chi_{2, +}(\tilde{\mathbf{k}})$ is well-defined near the south (north) pole of $\mathrm{FS}_2$, with the Dirac strings located at the respective antipodal points of these two Fermi surfaces. 
The Chern numbers of
$\chi_{1,+}$ on $\mathrm{FS}_1$ and $\chi_{2,+}$ on $\mathrm{FS}_2$ are $\mathcal{C}_1 = - 1$ and $\mathcal{C}_2 = + 1$ respectively.
The unitary transformations that diagonalize the Weyl band Hamiltonians $\mathcal{H}_{\mathrm{kin}}^{(\nu= \pm 1)}$ in Eq.~\eqref{continuum_models.c=1}  take the form
\begin{equation}
    U^{(\pm\mathbf{K}_0)}(\tilde{\mathbf{k}})
    =
    \sqrt{2\pi}
    \left(
    \begin{array}{cc}
         \mathcal{Y}_{\pm{\frac{1}{2}}; \frac{1}{2}, 
         - \frac{1}{2}}(\boldsymbol{\Omega}_{\tilde{k}})
         & \mathcal{Y}_{{\mp \frac{1}{2}}; \frac{1}{2}, 
         - \frac{1}{2}}(\boldsymbol{\Omega}_{\tilde{k}})
         \\
         \mp \mathcal{Y}_{\pm{\frac{1}{2}}; \frac{1}{2}, + \frac{1}{2}}(\boldsymbol{\Omega}_{\tilde{k}})
         & \mp \mathcal{Y}_{{\mp\frac{1}{2}}; \frac{1}{2}, + \frac{1}{2}}(\boldsymbol{\Omega}_{\tilde{k}})
    \end{array}
    \right).
    \label{c=1.unitary_transf.MH}
\end{equation}
As introduced below Eq.~\eqref{link.band_basis}, the unitary transformations can be written conveniently in terms of complex four-vectors, where here, the components are given by
\begin{align}
    u_{0}^{(\pm \mathbf{K}_0)} (\pm \tilde{\mathbf{k}}) 
    &= \nonumber
    \sqrt{\frac{\pi}{2}}\left[\mathcal{Y}_{\frac{1}{2}; \frac{1}{2}, - \frac{1}{2}}(\boldsymbol{\Omega}_{\tilde{k}})
    \mp
    \mathcal{Y}^*_{\frac{1}{2}; \frac{1}{2}, - \frac{1}{2}}(\boldsymbol{\Omega}_{\tilde{k}})\right],
    \\
    u_{1}^{(\pm \mathbf{K}_0)} (\pm \tilde{\mathbf{k}}) 
    &= \nonumber
    \sqrt{\frac{\pi}{2}}
    \left[
    \mp \mathcal{Y}_{\frac{1}{2}; \frac{1}{2}, \frac{1}{2}}(\boldsymbol{\Omega}_{\tilde{k}})
    -
    \mathcal{Y}^*_{\frac{1}{2}; \frac{1}{2}, \frac{1}{2}}(\boldsymbol{\Omega}_{\tilde{k}})
    \right],
    \\
    u_{2}^{(\pm \mathbf{K}_0)} (\pm \tilde{\mathbf{k}}) 
    &= \nonumber
    i \sqrt{\frac{\pi}{2}}
    \left[
    \pm \mathcal{Y}_{\frac{1}{2}; \frac{1}{2}, \frac{1}{2}}(\boldsymbol{\Omega}_{\tilde{k}})
    -
    \mathcal{Y}^*_{\frac{1}{2}; \frac{1}{2}, \frac{1}{2}}(\boldsymbol{\Omega}_{\tilde{k}})
    \right],
    \\
    u_{3}^{(\pm \mathbf{K}_0)} (\pm \tilde{\mathbf{k}}) 
    &= 
    \sqrt{\frac{\pi}{2}}
    \left[
    \mathcal{Y}_{\frac{1}{2}; \frac{1}{2}, - \frac{1}{2}}(\boldsymbol{\Omega}_{\tilde{k}})
    \pm
    \mathcal{Y}^*_{\frac{1}{2}; \frac{1}{2}, - \frac{1}{2}}(\boldsymbol{\Omega}_{\tilde{k}})
    \right],
    \label{qp=-1.unitary.four-vector}
\end{align}
where $\mathcal{Y}^*_{q; l ,l_z} = (-1)^{q + l_z} \mathcal{Y}_{-q; l, -l_z}$.

For a general form of inter-Fermi-surface pairing $[\Delta_\mathrm{inter}(\mathbf{k})]_{\sigma, \sigma'}$,
the effective pairing order takes the form
\begin{equation}
    \begin{aligned}
        &\Delta_{\mathrm{MSC}}^{(q_p=-1)}
                                                = 
        \sqrt{\frac{4\pi}{3}} \mathcal{Y}_{-1; 1, 1}(\boldsymbol{\Omega}_{\tilde{\mathbf{k}}}) [\Delta_\mathrm{inter}(\mathbf{k})]_{\uparrow, \uparrow}
        \\
        &
        - \sqrt{\frac{2\pi}{3}} \mathcal{Y}_{-1; 1, 0}(\boldsymbol{\Omega}_{\tilde{\mathbf{k}}}) 
        \Big([\Delta_\mathrm{inter}(\mathbf{k})]_{\uparrow, \downarrow}-[\Delta_\mathrm{inter}(\mathbf{k})]_{\downarrow, \uparrow}\Big).
        \\
        & - 
        \sqrt{\frac{4\pi}{3}} \mathcal{Y}_{-1; 1, +1}(\boldsymbol{\Omega}_{\tilde{\mathbf{k}}}) [\Delta_\mathrm{inter}(\mathbf{k})]_{\downarrow, \downarrow}.
        \label{qp=-1.pairing_order.general_inter_FS}
    \end{aligned}
\end{equation}
Here, 
the spin pairing channels in addition to 
the momentum-dependence of
$\Delta_\mathrm{inter}(\mathbf{k})$ determine the rotational irreps in the decomposition
of $\Delta_\mathrm{MSC}^{(q_p=-1)}$ but do not influence the underlying topology.
When the system has rotational symmetry, $l_z$ is a conserved quantity and further constrains the pairing channels.
Globally, the monopole superconducting order transforms under rotation $R_z$ according to its conserved angular momentum $l_z = l_{z, \mathrm{glob}}$.
However, the pairing order is a singular representation of $l_z$ and thus cannot be well-defined globally over the Fermi surface under a single gauge.
In other words, due to $\mathrm{U}(1)$ obstruction of the pairing phase, the monopole pairing order can no longer be described globally by conventional spherical harmonics.
Though, depending on the choice of gauge, $\lambda = \pm 1$, the $q_p = -1$ monopole pairing can be described locally with spherical harmonic symmetry, with $l_{z, \mathrm{loc}} = l_{z, \mathrm{glob}} + q_p \lambda$ denoting the local angular momentum.
Here, $l_{z, \mathrm{loc}}$ plays an essential role in the underlying topology of the system in momentum space. 
Namely, the shift in angular momentum reflects the nontrivial pair monopole charge 
\begin{math}
    |q_p| = |l_{z, \mathrm{loc}} - l_{z, \mathrm{glob}}|,
\end{math}
which, in addition to the discrepancy between $l_{z, \mathrm{loc}}$ near the gap nodes, is unique to monopole harmonic symmetry.
This shift between global and local angular momentum is a general feature of monopole superconducting order, which we shall later exploit in order to distinguish it from other pairing orders with spherical harmonic symmetry.

As an example, when there is inter-Fermi surface $s$-wave pairing, $\Delta_{\mathrm{inter}} = \Delta_0 i\sigma_y$, the effective low-energy pairing in Eq.~\eqref{qp=-1.pairing_order.general_inter_FS} 
resides in the lowest possible angular momentum $l=1$ channel as follows,
\begin{align}
        {\Delta_\mathrm{MSC}^{(q_p = -1, l_z = 0)}}
        (\tilde{\mathbf{k}}) 
        &= 
        -\Delta_0 \sqrt{\frac{8 \pi}{3}}
        \mathcal{Y}_{-1; 1, 0}
        (\theta_{\tilde{k}}, \varphi_{\tilde{k}})
        \label{q=-1.lz=0.effective_pairing}
        \\
        &= \nonumber
        - 4\pi  \Delta_0 
        \mathcal{Y}_{-\frac{1}{2}; \frac{1}{2}, +\frac{1}{2}} (\boldsymbol{\Omega}_{\tilde{k}})
        \mathcal{Y}_{-\frac{1}{2}; \frac{1}{2}, -\frac{1}{2}} (\boldsymbol{\Omega}_{\tilde{k}})
        \\
        &=  \nonumber
        - \Delta_0 \sin \theta_{\tilde{k}} e^{-i \lambda \varphi_{\tilde{k}}}.
\end{align}
Here, $\mathcal{H}_\mathrm{BdG}^{(q_p = -1)}(\mathbf{k})$ is invariant under rotation $R_z$, and the low-energy pairing order $\Delta_{\mathrm{MSC}}^{(q_p=-1)}$, if it does not break the symmetry spontaneously, is also invariant under $R_z$ and characterized by good quantum number $l_z = l_{z, \mathrm{glob}} = 0$ that labels the trivial irrep.
However, for given choice of gauge $\lambda = \pm 1$, the $\Delta_{\mathrm{MSC}}^{(q_p=-1, l_z=0)}$ pairing order locally has well-defined $p_x - \lambda i p_y$ symmetry with angular momentum $l_{z, \mathrm{loc}} = -\lambda$ near the gap node at the north ($\lambda = +1$) or south ($\lambda = -1$) pole of $\mathrm{FS}_1$.
As $\tilde{k}_z$ is varied between the gap nodes, the local symmetry changes from $p_x-ip_y$ to $p_x+ip_y$,  as evident in the group velocity of the surface modes (see Appendix~\ref{appendix:TBM}).
The shift between the global and local angular momenta is indicative of the pair monopole charge, $q_p=-1$. 
Moreover, this shift is generally true for any form of $\Delta_{\mathrm{inter}}$ and therefore for a general  $\Delta_{\mathrm{MSC}}^{(q_p=-1)}$ in Eq.~\eqref{qp=-1.pairing_order.general_inter_FS}.

We now derive the form factor in the first-order Josephson coupling between a spin singlet superconductor with a spin-independent band Hamiltonian
at the left side and the $q_p = -1$ monopole superconductor at the right side, where there is only bare spin-independent tunneling at the barrier.
In the band-projected basis, the monopole pairing order is described by an effective $d$-vector shown in Eq.~\eqref{msc_d-vector}.
The form factor hence takes the form
\begin{equation}
    \begin{aligned}
        &\mathfrak{F}_{\mathrm{sing}-\mathrm{MSC}}(\mathbf{k}, \mathbf{k}')
        =
        \\
        &2
        {d_{L,0}(\mathbf{k}')} {T^{(b)}_0(-\mathbf{k}, -\mathbf{k}')}
        [
        {\mathbf{d}^{(bp)*}_{R}(\mathbf{k})} \cdot {\mathbf{T}^{(b)}(\mathbf{k}, \mathbf{k}')}
        ]
        \\
        &-2 
        {d_{L,0}(\mathbf{k}')} {T_0^{(b)}(\mathbf{k}, \mathbf{k}')}
        [
        {\mathbf{d}^{(bp)*}_{R}(\mathbf{k})}
        \cdot
        {\mathbf{T}^{(b)}(-\mathbf{k}, -\mathbf{k}')}
        ]
        \\
        &-2i
        {d_{L,0}(\mathbf{k}')}  {\mathbf{d}^{(bp)*}_{R}(\mathbf{k})} \cdot 
        [
        {\mathbf{T}^{(b)}(\mathbf{k}, \mathbf{k}')}
        \times
        {\mathbf{T}^{(b)}(-\mathbf{k}, -\mathbf{k}')}
        ],
    \end{aligned}
\end{equation}
which is analogous to that of Eq.~\eqref{form_factor.triplet_singlet}, only now the effective $d$-vector and tunneling matrices are evaluated in the projected helical band basis.
The tunneling amplitudes that describe the tunneling between the states at the left, in the spin-$\uparrow$, $\downarrow$ representation, and the states at the right, in the helical band representation, are given by
\begin{subequations}
    \begin{align}
    T^{(b)}_0(\pm \mathbf{k}, \pm \mathbf{k}')
    &=
    T_0(\pm \mathbf{k}, \pm \mathbf{k}')
    [u_{R,0}^{(\pm \mathbf{K}_0)}(\pm \tilde{\mathbf{k}})]^*,
    \\
    \mathbf{T}^{(b)}(\pm \mathbf{k}, \pm \mathbf{k}')
    &=
    T_0(\pm \mathbf{k}, \pm \mathbf{k}')
    [\mathbf{u}_R^{(\pm \mathbf{K}_0)}(\pm \tilde{\mathbf{k}})]^*.
    \end{align}
\end{subequations}
Here,
$(u_{R,0}^{(\pm \mathbf{K}_0)}(\tilde{\mathbf{k}}),
\mathbf{u}_{R}^{(\pm \mathbf{K}_0)}(\tilde{\mathbf{k}})) $ are the complex four-vectors defined in Eq.~\eqref{qp=-1.unitary.four-vector},
related to the unitary transformation of the Weyl band Hamiltonian.
As the spinful bands of the $s$-wave superconductor are decoupled, we have $u_{L,0} = 1$ and $\mathbf{u}_L = 0$.

Consequently, the form factor describing the first-order Josephson coupling between a $q_p = -1$ monopole superconductor and a singlet superconductor simplifies to
\begin{equation}
    \begin{aligned}
        \mathfrak{F}_{\mathrm{sing}-\mathrm{MSC}}^{(q_p = -1)}(\tilde{\mathbf{k}}, \mathbf{k}')
        =
        & - 4\pi {\Delta_{L,0}(\mathbf{k}')}
        [{\Delta_{R, \mathrm{MSC}}^{{(q_p=-1)}}(\tilde{\mathbf{k}})}]^*
        \\
        &\times
        {T_0(-\mathbf{k}, -\mathbf{k}')}
        {T_0(\mathbf{k}, \mathbf{k}')}
        \\
        &\times
        \mathcal{Y}_{-\frac{1}{2}; \frac{1}{2}, - \frac{1}{2}}(\boldsymbol{\Omega}_{\tilde{k}})
        \mathcal{Y}_{-\frac{1}{2}; \frac{1}{2}, \frac{1}{2}}(\boldsymbol{\Omega}_{\tilde{k}}),
    \end{aligned}
    \label{form_factor.qp=-1.singlet}
\end{equation}
where $\Delta_{L,0}(\mathbf{k}')= d_{L,0}(\mathbf{k}')$,
$\Delta_{R, \mathrm{MSC}}^{{(q_p = -1)}}(\tilde{\mathbf{k}})$ is the band-projected effective monopole harmonic pairing order in Eq.~\eqref{qp=-1.pairing_order.general_inter_FS} characterized by the pair monopole charge $q_p=-1$, and $\tilde{\mathbf{k}} = \mathbf{k} - \mathbf{K}_0$ is the wave vector with respect to the Weyl node.
Despite the form factor expressed in Eq.~\eqref{form_factor.qp=-1.singlet} including multiple monopole harmonic functions, $\mathfrak{F}_{\mathrm{sing}-\mathrm{MSC}}$ is always non-singular, as it is related to the Josephson current, a physical observable. 
By employing the addition theorem of monopole harmonic functions~\cite{Wu1977} (see also Appendix~\ref{appendix:table_of_MHs}), 
which states that the product of two monopole harmonic functions with charges $q_1$ and $q_2$ can be expressed as a sum of monopole harmonic functions with monopole charge $q_3 = q_1 +q_2$, we find that the form factor in Eq.~\eqref{form_factor.qp=-1.singlet} has zero total monopole charge. In other words, it can be represented using standard spherical harmonic functions without singularities over the Fermi surface.

Similarly, we derive the form factor for a junction between the same $q_p = -1$ monopole superconductor and a standard spin triplet superconductor as follows, which is non-singular for the same reason as that of $\mathfrak{F}_{\mathrm{sing}-\mathrm{MSC}}^{(q_p = -1)}$,
\begin{eqnarray}
    &&\mathfrak{F}_{\mathrm{trip}-\mathrm{MSC}}^{(q_p = -1)} (\tilde{\mathbf{k}}, \mathbf{k}')
    =- 2 \pi
    {T_0(-\mathbf{k}, -\mathbf{k}')}
    {T_0(\mathbf{k}, \mathbf{k}')} 
    \nonumber 
    \\
    &&  \times {\Delta_{R, \mathrm{MSC}}^{{(q_p=-1)}^*}(\tilde{\mathbf{k}})} [(d_{L,x}(\mathbf{k}')-id_{L,y}(\mathbf{k}')) \mathcal{Y}_{-\frac{1}{2}; \frac{1}{2}, - \frac{1}{2}}^2(\boldsymbol{\Omega}_{\tilde{k}}) 
    \nonumber 
    \\
    &&+(d_{L,x}(\mathbf{k}')+id_{L,y}(\mathbf{k}')) \mathcal{Y}_{-\frac{1}{2}; \frac{1}{2}, \frac{1}{2}}^2(\boldsymbol{\Omega}_{\tilde{k}})].
    \label{form_factor.qp=-1.triplet}
\end{eqnarray}
It depends on, $\mathbf{d}_L(\mathbf{k}')$, the $d$-vector which encodes the spin structure of the triplet pairing order at the left. 
In contrast, the form factor $\mathfrak{F}_{\mathrm{sing}-\mathrm{MSC}}^{(q_p = -1)}$ in Eq.~\eqref{form_factor.qp=-1.singlet} takes a simpler form, as the symmetry of the singlet pairing order at the left is primarily determined by its symmetry in momentum space. 
For the microscopic models studied later in Secs.~\ref{sec:JG} and \ref{sec:distinguish_from_chiral_SCs}, to identify the momentum space topology of the monopole pairing order, we will focus on junctions with a spin singlet superconductor for simplicity.

\subsection{Symmetry and topological principles to probe monopole superconducting order}
\label{subsec:MSC.symmetry}

Symmetry principles have long been crucial in the design of Josephson experiments to reveal the pairing symmetry of unconventional superconductors~\cite{Geshkenbein1986, Geshkenbein1987, Harlingen1995, Tsuei2000}.
We now propose symmetry and topological principles to guide the design of a set of Josephson junctions which together can identify monopole superconducting order. 
This approach complements the discussion in Sec.~\ref{subsec:MSC.microscopic}, which was mainly based on microscopic derivations of the Josephson current. 

\begin{figure}
    \centering
    \includegraphics[width=\linewidth]{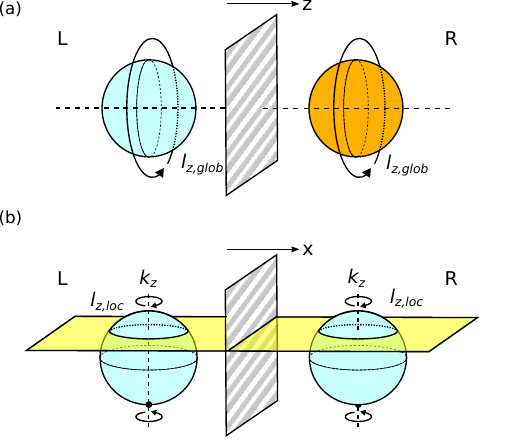}
    \caption{
    Josephson junction designs to probe monopole superconducting order, shown schematically in a hybrid picture of momentum and real space.  
    The junction barrier (gray, striped) is shown in real space, separating two superconductors at the left (L) and right (R).  
    Band Fermi surfaces are shown in momentum space at either side of the junction.
    (a)~Junction design to probe the global angular momentum, consisting of a junction oriented along the rotational axis ($z$ axis) between a monopole superconductor (cyan) and other known superconductor (orange).
    (b)~Junction design to probe the local angular momentum of the pairing order in momentum space. 
    The junction is between two identical monopole superconductors and is oriented perpendicular to the rotational $z$-axis along which the gap nodes lie. For conserved momentum $k_z$, it can be treated as an effective $2$D junction (yellow).    }
    \label{fig:junction_design}
\end{figure}

Monopole pairing order with pair monopole charge $q_p$ generally takes the form of a superposition of monopole harmonic functions $\mathcal{Y}_{q_p, l, l_{\hat{\mathbf{n}}, \mathrm{glob}}}$ for different integer values of $l \geq |q_p|$ and $|l_{\hat{\mathbf{n}}}| \leq l$.  
For $q_p \neq 0$, the monopole harmonic functions exhibit nodes pinned to the rotation quantization axis, $\hat{\mathbf{n}}$, which we often take as $\hat{z}$ without loss of generality.   
Furthermore, as discussed in Sec.~\ref{subsec:MSC.microscopic}, local angular momenta 
differ from the global angular momentum by the pair monopole charge, 
\begin{equation}
    l_{\hat{\mathbf{n}}, \mathrm{loc}} = l_{\hat{\mathbf{n}}, \mathrm{glob}} + \lambda q_p.   
\end{equation}
Here, $\lambda = \pm 1$ corresponds to different choices of gauge (see Appendix~\ref{appendix:table_of_MHs}).
To identify this exotic pairing order, we design a set of two Josephson junctions which respectively probe the global and local angular momentum of the pairing order. 
The nonzero shift between $l_{z, \mathrm{glob}}$ and $l_{z, \mathrm{loc}}$ indicates the pair monopole charge.

(i)~Globally, the monopole superconducting order remains a representation of the rotation group but resides in a topological sector characterized by its pair monopole charge $q_p$.
For two superconductors that transform identically under a given symmetry operation, which is also a symmetry of the junction, the Josephson energy phase relationship is invariant under this symmetry transformation. Hence, if rotation $R_{z}$ with respect to the $z$ axis is a symmetry of the junction, then the first-order Josephson coupling is allowed only if the superconducting orders at both sides of the junction transform in the same way under $R_{z}$, \textit{i.e.} they both form an irrep of $R_{z}$ labeled by the same $l_z$.
Monopole superconducting order given by, for example, $\mathcal{Y}_{q_p; l, l_{z, \mathrm{glob}}}$ transforms under rotation $R_{z}$ according to its globally conserved angular momentum, $l_{z, \mathrm{glob}}$. 
As such, to extract $l_{z, \mathrm{glob}}$,  we consider a junction which is oriented along a rotational $z$ axis, as shown in Fig.~\ref{fig:junction_design}(a).
By placing a monopole superconductor in junction with another superconductor with known pairing symmetry, one can extract the global angular momentum $l_{z, \mathrm{glob}}$ based on the existence of first-order $2\pi$-periodic Josephson current.

(ii)~ Locally in momentum space, the symmetry of monopole superconductor pairing order can be described by regular spherical harmonic symmetry. 
In the local representation, the pairing phase possesses nontrivial winding (see Appendix~\ref{appendix:table_of_MHs}), corresponding to chiral pairing order characterized by angular momentum $l_{z, \mathrm{loc}}$. 
Between the gap nodes pinned to the rotational axis, the local chirality changes by twice the pair monopole charge, as shown in the model in Sec.~\ref{subsec:MSC.microscopic}.
To probe  $l_{z, \mathrm{loc}}$, we consider the design shown in Fig.~\ref{fig:junction_design}(b), where two identical monopole superconductors form a Josephson junction oriented normal to the $z$ axis. 
For given conserved momentum $k_z$ along the rotational axis, the junction can be regarded effectively as a junction between two identical two-dimensional superconductors whose pairing orders are characterized by local angular momentum $l_{z, \mathrm{loc}}$. 
Consequently, the total Josephson current of the proposed junction is a superposition of contributions from these planar junctions, which can give signatures, for example the periodicity and phase shift of the current phase relation, that can indicate the local pairing angular momentum $l_{z, \mathrm{loc}}$.

In the following two sections, we analyze a set of Josephson junctions guided by the above design principles to probe the pairing symmetry of a prototypical monopole superconductor arising in a magnetic doped Weyl semimetal with inter-Fermi surface pairing in Eq.~\eqref{qp=-1.MSC.hamiltonian}. 
In Secs.~\ref{sec:JG} and \ref{sec:JL}, we propose junctions to probe the global angular momentum $l_{z, \mathrm{glob}}$ and local angular momentum $l_{z, \mathrm{loc}}$ respectively.
Together, these junctions can be used to extract the monopole charge, $|q_p| = |l_{z, \mathrm{glob}} - l_{z, \mathrm{loc}}|$,  and provide a foundation to guide efforts to experimentally uncover the monopole superconducting order.

\section{Josephson junctions to probe \texorpdfstring{$l_{z, \mathrm{glob}}$}{lz,glob} of the monopole superconducting order}
\label{sec:JG}

Following the symmetry principles presented in Sec.~\ref{subsec:MSC.symmetry}, we design Josephson junctions to extract the global angular momentum of monopole superconducting order, $l_{z, \mathrm{glob}}$.
We first examine a Josephson junction oriented along the $z$ direction between a monopole superconductor with pairing order $\Delta_\mathrm{MSC}^{(q_p=-1, l_z=0)}$ and a conventional $s$-wave superconductor, demonstrating the nonvanishing first-order Josephson current.
After, in Sec.~\ref{subsec:different_lz}, we study the Josephson coupling for the monopole pairing order $\Delta_{\mathrm{MSC}}^{(q_p=-1, l_z = 2)}$ in the same topological sector but with different global angular momentum, and in Sec.~\ref{subsec:qp=3}, monopole pairing order $\Delta_{\mathrm{MSC}}^{(q_p=3, l_z=0)}$ in a different topological sector of $q_p=3$.

\subsection{First-order Josephson current between \texorpdfstring{$s$}{s}-wave superconductor and  \texorpdfstring{$\Delta_\mathrm{MSC}^{(q_p = -1, l_z=0)}$}{(qp=-1, lz=0)} monopole superconductor}
\label{subsec:q=1.lz=0.all_subsections}

\subsubsection{Method 1: linear response theory}
\label{subsec:JG.GF}

\begin{figure}
    \includegraphics[width= \linewidth]{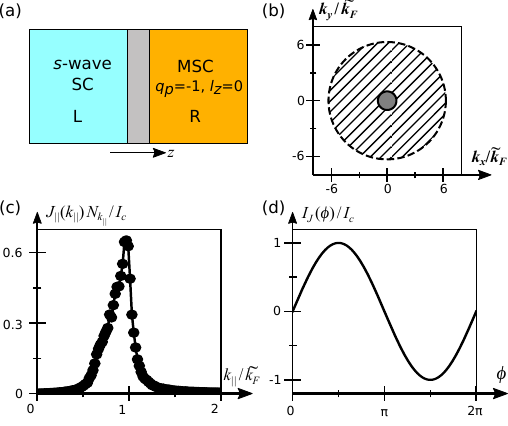}
    \centering
    \caption{
    (a) Schematic of the Josephson junction oriented along $z$ direction between an $s$-wave superconductor and a $\Delta_\mathrm{MSC}^{(q_p=-1, l_z=0)}$ monopole superconductor. 
    (b) Bulk Fermi surface projections on the $k_xk_y$ plane for $s$-wave superconductor (hatched) and monopole superconductor (solid) at two sides of the junction in (a). 
    (c) Relative magnitude of the contributions to the first-order Josephson current phase relation for different transverse momentum, $k_\parallel$, corresponding to Eq.~\eqref{msc_swave.GF_method.mag_for_transverse_k}.
    (d) first-order Josephson current between the monopole superconductor and $s$-wave superconductor at zero temperature and zero bias.
    Here, $v_F = 0.5$,  $\mu_R = 0.1$,  $\tilde{k}_F = 0.2$, and $\mu_L = 0.8$.  $\Delta_{L,0} = \Delta_{R,0} = \Delta_0 = 0.005$, $I_c =  0.62 \Delta_0e/\hbar$,
    and $N_{k_\parallel}$ is the number of transverse momentum cuts in the calculation.
    $T_0$, $A$, $\ell_L$, $\ell_R$, and $m$ are set to unity.
    }  \label{fig:WeylS.GF_method}
\end{figure}

We study Josephson effects between an $s$-wave superconductor 
and a monopole superconductor with pairing order $\Delta_\mathrm{MSC}^{(q_p = -1, l_z=0)}$
in a junction along the $z$ direction, shown schematically in Fig.~\ref{fig:WeylS.GF_method}(a).
Following the symmetry principles, as discussed in Sec.~\ref{subsec:MSC.symmetry}, the first-order Josephson coupling in this system is allowed due to the pairing order at two sides of the junction being characterized by the same global angular momentum $l_z = 0$.
To show this, we consider the continuum model of the monopole superconductor introduced in Eq.~\eqref{qp=-1.MSC.hamiltonian},
for which the effective pairing order $\Delta_\mathrm{MSC}^{(q_p = -1, l_z = 0)}$ in Eq.~\eqref{q=-1.lz=0.effective_pairing} is invariant under rotation $R_z$, having global angular momentum $l_z=0$.
Following the derivation based on linear response theory in Sec.~\ref{subsec:MSC.microscopic}, we demonstrate nonzero first-order Josephson coupling and the resulting critical current.

For a junction between an $s$-wave superconductor and $\Delta_\mathrm{MSC}^{(q_p=-1,l_z=0)}$ monopole superconductor, the form factor in Eq.~\eqref{form_factor.qp=-1.singlet} can be 
simplified to
\begin{align}
    &
    \mathfrak{F}_{s-\mathrm{MSC}}^{(q_p=-1, l_z=0)}
    (\tilde{\mathbf{k}}, \mathbf{k}')
    = 
    \frac{8\pi}{3}
    \Delta_{L,0}
    \Delta_{R,0}
    |Y_{1,1}(\boldsymbol{\Omega}_{\tilde{k}})|^2
    T_0^2 \delta_{\mathbf{k}_\parallel, \mathbf{k}'_\parallel},
                        \label{form_factor.qp=-1,lz=0.swave}
\end{align}
where $\Delta_{L,0}$ is the pairing amplitude of the $s$-wave superconductor, and $\Delta_{R,0}$ is the amplitude of the inter-Fermi surface pairing in the monopole superconductor. 
We model the tunneling amplitude through the junction barrier as a spin-independent delta function that preserves in-plane momenta 
with 
$T_0(\mathbf{k}, \mathbf{k}') = T_0 \delta_{\mathbf{k}_{\parallel}, \mathbf{k}'_\parallel}$ 
and $\mathbf{T}(\mathbf{k}, \mathbf{k}') = 0$.
Here, $T_0$ is a constant, and $\mathbf{k}_\parallel = \mathbf{k}_\parallel' = \tilde{\mathbf{k}}_\parallel$ is the wave vector in the $k_x k_y$ plane, which is conserved at the barrier.
Already, the nonvanishing first-order Josephson coupling can be seen from the form factor $\mathfrak{F}_{s-\mathrm{MSC}}^{(q_p=-1, l_z=0)}$, which encodes the symmetry of the junction as well as the pairing orders and itself is invariant under rotation $R_z$ and parity.

With the form factor and tunneling matrix above,
the Josephson current in Eq.~\eqref{Josephson_current} evaluates to
\begin{equation}
    \begin{aligned}
        I_J(\phi)
        =
        \frac{4e }{h}
        A T_0^2
                \Delta_{L,0}
        \Delta_{R,0} 
                        \sin \phi
        \int  \mathrm{d} {k}_\parallel
        \
        {k}_\parallel
        J({k}_\parallel),
    \end{aligned}
    \label{correlation_fnc.msc_swave}
\end{equation}
where
\begin{equation}
    \begin{aligned}
                                                        J ({k}_\parallel)         =
        k_\parallel^2 
        \frac{ \ell_L \ell_R}{(2\pi)^2}
        \int
        &\frac{\mathrm{d}\tilde{k}_z \mathrm{d}k'_z}{{k}_\parallel^2 + \tilde{k}_z^2}
        w(E_{L, \mathbf{k}'}, E_{R, \tilde{\mathbf{k}}}^{(bp)}; \beta)
    \end{aligned}
    \label{msc_swave.GF_method.mag_for_transverse_k}
\end{equation}
describes the contribution to the Josephson current for conserved transverse momentum, ${k}_\parallel$.
Here, $A$ is the surface area of the interface, and $\ell_L$ and $\ell_R$ are the system lengths along $z$ direction at the left and right sides respectively.
We consider the Josephson current at zero temperature, for which
\begin{math} w(E_{L, \mathbf{k}'}, E_{R, \tilde{\mathbf{k}}}^{(bp)}; \beta)
=
[2E_{L, \mathbf{k}'} E_{R, \tilde{\mathbf{k}}}^{(bp)}(E_{L, \mathbf{k}'}^2 - E_{R, \tilde{\mathbf{k}}}^{(bp)2})]^{-1}
\end{math} and is positive-definite.
The dispersions of the BdG quasiparticles for the $s$-wave superconductor and monopole superconductor are given by
$E_{L, \mathbf{k}'} = \sqrt{\xi_{L, \mathbf{k}'}^2
+ |\Delta_{L,0}|^2}$ and $E_{R, \tilde{\mathbf{k}}}^{(bp)} = \sqrt{\xi_{R, \tilde{\mathbf{k}}}^{(bp)2}
+ \left|\Delta_{R,0}\right|^2 {\tilde{k}^2_\parallel}/{\tilde{k}^2}}$ respectively.
Here,
$\xi_{L, \mathbf{k}'} = \hbar^2({k'_z}^2 + {k'}_\parallel^2)/2m - \mu_L$
and
$ \xi^{(bp)}_{R, \tilde{\mathbf{k}}} = \hbar v_F \sqrt{\tilde{k}_z^2 + \tilde{k}_\parallel^2} - \mu_R$
are the respective band dispersions.
The Fermi surfaces of $\xi_{L, \mathbf{k}'}$ and $\xi_{R, \tilde{\mathbf{k}}}^{(bp)}$ are shown projected to the $k_x k_y$ plane in Fig.~\ref{fig:WeylS.GF_method}(b).

In Fig.~\ref{fig:WeylS.GF_method}(c), we show $J(k_\parallel)$, the contributions to the first-order Josephson current, for different conserved transverse momentum $k_\parallel$. 
States near the equator of the Fermi surface of the doped Weyl semimetal, with $k_\parallel \approx \tilde{k}_F$, contribute the most to the Josephson tunneling, where the magnitude of $\Delta_\mathrm{MSC}^{(q_p=-1, l_z=0)}$ is at its maximum. 
As $k_\parallel$ approaches zero, the contributions to the Josephson current are suppressed as the magnitude of $\Delta_\mathrm{MSC}^{(q_p=-1, l_z=0)}$ reaches its minimum near its gap nodes along the $k_z$ axis.
For transverse momentum away from the Weyl Fermi surfaces, $k_\parallel > \tilde{k}_F+{\Delta_{R,0}}/{v_F}$, its contribution to the Josephson current vanishes due to vanishing low-energy density of states in $\Delta_\mathrm{MSC}^{(q_p=-1, l_z=0)}$ monopole superconductor.

Furthermore, in Fig.~\ref{fig:WeylS.GF_method}(d), we show the total first-order Josephson current phase relation given in Eq.~\eqref{correlation_fnc.msc_swave} numerically. 
It exhibits $2\pi$ periodicity and nonvanishing critical current $I_c=0.62\Delta_0e/\hbar$. 
In contrast, for a Josephson junction in the same geometry between an $s$-wave and a chiral $p_x\pm ip_y$-wave superconductor, the first-order Josephson current vanishes when there is no spin-orbit coupling in the bulk nor at the interface. 
Hence, though locally $\Delta_\mathrm{MSC}^{(q_p=-1, l_z=0)}$ exhibits chiral $p_x\pm ip_y$ pairing symmetry, the Josephson current in the above junction serves as an effective probe of the global angular momentum of the pairing order $l_{z, \mathrm{glob}}=0$. 
We discuss later the case when there exist spin-orbit interactions in Sec.~\ref{sec:distinguish_from_chiral_SCs}.

\subsubsection{Method 2: numerical results from tight-binding model}
\label{subsec:JG.tbm}
We further show the nonzero Josephson coupling in the same Josephson junction between an $s$-wave superconductor and $\Delta_\mathrm{MSC}^{(q_p=-1, l_z=0)}$ monopole superconductor, as shown in Fig.~\ref{fig:WeylS.GF_method}(a),
using a tight-binding model on a cubic lattice (see Appendix~\ref{appendix:TBM_JJ} and \ref{appendix:TBM} for details). 
As $k_x$ and $k_y$ are assumed to be conserved across the junction, we take periodic boundary conditions in the $x$ and $y$ directions. 
We consider $2N_z$ sites along $z$ direction and take open boundary conditions at the two ends at $n_z=1$ and $2N_z$. 
The junction barrier is located between $n_z=N_z$ and $n_z=N_z +1$, where only spin-independent nearest neighbor hopping within the barrier is considered. 
At given conserved momenta $k_x$ and $k_y$, we model the barrier along $z$ direction as 
\begin{equation}
    \begin{aligned}
        H_{\mathrm{link}} (k_x, k_y)
        =
        t_0 \sum_{\sigma = \uparrow, \downarrow}
        ( c^\dagger_{k_x, k_y;\sigma, N_z + 1}
        c_{k_x, k_y;\sigma, N_z}+ \mathrm{h.c.})
    \end{aligned}
    \label{Hlink.tbm.one_site}
\end{equation}
where $t_0$ is the spin-independent hopping amplitude, and $c_{k_x, k_y;\sigma, n_z}$ annihilates a particle at site $n_z$ with momentum $k_x$ and $k_y$ and spin $\sigma$.  
If we increase the thickness of the barrier to a few sites, there is no qualitative change in Josephson current. 

\begin{figure}
    \centering
    \includegraphics[width=\linewidth]
    {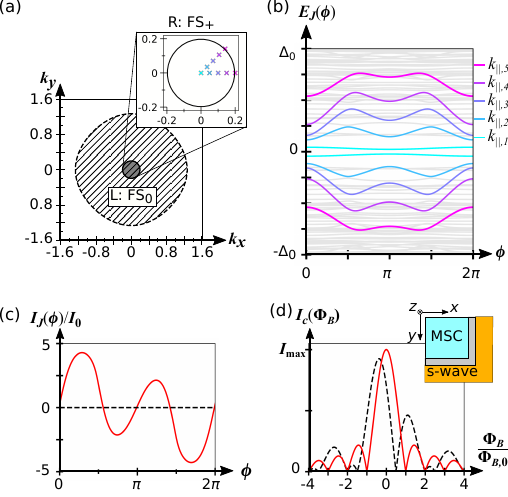}
    \label{fig:WeylS.zjunction.epr}
    \caption{Numerical results for a tight-binding model of the Josephson junction along $z$ direction between an $s$-wave superconductor and $\Delta_\mathrm{MSC}^{(q_p=-1, l_z=0)}$ monopole superconductor. 
    (a)~Maximal cross section of bulk Fermi surface $\mathrm{FS}_+$ of the monopole superconductor at $k_z=K_{0,z}$ (solid) and that of the Fermi surface  $\mathrm{FS}_0$ of the $s$-wave superconductor at $k_z=0$ (hatched). Here, $\mu_L/t_0 = 3.88$ and $\mu_R /t_0 = 0.05$. 
    (b)~The Josephson energy phase relation contributed by different momenta in (a), $k_{\parallel,n} = (n-1)\tilde{k}_F/4$, $n=1, \cdots, 5$, and $\tilde{k}_F=0.2$. 
    Here, the relative magnitude is given by $\Delta^{(n)}_{0} = \Delta_0 k_{\parallel,n}/\tilde{k}_F$, with $\Delta_0/|t_0| = 0.04$.
    Lowest {energy} states for a given radial cut correspond to the cuts shown in (a) and are shown in color, while the gray bands represent higher-{energy} contributions.
    (c)~The total Josephson current phase relation (red, solid) compared to that of a $z$ direction junction between a $p_x - ip_y$ superconductor and an $s$-wave SC (black, dashed).
    (d)~The Fraunhofer pattern of a corner junction (inset) in the $xy$-plane of a monopole superconductor (red, solid), compared to that of a $p_x + ip_y$ superconductor (black, dashed). 
    Here, $I_0 = {e \Delta_0}/{\hbar}$,
        and the system size is $2N_z = 300$. 
    }
    \label{fig:WeylS.zjunction}
\end{figure}

In Fig.~\ref{fig:WeylS.zjunction}, we present our numerical results of the Josephson current for the junction between an $s$-wave superconductor and a $\Delta_\mathrm{MSC}^{(q_p=-1, l_z=0)}$ monopole superconductor. 
Since the junction breaks translational symmetry along $z$ direction, longitudinal momentum $k_z$ is not conserved.
This effectively allows Cooper pairs with different $k_z$ components on either side of the junction to tunnel across the interface. 
Consequently, even with mismatched $3$D Fermi surfaces, a nonzero Josephson current can still occur as long as their $2$D projections in the transverse momentum plane overlap sufficiently, as illustrated in Fig.~\ref{fig:WeylS.zjunction}(a). 
Although the lattice has fourfold rotation symmetry with respect to the $z$ axis, the projected Fermi surfaces are nearly spherical in the $k_xk_y$ plane. 
In Fig.~\ref{fig:WeylS.zjunction}(b), contributions to the Josephson energy phase relation are calculated for representative momenta ${\mathbf{k}}_\parallel$ at different radii $k_\parallel=\sqrt{{{k}_x^2 + {k}_y^2}}$. 
At a given value of $k_\parallel$, different azimuthal angles, $\varphi_{\tilde{k}}$, contribute nearly the same energy phase relation.
Hence, the contributions from states at two representative azimuthal angles, $\varphi_{\tilde{k}} =0$ and $\pi/4$, shown in the inset of Fig.~\ref{fig:WeylS.zjunction}(a), overlap with each other in Fig.~\ref{fig:WeylS.zjunction}(b). 
At $k_{\parallel}=k_{\parallel,1}=0$, the states close to north and south poles of the Fermi surfaces contribute to the Josephson coupling. 
However, the magnitude of the pairing gap of the monopole superconductor, $\Delta_\mathrm{MSC}^{(q_p=-1, l_z=0)}$, vanishes at the poles of the Weyl Fermi surfaces. 
As such, states at $k_{\parallel,1}$ correspond to vanishingly small Josephson energy and therefore flat energy phase relation. 
As $k_\parallel$ increases to nonzero values of $k_{\parallel,i}$ with $i \ge 2$, the corresponding contribution to Josephson energy shows significant $2\pi$ periodicity. 
Furthermore, the overall magnitude of Josephson energy increases as the magnitude of pairing order $\Delta_\mathrm{MSC}^{(q_p=-1, l_z=0)}$ at $k_{\parallel,i}$ increases until reaching its maximum at  $k_{\parallel, 5} \approx \tilde{k}_F$.

In addition to the net $2\pi$-periodic Josephson energy phase relation, there are contributions from second-order Josephson tunneling, where multiple scatterings from the junction interface lead to $\pi$-periodic contributions to the energy phase relation. 
Such higher-order contributions are most significant for states near the gap nodes of the $\Delta_{\mathrm{MSC}}^{(q_p=-1, l_z=0)}$ monopole pairing order, for example, at $k_\parallel  = k_{\parallel, 2}$ in Fig.~\ref{fig:WeylS.zjunction}(b). We provide further detailed explanation on this in Appendix~\ref{appendix:effective_pairing_channels}.

We calculate the total Josephson current  at zero temperature,
\begin{math}
    \left.
    I_J(\phi)
    \right|_{T=0} 
    = 
    (-{2e}/{\hbar})
    \sum_{\mathbf{k}_\parallel} 
    {\partial E_{J}(\mathbf{k}_\parallel, \phi)}/{\partial \phi}.
    \end{math}
Here, $E_{J}(\mathbf{k}_\parallel, \phi)$ is the Josephson energy phase relation for occupied states with transverse momentum $\mathbf{k}_\parallel$. 
In Fig.~\ref{fig:WeylS.zjunction}(c), we show our numerical result of Josephson current with the summation over transverse momentum $k_\parallel \leq \tilde{k}_F$ at representative values, $k_{\parallel, i=1,\cdots,5}$, and taking account of the measure in polar coordinates. 
For $k_\parallel > \tilde{k}_F+{\Delta_{R,0}}/{v_F}$, there are no in-gap Andreev states contributing to the Josephson current as discussed in Sec.~\ref{subsec:JG.GF}.
Moreover,
comparing with Fig.~\ref{fig:WeylS.GF_method}(c), 
the total Josephson current phase relation in Fig.~\ref{fig:WeylS.zjunction}(c) is modulated by $\pi$-periodic contributions from higher-order tunneling processes.
However, the total Josephson current, obtained by superimposing contributions from different $k_\parallel$, is nonetheless $2\pi$-periodic.

The Josephson current can be further enhanced when the Weyl points of the bulk monopole superconductor are close to the Fermi surface of the bulk s-wave superconductor. 
As an additional example, we keep the overlap area of the $2$D projections of the Fermi surfaces in the transverse momentum plane identical to that in Fig.~\ref{fig:WeylS.zjunction}(a). 
However, we reduced the $s$-wave superconductor’s Fermi wavevector from $k_F=1.63$ ($\mu_L/t_0 =  3.88$) to $k_F=1.05$ ($\mu_L/t_0 = 5$), ensuring that the bulk Fermi surfaces at two sides of the junction also overlap in $3$D momentum space. 
We observed $28$\% enhancement in the critical Josephson current. 
Nonetheless, the qualitative features remained unchanged, including the nonvanishing first-order Josephson current as well as the periodicity of the Josephson current phase relations.

the first-order $2\pi$-periodic Josephson current between an $s$-wave superconductor and $\Delta_{\mathrm{MSC}}^{(q_p=-1, l_z=0)}$ monopole superconductor is nonvanishing in this tight-binding model for the same symmetry reasons discussed for the continuum model in Sec.~\ref{subsec:JG.GF}.
The tight-binding model preserves the two parity-related Fermi surfaces which have opposite Chern numbers.
Although the Fermi surfaces are not perfectly spherical, the topology of the effective pairing remains invariant and is still classified by the pair monopole charge, $q_p = -1$.
Additionally, the global angular momentum of the pairing order is constrained to the $l_z = 0$ channel, as both the kinetic and the proximitized pairing parts of the tight-binding Hamiltonian are invariant under lattice rotations $R_z$. 
Therefore, the effective pairing order likewise transforms according to conserved $l_{z, \mathrm{glob}} = 0$.  
However, compared to the effective pairing derived in the continuum model in Eq.~\eqref{q=-1.lz=0.effective_pairing}, there are small but nonzero contributions to the pairing order from monopole harmonics in the $l = 2$ partial wave channel as a result of the lattice symmetry.
Nonetheless, these contributions from higher partial wave channels do not significantly affect the nonzero first-order Josephson coupling. 
Hence, because the tight-binding model preserves the symmetry and topology of the system, the essential features of the Josephson current, for example its $2\pi$-periodicity, agree with the previous result based on continuum model.
Furthermore, these  features persist in the long Josephson junction limit, as discussed in Sec.~\ref{appendix:long_SNS}.

We additionally examine a corner junction between a $\Delta_\mathrm{MSC}^{(q_p=-1, l_z=0)}$ monopole superconductor and $s$-wave superconductor.
The corner junction is symmetrically aligned with two faces along the $x$ and $y$ directions, as shown schematically in the inset in Fig.~\ref{fig:WeylS.zjunction}(d). 
Because this geometry is sensitive to pairing phase difference along the two orthogonal crystal axes~\cite{Harlingen1995},
it can be used as an additional probe of global angular momentum $l_{z, \mathrm{glob}}$. 
Based on the same tight-binding model of $\Delta_\mathrm{MSC}^{(q_p=-1, l_z=0)}$ monopole superconductor, we study the resulting critical current in the presence of an external magnetic field along the $z$ direction. 
As the external flux $\Phi_B$ through the barrier of the corner junction varies,
we obtain the Fraunhofer pattern of the corner junction, as shown in Fig.~\ref{fig:WeylS.zjunction}(d).
The Fraunhofer pattern is qualitatively the same as that of an SIS junction between two $s$-wave superconductors, with nodes at integer flux quantum, $\Phi_0 = h/2e$.
This is in contrast to, for example, a corner junction between an $s$-wave and chiral $p$-wave superconductor, which gives rise to an interference pattern that is asymmetric with respect to zero flux.

In alignment with the findings presented in the preceding section, where the Josephson current was analyzed through linear response theory,
the results from the tight-binding model further reinforce that, globally, the $\Delta_\mathrm{MSC}^{(q_p=-1, l_z=0)}$ monopole superconducting pairing order transforms according to $l_{z, \mathrm{glob}} = 0$.  
Consequently, the monopole superconductor has nonzero Josephson coupling with an $s$-wave superconductor in a junction along the $z$ axis, leading to an overall nonvanishing $2\pi$-periodic first-order Josephson current.
We next discuss similar junction designs to probe monopole superconductors with nonzero global angular momentum $z$ component, or when the pairing order is in a different topological sector.

\subsection{Probing monopole superconducting orders with nonzero global angular momenta \texorpdfstring{$l_{z, \mathrm{glob}}$}{lz,glob}}
\label{subsec:different_lz}

We now demonstrate that the $z$ direction Josephson junction design introduced in Sec.~\ref{subsec:MSC.symmetry} can be generalized to probe monopole pairing orders with nonzero $z$-component global angular momentum. 
As an example, we consider the monopole pairing order in the same topological sector as that in Sec.~\ref{subsec:q=1.lz=0.all_subsections} given by the pair monopole charge $q_p=-1$, but in a different angular momentum channel with $l=l_z=2$.
Following the symmetry argument, the first-order Josephson coupling is allowed only when coupled to another superconductor with pairing order in the $l_z=2$ channel, for example, a $d_
{x^2-y^2} + id_{xy} $ wave superconductor, in an SIS junction with geometry preserving the rotational symmetry about $z$ axis.
To demonstrate the symmetry principle, here, we focus on deriving the form factor $\mathfrak{F}_{\mathrm{sing}-\mathrm{MSC}}^{(q_p=-1)}$  in this system and show that it leads to nonzero first-order Josephson coupling. 
In Appendix~\ref{appendix:dwave}, we also show nonvanishing first-order Josephson current obtained numerically from a tight-binding model of this system.

We first introduce the model of a monopole superconductor for which the pairing order is in the topological sector defined by pair monopole charge $q_p=-1$ yet now transforms under rotation $R_z$ according to global angular momentum $l_z = 2$.
For simplicity, we start from the same band Hamiltonian 
of the doped magnetic Weyl semimetal, as shown in Eq.~\eqref{continuum_models.c=1}.
We now consider the case of inter-Fermi surface $p_x + ip_y$ pairing in the ${|\uparrow \uparrow\rangle}$ channel, given by
$\Delta_\mathrm{inter}(\mathbf{k}) = (\Delta_0/2\tilde{k}_F)( k_x + i k_y) (\sigma_0 + \sigma_z)$. 
The band-projected superconducting order in Eq.~\eqref{qp=-1.pairing_order.general_inter_FS}  describing the low-energy pairing between helical Fermi surfaces enclosing the Weyl nodes at $\pm \mathbf{K}_0$ is now given by
\begin{align}
    \Delta_\mathrm{MSC}^{\black{(q_p = -1, l_z = 2)}}(\tilde{\mathbf{k}}) 
    &= \Delta_0 \sqrt{\frac{4\pi}{5}} \mathcal{Y}_{-1; 2, 2}(\theta_{\tilde{k}}, \varphi_{\tilde{k}})
    \label{eff_pairing.lz=2}
    \\
    \nonumber
    &=
    \Delta_0 \cos^2\frac{\theta_{\tilde{k}}}{2} \sin \theta_{\tilde{k}} e^{i(2- \lambda) \varphi_{\tilde{k}}},
\end{align}
where, in consistency with prior notation, $\lambda = \pm 1$ corresponds to a gauge where the Dirac string is located at $\theta_{\tilde{k}} = \pi (1 + \lambda)/2$. 
As the Chern numbers of the Fermi surfaces that participate in the inter-Fermi surface pairing are unchanged, the effective pairing order remains
in the same topological sector characterized by the pair monopole charge, $q_p = -1$.
However, the pairing order now transforms according to its global angular momentum, $l_z = l_{z, \mathrm{glob}} = 2$.
The nonzero global angular momentum can be attributed to the inter-Fermi surface $(p_x +ip_y) {| \uparrow \uparrow \rangle}$ pairing, which transforms according to its total angular momentum $j_z = 2$ in the presence of spin-orbit coupling.
When coupled to the spin-texture of the Weyl semimetal Fermi surface, this induces the global $l_z = 2$ winding of the monopole superconducting order in momentum space.

\begin{figure}
    \centering
    \includegraphics[width=\linewidth]{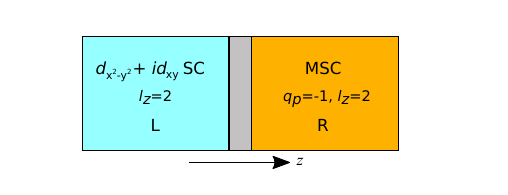}
    \caption{Junction along $z$ direction between a $d_{x^2-y^2} + id_{xy}$ superconductor (left) and $\Delta_{\mathrm{MSC}}^{(q_p=-1, l_z=2)}$ monopole superconductor (right).
    }
    \label{fig:MSC.lz=2.dwave.junction}
\end{figure}
We now consider  a Josephson junction along the $z$ direction between a $d_{x^2-y^2} + id_{xy}$ superconductor and $\Delta_{\mathrm{MSC}}^{(q_p = -1, l_z = 2)}$ monopole superconductor, as shown in Fig.~\ref{fig:MSC.lz=2.dwave.junction}.
We employ the form factor for a junction between a $q_p = -1$ monopole superconductor and spin singlet superconductor, $\mathfrak{F}_{\mathrm{sing}-\mathrm{MSC}}^{(q_p = -1)}$, as shown in Eq.~\eqref{form_factor.qp=-1.singlet}.
Only now, $\Delta_{L,0}(\mathbf{k}') = (\Delta_{L,0}/k_F^2)(k'_x + ik'_y)^2$ describes the $d_{x^2-y^2} + id_{xy}$ pairing order and $\Delta_{R, \mathrm{MSC}}^{(q_p=-1)} = \Delta_{\mathrm{MSC}}^{(q_p=-1, l_z=2)}(\tilde{\mathbf{k}})$ is the monopole superconducting order in Eq.~\eqref{eff_pairing.lz=2}.
For simplicity, we consider spin-indpendent tunneling that trivially respects the junction geometry, given by constant $T_0(\mathbf{k}, \mathbf{k}) = T_0$ and $\mathbf{T}(\mathbf{k}, \mathbf{k}') = 0$.
The form factor for the junction between a $d_{x^2-y^2} + id_{xy}$ superconductor and $\Delta_{\mathrm{MSC}}^{(q_p = -1, l_z = 2)}$ monopole superconductor simplifies to
\begin{align}
    &\mathfrak{F}_{d+id, \mathrm{MSC}}^{(q_p = -1, l_z = 2)}(\mathbf{k}, \mathbf{k}')        
    = 
            \Delta_{L,0} \Delta_{R,0} T_0^2  \frac{{k'}^2}{k_F^2}
    \cos^2 \frac{\theta_{\tilde{k}}}{2} \sin^2 \theta_{\tilde{k}},
        \end{align}
with $\Delta_{R,0}$ being the amplitude of the inter-Fermi surface pairing.
Because both the $d_{x^2-y^2}+id_{xy}$ superconductor and $\Delta_\mathrm{MSC}^{(q_p=-1, l_z=2)}$ monopole superconductor transform under rotation $R_z$ according to their global angular momentum $l_z = 2$, the form factor itself is invariant under $R_z$.
Hence, per the same symmetry arguments in Sec.~\ref{subsec:JG.GF}, this leads to nonvanishing first-order Josephson coupling.

\subsection{Generalization to monopole pairing orders with higher monopole charges}
\label{subsec:qp=3}

We further demonstrate the probe of global angular momentum for monopole superconductors in different topological sectors characterized by higher monopole charges, $|q_p|>1$. 
Although the topological sector is different, the same symmetry principles to extract $l_{z, \mathrm{gob}}$ outlined in Sec.~\ref{subsec:MSC.symmetry} still hold.
We begin with the general model of the monopole superconductor in Eq.~\eqref{MSC.general} but now consider a multi-Weyl semimetal with inter-Fermi surface pairing between two parity-related Fermi surfaces.
For three-dimensional Weyl semimetals on a lattice with $n$-fold rotational symmetry ($n=2,3,4,6$),
there can exist single, double, or triple Weyl nodes~\cite{Fang2012, Tsirkin2017}.
The lattice point group symmetry sets an upper bound to the chirality of the Weyl node, with $|\nu| \leq [n/2]$, and similarly for the 
pair monopole charge, $|q_p| \leq n/2$.

As an example, we study monopole pairing order $\Delta_\mathrm{MSC}^{(q_p=3, l_z=0)}$ that has pair monopole charge $q_p=3$ and transforms according to global angular momentum $l_z = 0$.
Such pairing order can arise in a doped triple Weyl semimetal with inter-Fermi surface $s$-wave pairing.
We consider a minimal model of a triple Weyl semimetal on a three-dimensional hexagonal lattice, which hosts triple Weyl nodes protected by $C_6$ symmetry (see Appendix~\ref{appendix:TBM.qp=3}).
At appropriate doping, the system has two parity-related Fermi surfaces, $\mathrm{FS}_1$ and $\mathrm{FS}_2$, with Chern numbers $\mathcal{C}_1=3$ and $\mathcal{C}_2 = -3$.

From the tight-binding model,
we consider the continuum model of the band Hamiltonians $\mathcal{H}_{\mathrm{kin}, 1}^{(\nu=-3)}$ and $\mathcal{H}_{\mathrm{kin}, 2}^{(\nu=+3)}$ describing Fermi surfaces $\mathrm{FS}_1$ and $\mathrm{FS}_2$ surrounding triple Weyl nodes at $\pm \mathbf{K}_0 = (0,0,\pm K_{0,z})$.
The low-energy band Hamiltonian defined near $\mathbf{K}_0$ is given by
\begin{align}
    & 
    \mathcal{H}^{(\nu= -3)}_{\mathrm{kin},1} (\tilde{\mathbf{k}} + \mathbf{K}_0) 
    = -\mu + \mathbf{h}_1^{(\nu= -3)}(\tilde{\mathbf{k}})\cdot \boldsymbol{\sigma}
    \label{C=3.kin.continuum}
    \\
    &= \nonumber
    -\mu + \hbar v_F \tilde{k}_z \sigma_z + \hbar v_F  \left[ (\tilde{k}_x +i     \tilde{k}_y)^3 \sigma_+ + \mathrm{h.c.}\right],
\end{align}
where $\sigma_\pm = \left( \sigma_x \pm i \sigma_y \right)/2$, and is related to the other Fermi surface by parity, $\mathcal{H}_{\mathrm{kin}, 1}^{(\nu= -3)}(
\mathbf{k}) 
= \sigma_z \mathcal{H}_{\mathrm{kin}, 2}^{(\nu= +3)}(-\mathbf{k})\sigma_z$.
The helical eigenstate at $\mathrm{FS}_1$
is $\chi_{1,+}(\tilde{\mathbf{k}}) = ( \cos ({\theta_{h_1}}/{2}) e^{-i (\lambda -1)\varphi_{h_1} /2},
\sin ({\theta_{h_1}}/{2}) e^{-i(\lambda +1)\varphi_{h_1} /2}
)$, and at $\mathrm{FS}_2$, it is $\chi_{2,+}(\tilde{\mathbf{k}}) = \sigma_z \chi_{1,+}(-\tilde{\mathbf{k}})$.
Here, consistent with previous notation, $\lambda=\pm1$ denotes the choice of gauge, and $\theta_{h_1}$ and $\varphi_{h_1}$ are the polar and azimuthal angles defined in $\mathbf{h}_1^{(\nu= -3)}$ spherical coordinates by  $\mathbf{h}_1^{(\nu= -3)}\equiv |h_1^{(\nu= -3)}| (\sin \theta_{h_1} \cos \varphi_{h_1}, \sin \theta_{h_1} \sin \varphi_{h_1} ,  \cos \theta_{h_1})$. 
They are related to momentum space polar and azimuthal angles by
$\sin \theta_{h_1}= \sin^3 \theta_{\tilde{k}}/{(\cos^2 \theta_{\tilde{k}} + \sin^6 \theta_{\tilde{k}})^{1/2}}$ and $\varphi_{h_1} = - 3 \varphi_{\tilde{k}}$.
We study the case of inter-Fermi surface $s$-wave pairing in Eq.~\eqref{MSC.general}, where $\Delta_\mathrm{inter}(\mathbf{k}) = i\Delta_0\sigma_y$, with $\Delta_0$ being the pairing amplitude.

In the weak-coupling regime, we obtain the low-energy band-projected pairing order from Eq.~\eqref{MSC.effective_gap_function.weak_coupling} which takes the following form:
\begin{align}
            \Delta_\mathrm{MSC}^{{(q_p=3, l_z=0)}}(\tilde{\mathbf{k}})
        &=
        - \Delta_0 
        \sqrt{\frac{8\pi}{3}}
        \mathcal{Y}_{-1; 1,0}(\theta_{h_1}, \varphi_{h_1})
        \label{q=3.eff_pairing}
        \\
        \nonumber
        &= - \Delta_0
        \frac{\sin^3 \theta_{\tilde{k}}}{\sqrt{\cos^2 \theta_{\tilde{k}} + \sin^6 \theta_{\tilde{k}}}}
        e^{i 3 \lambda \varphi_{\tilde{k}}}.
    \end{align}
As the mapping from the spherical coordinates in $\mathbf{h}_1^{(\nu=-3)}$ space to those in $\tilde{\mathbf{k}}$ space is characterized by a winding number of $-3$, the $q = -1$ monopole harmonic order defined here in $\mathbf{h}_1^{(\nu=-3)}$ spherical coordinates implies that the pairing order in momentum space is characterized by $q_p = 3$ pair monopole charge, which agrees with $q_p = (\mathcal{C}_1 - \mathcal{C}_2)/2= 3$. 
As a result, local angular momentum is  given by $l_{z, \mathrm{loc}} = \pm 3$, while the global angular momentum is $l_z = 0$.
Furthermore, the pair monopole charge sets the lower bound of the partial wave channels,  $l\geq 3$. 
Different partial wave channels of $l\ge 3$ contribute to the pairing order, $\Delta_\mathrm{MSC}^{{(q_p=3, l_z=0)}}(\tilde{\mathbf{k}}) = \sum_{l\geq 3} \alpha_l \mathcal{Y}_{q_p=3; l; l_z=0}(\theta_{\tilde{k}}, \varphi_{\tilde{k}})$ with decreasing values of superposition coefficients $\alpha_l$ as $l$ increases from $3$.

We now examine the Josephson coupling between the above $q_p=3$ monopole superconductor and a spin singlet superconductor, following the methods introduced in Sec.~\ref{subsec:MSC.microscopic}.
For comparison,
we consider the case where there is bare spin-independent tunneling at the junction barrier.
The form factor for this system evaluates to
\begin{equation}
    \begin{aligned}
        &
        \mathfrak{F}^{{(q_p = 3)}}_{\mathrm{sing}-\mathrm{MSC}}(\tilde{\mathbf{k}}, \mathbf{k}')
        =
        - 4\pi {\Delta_{L,0}}(\mathbf{k}')
        [{\Delta_{R, \mathrm{MSC}}^{{(q_p = 3)}}(\tilde{\mathbf{k}})}]^*
        \\
        &\times
        {T_0(-\mathbf{k}, -\mathbf{k}')}
        {T_0(\mathbf{k}, \mathbf{k}')}
        \mathcal{Y}_{-\frac{1}{2}; \frac{1}{2}, - \frac{1}{2}}(\boldsymbol{\Omega}_{h_1})
        \mathcal{Y}_{-\frac{1}{2}; \frac{1}{2}, \frac{1}{2}}(\boldsymbol{\Omega}_{h_1}),
    \end{aligned}
\end{equation}
where $\boldsymbol{\Omega}_{h_1} = (\theta_{h_1}, \varphi_{h_1})$ is the solid angle in $\mathrm{h}_1^{(\nu=3)}$-space and $\Delta_\mathrm{MSC}^{(q_p=3}(\tilde{\mathbf{k}})$ is the effective pairing order of the $q_p=3$ monopole superconductor.
Similar to the case of the $q_p = -1$ monopole superconductor, whose form factor is given in Eq.~\eqref{form_factor.qp=-1.singlet}, the half-integer monopole harmonics arise from the effective spin-dependent tunneling arising from the spin-orbit coupling in the triple Weyl semimetal.
The above form factor is non-singular and can be used to determine the first-order Josephson coupling of this system.

Similar to Sec.~\ref{subsec:q=1.lz=0.all_subsections}, we further consider a simple case that the junction is aligned along $z$ direction, the singlet superconductor is $s$-wave, and the bare spin-independent tunneling is a constant, $T_0(\mathbf{k}, \mathbf{k}') = T_0$. 
Then, the form factor for this junction simplifies to
\begin{equation}
    \mathfrak{F}_{s - \mathrm{MSC}}^{(q_p=3, l_z = 0)}(\tilde{\mathbf{k}}, \mathbf{k}')
    =
    T_0^2 \Delta_{L,0} \Delta_{R,0}
    \frac{\sin^6 \theta_{\tilde{k}}}{\cos^2 \theta_{\tilde{k}} + \sin^6 \theta_{\tilde{k}}},
\end{equation}
where $\Delta_{L,0}$ and $\Delta_{R,0}$ are the pairing amplitudes of the $s$-wave superconductor and $\Delta_{\mathrm{MSC}}^{(q_p=3, l_z=0)}$ monopole superconductor respectively.
$\mathfrak{F}_{s - \mathrm{MSC}}^{(q_p=3, l_z = 0)}$ itself is invariant under rotation $R_z$ and parity, as both the $s$-wave superconductor and $\Delta_{\mathrm{MSC}}^{(q_p=3, l_z=0)}$ monopole superconductor both form irreps of $l_z =0$.
As discussed in Sec.~\ref{subsec:JG.GF}, this leads to nonvanishing first-order Josephson current.
Hence, following the junction design principles introduced in Sec.~\ref{subsec:MSC.symmetry}, we may similarly design probes of monopole pairing order in different topological sectors.


\section{Josephson junctions to probe \texorpdfstring{$l_{z, \mathrm{loc}}$}{lz,\mathrm{loc}}  of the monopole superconducting order}

\label{sec:JL}

We now explore how to design Josephson junctions to probe local angular momentum $l_{z, \mathrm{loc}}$ of monopole superconducting order, following the second junction design introduced in Sec.~\ref{subsec:MSC.symmetry}. We consider two identical monopole superconductors forming a Josephson junction oriented along, for example, the $x$ direction, which is perpendicular to the high-symmetry rotational axis along $z$. 
As discussed in Sec.~\ref{subsec:MSC.symmetry}, the total Josephson current can be regarded as a superposition of contributions from a family of Josephson junctions between two $2$D chiral superconductors, corresponding to different values of conserved momentum $k_z$.
It has been studied extensively in, for example, Refs.~\citenum{Kwon2004, Alicea2012} that a planar junction between two chiral superconductors can host $4\pi$-periodic Josephson current arising from the tunneling of chiral boundary modes. 
Therefore, by studying the Josephson energy phase relation in our system, we can obtain the information about local symmetry of monopole pairing order. 
In the continuum model of a $\Delta_\mathrm{MSC}^{(q_p=-1, l_z=0)}$ monopole superconductor, its pairing order, described by Eq.~\eqref{qp=-1.MSC.hamiltonian}, exhibits local chiral $p$-wave pairing with $l_{z, \mathrm{loc}} = \mp 1$ for a given value of $\tilde{k}_z$, 
\begin{equation}
    \Delta_{\mathrm{MSC}}^{{(q_p=-1, l_z = 0)}}(\tilde{\mathbf{k}})
    = 
        \tilde{\Delta}(\tilde{k}_z)
    ( \tilde{k}_x - i \lambda \tilde{k}_y).
    \label{qp=1.jz=0.expansion.k}
\end{equation}
Here, the local pairing amplitude is modulated by $\tilde{k}_z$ as $\tilde{\Delta}(\tilde{k}_z) = -\Delta_0 \sqrt{1 - (\tilde{k}_z/\tilde{k})^2}$, and $\lambda=\pm 1$ describes the local chirality of the pairing order, which changes sign as $\tilde{k}_z$ varies from the north pole to south pole, indicative of the nontrivial topological charge of the pairing order, $q_p = -1$.

\begin{figure}
    \centering
    \includegraphics[width=\linewidth]{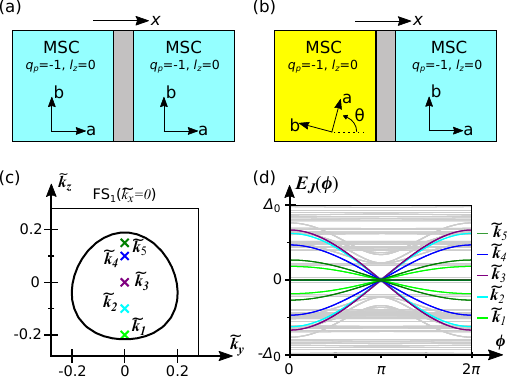}
    \caption{
    (a)~Single Josephson junction along the $x$ direction between two identical monopole superconductors with pairing order $\Delta_{\mathrm{MSC}}^{(q_p = -1, l_z = 0)}$ and the same crystal orientation.
    (b)~Same Josephson junction as in (a), only now that the superconductor at the left (yellow) is rotated by an angle $\theta$ about the $z$ axis pointing out of plane.
    (c)~Bulk Fermi surface $\mathrm{FS}_1$ of the doped Weyl semimetal on a cubic lattice.
    Representative momenta $\tilde{\mathbf{k}}_{i=1, \cdots, 5}$ are shown, with $\tilde{k}_y=0$ and $\tilde{k}_z=-0.2, -0.1, 0.0, 0.1, 0.15$, respectively.
    As a reference, we plot the largest $\tilde{k}_y\tilde{k}_z$ cross sections ($\tilde{k}_x=0$) of bulk Fermi surfaces, which are identical for the two sides of the junction.
    (d)~Josephson energy phase relations for states at the representative transverse momenta (in color) and states at the same values of $\tilde{k}_z$ but $k_y=\pm 0.1,\pm 0.2$ (in gray). 
    Here, the system size and other parameters are the same as those taken in Fig.~\ref{fig:WeylS.zjunction}. }
    \label{fig:MSC_MSC.local_probe.xjunction.epr}
\end{figure}

We next consider a Josephson junction along the $x$ direction between two monopole superconductors with identical pairing order $\Delta_\mathrm{MSC}^{(q_p = -1, l_z=0)}$, shown schematically in Fig.~\ref{fig:MSC_MSC.local_probe.xjunction.epr}(a).
Periodic boundary conditions are taken in $y$ and $z$ directions so that momenta $\tilde{k}_y$ and $\tilde{k}_z$ transverse to the junction barrier are conserved. 
At $\tilde{k}_y = 0$, the above monopole superconducting pairing order can be reduced to a family of $p_x$-wave pairing orders locally at different values of $\tilde{k}_z$ with pairing amplitudes $\tilde{\Delta}(\tilde{k}_z)$, each of which support a Majorana state localized at the junction interface contributing $4\pi$-periodic Josephson energy phase relation, $E_J \propto \cos(\phi/2)$.
Therefore, we expect the total Josephson current phase relation in our system  to exhibit $4\pi$-periodicity as a result of the local chiral $p$-wave pairing symmetry of the $\Delta_\mathrm{MSC}^{(q_p=-1, l_z=0)}$ monopole pairing order. 

\begin{figure}
    \centering
    \includegraphics[width=\linewidth]{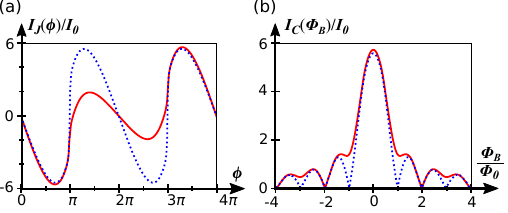}
    \caption{
    Josephson current of the junction shown in Fig.~\ref{fig:MSC_MSC.local_probe.xjunction.epr} without and with magnetic flux through the junction interface.
    (a)~The total Josephson current phase relation and (b)~the Fraunhofer pattern when conserving (red, solid) or breaking (blue, dashed) fermion parity.  
    Here, $I_0 = {e\Delta_0}/{\hbar}$,  $\Phi_{0} = {h}/{2e}$,  and $\Delta_0/t = 0.04$.
    }
    \label{fig:MSC_MSC.local_probe.xjunction.cpr}
\end{figure}

We study a tight-binding model of the Josephson junction between two identical $\Delta_\mathrm{MSC}^{(q_p = -1, l_z = 0)}$ monopole superconductors on a cubic lattice, similar to that discussed in Sec.~\ref{subsec:JG.tbm}, but with the junction now aligned along the $x$ direction, as shown schematically in Fig.~\ref{fig:MSC_MSC.local_probe.xjunction.epr}(a). 
Due to the mirror $M_z$ symmetry of the system, there are identical contributions to the Josephson energy phase relation from $\mathrm{FS}_1$ and $\mathrm{FS}_2$. 
Thus, we first examine states at five representative transverse momentum $\tilde{\mathbf{k}}_i = (0, \tilde{k}_i)$ ($i=1,\dots,5$) within $\mathrm{FS}_1$ in the $k_yk_z$ plane, as shown in Fig.~\ref{fig:MSC_MSC.local_probe.xjunction.epr}(c). 
At each conserved $\tilde{\mathbf{k}}_i$, the effective $1$D Josephson junction along $x$ direction hosts midgap states localized at the junction interface, which contribute to $4\pi$-periodic Josephson energy phase relation, as shown in Fig.~\ref{fig:MSC_MSC.local_probe.xjunction.epr}(d). 
As the magnitude of $\tilde{k}_z$ increases towards the locations of pairing gap nodes, the effective pairing amplitude, $\tilde{\Delta}(\tilde{k}_z)$, decreases. 
Furthermore, at phase difference $\phi = \pi$, the junction interface plays the role of a domain wall, where the pairing order changes its sign at the junction barrier. 
For each representative transverse momentum at $\tilde{k}_y=0$, the pairing order is odd under mirror $M_x$ reflection, ${\Delta_\mathrm{MSC}^{(q_p=-1, l_z=0)}(\tilde{k}_x, \tilde{k}_y=0, \tilde{k}_z)} = {- \Delta_\mathrm{MSC}^{(q_p=-1, l_z=0)}(-\tilde{k}_x, \tilde{k}_y=0, \tilde{k}_z)}$. 
Together with the BdG particle-hole symmetry, this system is analogous to the Jackiw-Rebbi model~\cite{Jackiw1976} and exhibits zero energy Andreev states localized at junction interface when $\phi = \pi$. 
When fermion parity is conserved at $\phi=\pi$, the total energy phase relation is $4\pi$-periodic and leads to Josephson current phase relation of the same periodicity, shown in Fig.~\ref{fig:MSC_MSC.local_probe.xjunction.cpr}(a) in red. 
As a result, when a nonzero external magnetic flux is varied through the junction interface, the Josephson Fraunhofer pattern exhibits lifted nodes at odd integer flux quanta, shown in red in Fig.~\ref{fig:MSC_MSC.local_probe.xjunction.cpr}(b). 
The conserved fermion parity can be achieved experimentally, for example, by taking a dynamic measurement of the Josephson current while maintaining a fixed voltage to prevent parity transitions~\cite{Kurter2015,Abboud2022,Yue2024}.

The existence of $4\pi$-periodic Josephson energy phase relation in this single Josephson junction oriented along the $x$ direction stems from the local spatial symmetry property of the monopole pairing order, which contains nonzero component of $p_x$-wave pairing, or say, nonzero components of $l_{z, \mathrm{loc}} = \pm 1$. 
Furthermore, due to spin-orbit coupling from Weyl band Hamiltonian, we examine the pairing order under combined spin and spatial rotations. 
The $\Delta_\mathrm{MSC}^{(q_p=-1, l_z=0)}$ monopole superconductor is invariant under simultaneous spin and spatial rotations; hence, the local $z$-component angular momentum, $l_{z, \mathrm{loc}}$, in momentum space and of spin, $s_{z}$, of the pairing order satisfies $l_{z, \mathrm{loc}} + s_z =0$.  
However, the existence of $4\pi$-periodic Josephson current constrains $l_{z, \mathrm{loc}} \neq 0$, 
\textit{i.e.} the monopole superconducting order has the symmetry of a chiral superconducting order locally in momentum space. 
Specifically,
at conserved momentum $\tilde{k}_z$, 
the above junction can only be described in analogy to that between two superconductors with spin-polarized chiral $p$-wave pairing order, where the Cooper pair is locally described by $(p_x+ip_y)|\downarrow \downarrow\rangle$ or $(p_x - ip_y)|\uparrow \uparrow\rangle$ pairing, satisfying $l_{z, \mathrm{loc}} = -s_{z}$.   
Later, in Sec.~\ref{sec:distinguish_from_chiral_SCs}, we show that the $\Delta_\mathrm{MSC}^{(q_p=-1, l_z=0)}$ monopole superconducting order is indeed unique and different from these chiral superconductors globally. 
Nonetheless, though the monopole superconducting order is inherently different from any pairing based on spherical harmonic symmetry, at given $\tilde{k}_z$, one can describe the symmetry of the monopole pairing order locally in analogy to that of a $2$D chiral superconducting pairing order. 
As such, for given conserved $\tilde{k}_z$,
the $\Delta_\mathrm{MSC}^{(q_p=-1, l_z=0)}$ superconductor can be described locally in momentum space as having chiral $p$-wave symmetry. 
Furthermore, the qualitative features of the Josephson current phase relation in Fig.~\ref{fig:MSC_MSC.local_probe.xjunction.cpr}(a), 
specifically the periodicity, are unchanged in the long junction limit, as discussed in Sec.~\ref{appendix:long_SNS}.

If the symmetry of the spin-orbit coupled system is not known \textit{a priori}, we additionally examine a second Josephson junction to demonstrate the local symmetry of pairing order. 
Consider the junction geometry where the monopole superconductor to the left of the junction barrier is rotated by an angle $\theta$ about the $z$ axis, as shown in Fig.~\ref{fig:MSC_MSC.local_probe.xjunction.epr}(b). 
If the Josephson energy phase relation is invariant under the rotation, exhibiting $4\pi$-periodicity with no phase shift, the pairing order lies in the zero total angular momentum sector with a nonzero orbital angular momentum $l_{z,\mathrm{loc}}$. 
We check this numerically using the tight-binding model, where the fourfold lattice rotation takes discrete values of $\theta = n\pi/2$ with $n \in \mathbb{Z}$, and obtain the same Josephson energy phase relation as in Fig.~\ref{fig:MSC_MSC.local_probe.xjunction.epr}(d).  
Then, following the same reasoning as in the paragraph above, 
this additional rotated junction demonstrates the local chiral $p$-wave symmetry of the $\Delta_{\mathrm{MSC}}^{(q_p=-1,l_z=0)}$ monopole pairing order. 

Generally, the monopole superconducting order is always nodal, and, locally in momentum space, its pairing order can always be described by chiral spherical harmonic pairing symmetry with angular momentum $l_{z, \mathrm{loc}}$ (see Appendix~\ref{appendix:table_of_MHs} for details). 
We can consider the same Josephson junction geometry to identify the local pairing symmetry of other monopole orders that have a different pair monopole charge or a different angular momentum $l_{z,\mathrm{glob}}$. 


\section{Long Josephson junctions}
\label{appendix:long_SNS}

\begin{figure}
    \centering
    \includegraphics[width=0.95\linewidth]{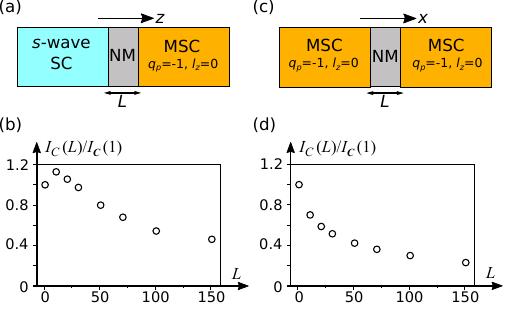}
    \caption{(a)~SNS Josephson junction along the $z$ direction between and $s$-wave and $\Delta_{\mathrm{MSC}}^{(q_p=-1, l_z =0)}$ monopole superconductor, analogous to that of 
    Fig.~\ref{fig:WeylS.GF_method}(a) but consisting of a normal metal (NM) region of length $L$ shown in gray.
    (b)~Josephson critical current as a function of $L$ in (a), normalized to the critical current for the junction of $L=1$. 
    Parameters are identical to those in Fig.~\ref{fig:WeylS.zjunction}. 
    (c)~SNS Josephson junction along the $x$ direction between identical $\Delta_{\mathrm{MSC}}^{(q_p=-1, l_z =0)}$ monopole superconductors, analogous to that of
    Fig.~\ref{fig:MSC_MSC.local_probe.xjunction.epr}(a) 
    but consisting of a normal metal (NM) region of length $L$ shown in gray.
    (d)~Josephson critical current as a function of $L$ in (c), normalized to the critical current for the junction of $L=1$. 
    Parameters are identical to those in Fig.~\ref{fig:MSC_MSC.local_probe.xjunction.epr}. 
    }
    \label{fig:long_junctions}
\end{figure}

In this section, we explore the Josephson effect in long superconductor-normal metal-superconductor (SNS) junctions using tight-binding models,  where the width of the normal metal region exceeds the superconducting coherence length. Our numerical calculations show that both the nonzero Josephson coupling and the $2\pi$- and $4\pi$-periodic Josephson current phase relations, as discussed in Secs.~\ref{sec:JG} and \ref{sec:JL}, are not exclusive to the short junction regime. Instead, these fundamental features arise from the symmetry of the superconducting order and the presence of topological boundary states, which remain robust for varying junction lengths as long as ballistic transport is maintained in the normal metal regime. However, the critical Josephson current magnitude decreases with increasing junction length.

We consider the tight-binding model Hamiltonian, analogous to the setup in Appendix~\ref{appendix:TBM_JJ},
\begin{math}
    H_\mathrm{JJ}(\mathbf{k}_\parallel) = H_{\mathrm{BdG},L}(\mathbf{k}_\parallel) + H_{\mathrm{link}}(\mathbf{k}_\parallel) + H_{\mathrm{BdG},R}(\mathbf{k}_\parallel). 
\end{math}
Here, the junction is oriented along the $\hat{\mathbf{n}}$ direction ($\hat{\mathbf{n}} = \hat{x}, \hat{z}$), and transverse momentum $\mathbf{k}_\parallel$ is conserved. The bulk superconductor BdG Hamiltonians on the left and right sides follow the form of Eq.~\eqref{JJ.BdG_hams}. 
For the normal metal region, the tight-binding Hamiltonian is defined as
\begin{equation}
    \begin{aligned}
        H_\mathrm{link} (\mathbf{k}_\parallel) = 
            \sum_{\langle n, n' \rangle}
            \sum_{\sigma
            = \uparrow, \downarrow}
            t_0 \delta_{n, n'+1}
             c^\dagger_{n,\mathbf{k}_\parallel; \sigma}
            c_{n', \mathbf{k}_\parallel; \sigma} + \mathrm{h.c.},
    \end{aligned}
    \label{Hlink.SNS.appendix}
\end{equation}
where the nearest neighbor sum $\langle n,  n'\rangle$ runs from site $N$ to site $N + L$ along $\hat{\mathbf{n}}$ at the two ends of the normal metal region. For $L=1$,
this reduces to Eq.~\eqref{Hlink.tbm.one_site}.
Here, we assume spin-independent and transverse momentum-independent nearest-neighbor hopping with constant amplitude $t_0= -0.5$.

We first study a long Josephson junction oriented along the $z$ direction, between an $s$-wave and a $\Delta_{\mathrm{MSC}}^{(q_p=-1, l_z =0)}$ monopole superconductor, as shown schematically in Fig.~\ref{fig:long_junctions}(a). The Fermi velocities for the left and right superconductors are $v_{F,L}= 1.00$ and  $v_{F,R}= 0.90$, with a pairing amplitude $\Delta_0= 0.02$ which corresponds to coherence lengths $\xi_L\approx 50$ and  $\xi_R \approx 45$ in the unit of lattice constant. As the length of normal metal region $L$ increases from $1$ to $151$, the first-order Josephson coupling remains nonvanishing, with a $2\pi$ periodic Josephson energy phase relation similar to that shown in Fig.~\ref{fig:WeylS.zjunction}(c) for the short junction studied in Sec.~\ref{subsec:JG.tbm}. The critical Josephson current $I_C$, shown in Fig.~\ref{fig:long_junctions}(b), initially increases for $L = 11$ and then decays exponentially as $L$ increases from $11$ to $151$. 

Next, we examine an SNS junction along the $x$ direction between identical $\Delta_{\mathrm{MSC}}^{(q_p=-1, l_z =0)}$ monopole superconductors, as described in Sec.~\ref{sec:JL}, with Fermi velocity $v_F = 0.51$ and coherence length $\xi \approx 25$. We find nonvanishing Josephson current and $4\pi$-periodic Josephson energy phase relation, which is qualitatively the same as that shown in Fig. \ref{fig:MSC_MSC.local_probe.xjunction.cpr}(a), as the junction length increases from $L=1$
to $L=151$. The zero-energy Majorana bound states still contribute $4\pi$ periodic Josephson current, as discussed in Sec.~\ref{sec:JL}. The Josephson critical current decays exponentially for increasing $L$, as presented in Fig.~\ref{fig:long_junctions}(d).

Overall, for both junctions in Figs.~\ref{fig:long_junctions}(a) and \ref{fig:long_junctions}(b), the qualitative properties of Josephson current and its current phase relation remain unchanged despite an exponential decay in the critical current for long junctions. These results affirm that even in the long SNS junction geometry, our design principles remain valid to identify the monopole superconducting order via probing the global and local angular momenta of the monopole superconducting order through the Josephson coupling.


\section{Josephson junctions to probe monopole superconducting order in the presence of spin-orbit interactions}
\label{sec:distinguish_from_chiral_SCs}

Lastly, in the presence of spin-orbit interactions at the junction interface or spin-orbit coupling in the bulk, the spin and orbital angular momentum are no longer conserved separately. The Josephson coupling is allowed when the pairing orders at two sides of the junction share the same value of total angular momentum along the common rotational symmetry axis following the junction geometry. 
For example, consider an SIS Josephson junction along $z$ axis between an $s$-wave superconductor with pairing $\Delta_{L,0} | \uparrow \downarrow -\downarrow \uparrow \rangle$ and a chiral $p$-wave triplet superconductor with zero total angular momentum pairing, for example $\Delta_{R,0} [(p_x - ip_y){|\uparrow \uparrow\rangle} + (p_x+ip_y){|\downarrow \downarrow\rangle}]$. 
In the presence of spin-orbit interactions, it shows nonzero first-order Josephson coupling, similar to that in the Josephson junction of the same geometry between an $s$-wave superconductor and a monopole superconductor with pairing order $\Delta_{\mathrm{MSC}}^{(q_p = -1, l_z=0)}$. 
Consequently, junction design (I) introduced in Sec.~\ref{sec:JG} cannot distinguish $\Delta^{(q_p = -1, l_z=0)}_\mathrm{MSC}$ monopole pairing order from chiral $p$-wave pairing with $j_z=0$. Furthermore, junction design (II) introduced in Sec.~\ref{sec:JL} similarly cannot distinguish them, as the Majorana surface modes localized at the junction interface between chiral $p$-wave superconductors also give rise $4\pi$-periodic Josephson current phase relation. 
Hence, we must resort to a third junction design to discern the $\mathrm{\Delta}_\mathrm{MSC}^{(q_p=-1, l_z=0)}$ superconducting order in the presence of spin-orbit interaction at the barrier.

As an example, we consider the effect of bare interface Rashba-type spin-orbit interaction. 
Generally, when spin-orbit interaction preserves time-reversal symmetry, the form factor between a spin singlet and a $q_p=-1$ monopole superconductor takes the following form:
\begin{equation}
    \mathfrak{F}_\mathrm{\mathrm{sing}-\mathrm{MSC}}^{(q_p = -1)} (\tilde{\mathbf{k}}, \mathbf{k}')
    =
    \mathfrak{f}_0(\tilde{\mathbf{k}}, \mathbf{k}')
    +
    \mathfrak{f}_1(\tilde{\mathbf{k}}, \mathbf{k}')
    +
    \mathfrak{f}_2(\tilde{\mathbf{k}}, \mathbf{k}'),
    \label{form_factor.qp=-1.singlet.with_SOC}
\end{equation}
where
\begin{subequations}
    \begin{equation}
        \begin{aligned}
            &\mathfrak{f}_0(\tilde{\mathbf{k}}, \mathbf{k}')
            = 
            - 4\pi \Delta_{L,0}(\mathbf{k}') [\Delta_{R, \mathrm{MSC}}^{(q_p = -1)}(\tilde{\mathbf{k}}) ]^*
            \\
            &
            \times
            T_0^2
            (\mathbf{k}, \mathbf{k}')
            \mathcal{Y}_{-\frac{1}{2}; \frac{1}{2}, - \frac{1}{2}}(\boldsymbol{\Omega}_{\tilde{k}})
            \mathcal{Y}_{-\frac{1}{2}; \frac{1}{2}, \frac{1}{2}}(\boldsymbol{\Omega}_{\tilde{k}}),
        \end{aligned}
    \end{equation}
    \begin{equation}
        \begin{aligned}
            &
            \mathfrak{f}_1(\tilde{\mathbf{k}}, \mathbf{k}')
            = 
            - 4\pi \Delta_{L,0}(\mathbf{k}') [\Delta_{R, \mathrm{MSC}}^{(q_p = -1)}(\tilde{\mathbf{k}}) ]^*
            T_0
            (\mathbf{k}, \mathbf{k}')
            \\
            &
            \times
            \Big\{
            \mathcal{Y}_{-\frac{1}{2}; \frac{1}{2}, + \frac{1}{2}}^2(\boldsymbol{\Omega}_{\tilde{k}})
            [
            T_{x}
            (\mathbf{k}, \mathbf{k}') - iT_{y}
            (\mathbf{k}, \mathbf{k}')
            ]
            \\
            &
            \hspace{1em} +
            \mathcal{Y}_{-\frac{1}{2}; \frac{1}{2}, - \frac{1}{2}}^2(\boldsymbol{\Omega}_{\tilde{k}})
            [
            T_{x}
            (\mathbf{k}, \mathbf{k}')
            + i T_{y}
            (\mathbf{k}, \mathbf{k}')
            ]
            \Big\},
        \end{aligned}
    \end{equation}
    and
    \begin{equation}
        \begin{aligned}
            &\mathfrak{f}_2(\tilde{\mathbf{k}}, \mathbf{k}')
            = 
            - 4\pi \Delta_{L,0}(\mathbf{k}') [\Delta_{R, \mathrm{MSC}}^{(q_p = -1)}(\tilde{\mathbf{k}}) ]^*
            \\
            &
            \times
            |\mathbf{T}
            (\mathbf{k}, \mathbf{k}')
            |^2
            \mathcal{Y}_{-\frac{1}{2}; \frac{1}{2}, - \frac{1}{2}}(\boldsymbol{\Omega}_{\tilde{k}})
            \mathcal{Y}_{-\frac{1}{2}; \frac{1}{2}, \frac{1}{2}}(\boldsymbol{\Omega}_{\tilde{k}}).
        \end{aligned}
    \end{equation}
\end{subequations}
Here, $\mathfrak{f}_0$, $\mathfrak{f}_1$, and  $\mathfrak{f}_2$ denote the contributions to the form factor that are independent of, linear in, and quadratic in spin-orbit interaction amplitude, respectively. 
We take $\mathbf{T}(\mathbf{k}, \mathbf{k}') = - \mathbf{T}(-\mathbf{k}, -\mathbf{k}')$ and $T_0(\mathbf{k}, \mathbf{k}') = T_0(-\mathbf{k}, -\mathbf{k}')$ so that the tunneling Hamiltonian in Eq.~\eqref{tunneling_hamiltonian.spin_basis} has time-reversal symmetry.

\begin{figure}
    \centering\includegraphics[width=\linewidth]{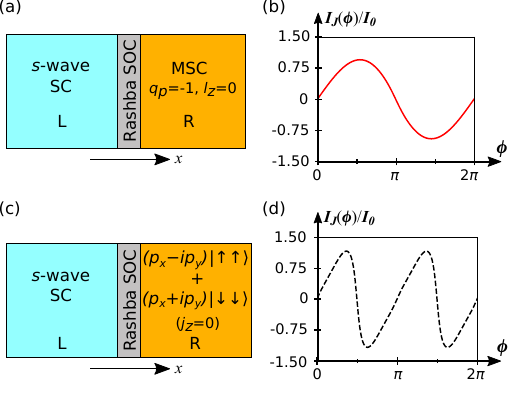}
    \caption{
    Tight-binding results of Josephson current for a junction with an $s$-wave superconductor in the presence of Rashba-type spin-orbit interaction at the junction interface.
    (a)~Schematic of single Josephson junction between an $s$-wave superconductor and $\Delta_\mathrm{MSC}^{(q_p = -1, l_z=0)}$ monopole superconductor and  along the $x$ direction.
    (b)~$2\pi$-periodic Josephson current phase relation (red, solid) for 
    the junction shown in (a).
    (c)~Schematic of junction between an $s$-wave superconductor and  $(p_x - ip_y)\left|\uparrow \uparrow\right \rangle + (p_x + ip_y)\left|\downarrow \downarrow\right \rangle$ superconductor.
    (d)~$\pi$-periodic Josephson current phase relation (black, dashed) of the junction shown in (c).
                Results are computed using a system of $2N_x = 300$ sites and with the same parameters of the tight-binding band Hamiltonians in Fig.~\ref{fig:WeylS.zjunction}.
    For simplicity, $t_{\mathrm{SO}} = t_0$ at the interface.
        }
    \label{fig:jz=0.chiralp_junction}
\end{figure}

Now we consider a Josephson junction along $x$ direction between an $s$-wave superconductor and the $\Delta_\mathrm{MSC}^{(q_p=-1, l_z=0)}$ monopole superconductor, as shown schematically in Fig.~\ref{fig:jz=0.chiralp_junction}(a). 
In the presence of Rashba-type spin-orbit interaction at the junction interface, 
$\mathbf{T}(\mathbf{k},\mathbf{k}') = T_\mathrm{SO} \delta_{\mathbf{k}_{\parallel}, \mathbf{k}'_{\parallel}} \hat{\mathbf{x}} \times \hat{\mathbf{k}} $. 
For simplicity, we take the spin-independent tunneling amplitude as a constant, $T_0(\mathbf{k}, \mathbf{k}') = T_0$.
Terms in the form factor in Eq.~\eqref{form_factor.qp=-1.singlet.with_SOC} that  contribute to nonvanishing first-order Josephson coupling are given by
\begin{math}
    \mathfrak{f}_0(\tilde{\mathbf{k}}, \mathbf{k}') = - T_0^2 \Delta_{L,0} \Delta_{R,0} \sin^2 \theta_{\tilde{k}}
\end{math}
and
\begin{math}
    \mathfrak{f}_2(\tilde{\mathbf{k}}, \mathbf{k}') = - T_\mathrm{SO}^2 
    |\hat{\mathbf{x}} \times \hat{\mathbf{k}}|^2 \Delta_{L,0} \Delta_{R,0} \sin^2 \theta_{\tilde{k}} \delta_{\mathbf{k}_{\parallel}, \mathbf{k}'_{\parallel}},
\end{math}
with $\Delta_{L,0}$ and $\Delta_{R,0}$ being the pairing amplitudes of the $s$-wave  and $\Delta_\mathrm{MSC}^{(q_p=-1, l_z=0)}$ monopole superconductor respectively.
Specifically, contributions from $\mathfrak{f}_0$ and $\mathfrak{f}_2$  are invariant under rotation $R_x$ and thus lead to 
nonzero first-order Josephson current between the $\Delta_\mathrm{MSC}^{(q_p=-1, l_z=0)}$ monopole superconductor and $s$-wave superconductor in a junction along the $x$ direction.

In contrast,
a $(p_x - ip_y){|\uparrow \uparrow \rangle} + (p_x +ip_y){|\downarrow \downarrow\rangle}$ chiral $p$-wave superconductor has vanishing first-order Josephson coupling with an $s$-wave superconductor in the same junction geometry, shown schematically in Fig.~\ref{fig:jz=0.chiralp_junction}(c), and with the same Rashba-type spin-orbit interaction at the interface. 
The form factor in Eq.~\eqref{single_triplet_form_factor} simplifies to $\mathfrak{F}_{\mathrm{sing}-\mathrm{trip}}
\propto
T_0 T_\mathrm{SO} \hat{x} \cdot (\hat{\mathbf{k}} \times \mathbf{d}_R^* (\mathbf{k}) )$.
Here, the $d$-vector in consideration is  $\mathbf{d}_R(\mathbf{k}) \propto (k_y, -k_x, 0)$.
The form factor reduces to
$\mathfrak{F}_{\mathrm{sing}-\mathrm{trip}}(\mathbf{k}, \mathbf{k}') \propto 
k_x k_z$, which is odd under mirror $M_z$ reflection, $k_z \mapsto -k_z$.
Therefore, the net first-order Josephson current of this system is zero. 
As such, the presence of first-order Josephson current in this additional junction design can be used to distinguish $(p_x - ip_y){|\uparrow \uparrow \rangle} + (p_x +ip_y){|\downarrow \downarrow\rangle}$ chiral superconductor from the $\Delta_\mathrm{MSC}^{(q_p=-1, l_z=0)}$ monopole superconductor.
For other monopole superconducting pairing orders in different $q_p$ or $l_z$ sectors, one can design similar junctions to distinguish monopole superconducting order from chiral pairing orders in the presence of bare spin-orbit interactions at the interface.

We demonstrate the analytic results above using numerical calculation of the Josephson current based on tight-binding models. The system is analogous to that introduced in Sec.~\ref{subsec:JG.tbm}, only now the junction is oriented along $x$ direction and includes spin-dependent tunneling at the interface.
For given conserved transverse momenta $k_y$ and $k_z$, we model the junction barrier as
\begin{equation}
    \begin{aligned}
        &H_\mathrm{link} (k_y, k_z) 
        =
        \\
        &         \sum_{\sigma, \sigma' = \uparrow, \downarrow} 
                c^\dagger_{k_y, k_z, \sigma, N_x+1}
        \left[ \hat{t}(k_y, k_z)\right]_{\sigma, \sigma'}
        c_{k_y, k_z, \sigma', N_x}
        + \mathrm{h.c.}
    \end{aligned}
\end{equation}
Here, $\hat{t}(k_y, k_z) = t_0 \sigma_0 + \mathbf{t}(k_y, k_z) \cdot \boldsymbol{\sigma}$,
where $t_0$ is the spin-independent hopping amplitude, and
$\mathbf{t}(k_y, k_z) = t_{\mathrm{SO}}(\sin k_y \hat{z} - \sin k_z \hat{y})$ is the Rashba-type spin-orbit coupling at the interface, with $t_\mathrm{SO}$ being the spin-dependent hopping amplitude.

The Josephson current for the junction along $x$ direction between a $\Delta_\mathrm{MSC}^{(q_p=-1, l_z=0)}$ monopole superconductor and $s$-wave superconductor is shown in Fig.~\ref{fig:jz=0.chiralp_junction}(b).
In agreement with the above arguments using the form factor $\mathfrak{F}_{\mathrm{sing}-\mathrm{MSC}}^{(q_p=-1)}$, there is an overall $2\pi$-periodic Josephson current corresponding to nonvanishing first-order Josephson coupling.
In contrast, for a junction between an $s$-wave superconductor and $(p_x - ip_y){|\uparrow \uparrow\rangle}+ (p_x + ip_y)\left|\downarrow \downarrow\right \rangle$ chiral superconductor,
there is no first-order Josephson coupling, as shown in Fig.~\ref{fig:jz=0.chiralp_junction}(d).
Rather, the leading order Josephson coupling between the $s$-wave and chiral $p$-wave superconductor is a second-order process, corresponding to $\pi$-periodic Josephson current.
It is to be noted that because we have taken the spin-independent and spin-dependent tunneling amplitudes to both be of order $1$ in the tight-binding model, the magnitude of the Josephson critical current for the two junctions is similar; however, this does not change the qualitative aspects, specifically the periodicity. 
The numerical results are in agreement with the above microscopic analysis and demonstrate that this additional junction can be used to distinguish $\Delta_\mathrm{MSC}^{(q_p=-1, l_z=0)}$ pairing order from other chiral $p$-wave superconductors in the presence of spin-orbit interaction at the interface.
Similar results hold for a chiral $(p_x + ip_y)|\downarrow \downarrow\rangle$ superconductor, $(p_x - ip_y)\left|\uparrow \uparrow\right \rangle$ superconductor, or linear superposition of the two.
In other words, the monopole superconducting order cannot be described as a simple linear superposition of spin-polarized chiral superconducting pairing orders globally but rather exists in its own distinct topological class.

\section{Conclusion}

We have proposed using a set of phase-sensitive Josephson junctions to probe monopole superconducting orders and to distinguish them from other known pairing orders with spherical harmonic symmetry. We develop the designs based on two approaches. Firstly, we employed the linear response approach to derive the first-order Josephson current between two uniform superconductors. 
In addition to accounting for the microscopic details of the tunneling processes, including spin-orbit interactions at the junction interface and spin-orbit coupling in the bulk superconductors, we emphasize that the symmetry of superconducting pairing orders and the junction geometry is encoded in a form factor $\mathfrak{F}(\mathbf{k}, \mathbf{k}')$, which is nonsingular and determines the symmetry selection rules of the first-order Josephson coupling. 
Complementing the first approach, we have also proposed symmetry and topological principles in the design of phase-sensitive probes of monopole superconducting order. 
We noted that the $\mathrm{U}(1)$ topological obstruction of the pairing phase can be exhibited as a nonzero shift between the global and local angular momentum of the monopole superconducting order.  
Then, following symmetry principles, we proposed two classes of Josephson junctions of monopole superconductors to probe the global and local angular momentum of the superconducting order.

We considered the first class of Josephson junctions aligned along the rotational axis ($z$ axis) between a monopole superconductor and other superconductors with known pairing symmetry. 
As the first-order $2\pi$-periodic Josephson current is allowed only when superconducting orders at both sides of the junction transform under rotation $R_z$ in the same way, we can use this to probe $l_{z, \mathrm{glob}}$, the global angular momentum of the monopole pairing order. We have studied Josephson currents in this class of junctions between monopole superconductors with different pair monopole charges and different $l_{z, \mathrm{glob}}$ and superconductors with standard $s$- and $d$-wave spherical harmonic symmetry, using both linear response and tight-binding calculations. 
The second class of junctions, designed to extract the local angular momentum $l_{z, \mathrm{loc}}$ of the pairing order, consists of two identical monopole superconductors in a junction aligned perpendicular to the rotational $z$ axis. 
The Josephson current can be viewed as a superposition of contributions from different values of conserved momentum $k_z$ local in momentum space. By analyzing the existence of $4\pi$-periodic Josephson current at different orientations of the monopole superconductor based on tight-binding models, 
we have identified the local chiral symmetry of monopole pairing order characterized by angular momentum $l_{z, \mathrm{loc}}$. The shift between $l_{z, \mathrm{glob}}$ and $l_{z, \mathrm{loc}}$ indicates the nonzero pair monopole charge of the exotic superconducting order. The underlying symmetry and topological principles of this work provide useful guidance for the ongoing experimental efforts toward realizing and identifying exotic monopole superconducting pairing orders in doped topological semimetal materials. 

\section*{Acknowledgments}
We acknowledge the support of the NSF CAREER Grant No.~DMR-1848349. In addition, YL acknowledges the Institute for Quantum Matter, an Energy Frontier Research Center funded by the U.S. Department of Energy, Office of Science, Office of Basic Energy Sciences, under Award DESC0019331, for support during the initial stage of this work. JYZ acknowledges partial support from the Johns Hopkins University Theoretical Interdisciplinary Physics and Astronomy Center.

\appendix

\section{Linear response theory of Josephson coupling in the presence of spin-orbit interactions}
\label{appendix:GF_method}
We review the derivation of the first-order Josephson current in an SIS junction via linear response theory, accounting for a general form of spin-orbit interaction in the bulk superconductors or at the junction interface.
We consider the Hamiltonian of the Josephson junction presented in Eq.~\eqref{BdG_JJ}, where two superconductors are connected by a weak link modeled by the tunneling Hamiltonian in Eq.~\eqref{tunneling_hamiltonian.spin_basis}. 

\subsection{First-order Josephson current and form factor in the presence of spin-dependent tunneling}
\label{appendix.subec:josephson_coupling}
The total electric tunneling current through the junction is
$I(t) = |e| \langle \dot{N}_{L}(t) \rangle$, where $N_{L} = \sum_{\sigma, \mathbf{k}} c^\dagger_{L, \mathbf{k}, \sigma} c_{L, \mathbf{k}, \sigma}$ is the number of particles at the left side of the barrier. 
We evaluate using linear response theory, where $H_\mathrm{link}$ in Eq.~\eqref{tunneling_hamiltonian.spin_basis} is treated perturbatively in the thermal weight expansion to the first-order.
the first-order current is given by 
$I(t) = 
{-i (e/\hbar) \int_{-\infty}^{+\infty} \mathrm{d}t' \Theta(t-t') \langle [ \dot{N}_L(t), H_{\mathrm{link}}(t')]\rangle}$, 
where 
$\Theta(t-t')$ is the Heaviside function, and $\dot{N}_L(t)$ and $H_\mathrm{link}(t')$ are in the interaction picture.
It follows that the current is a superposition of the normal current from quasiparticle tunneling and the Josephson supercurrent from the tunneling of Cooper pairs~\cite{Mahan2000}.
The Josephson tunneling current is given by the imaginary part of the current-current correlation function,
\begin{equation}
I_J (t, \phi)
=
-\frac{2e}{\hbar}
\mathrm{Im}
[
e^{-2ieVt}
\Phi_\mathrm{ret}(eV, \phi)
],
\end{equation}
where $eV = \mu_L - \mu_R$ is the voltage bias, and $\phi = \phi_R - \phi_L$ is the overall $\mathrm{U}(1)$ phase difference of the superconducting order parameters.
Note that in the main text, we work at zero bias, $eV = 0$.  
Here, the correlation function is given by
$\Phi_\mathrm{ret}(t-t', \phi) = 
{-i \Theta(t-t')}
\langle[A(t), A(t')]\rangle$, where
\begin{equation}
    A(t)
    =
    \sum_{\mathbf{k}, \sigma; \mathbf{k}', \sigma'} T_{\sigma, \sigma'}(\mathbf{k}, \mathbf{k}') c^\dagger_{R, \mathbf{k}, \sigma} (t)
    c_{L, \mathbf{k}', \sigma'}(t)
\end{equation}
is the current operator and $T(\mathbf{k}, \mathbf{k}')$ is the tunneling matrix.
It is assumed that
the two states located at the left and right sides satisfy $\{c_{L, \mathbf{k}', \sigma'}, c_{R, \mathbf{k}, \sigma}\} = \{c_{L, \mathbf{k}',\sigma'}, c^\dagger_{R, \mathbf{k}, \sigma}\} =0$.
Although the above correlation function is derived from first-order perturbation theory, it corresponds to a second-order process in $T_{\sigma, \sigma'}$, which can be regarded as the tunneling of a Cooper pair.

We evaluate the corresponding correlation function in imaginary frequency space, $\Phi(i\omega, \phi) = -\int_{0}^\beta e^{i\omega \tau} \mathrm{d}\tau \langle \mathcal{T}_\tau A(\tau) A(0)\rangle$ and expand with Wick's theorem:
\begin{equation}
    \begin{aligned}
        \Phi(i\omega, \phi)
        = 
        &-
        \sum_{\mathbf{k}, \mathbf{k}'}
        \sum_{\sigma, \sigma', \rho, \rho'}  
        T_{\sigma, \sigma'}(\mathbf{k}, \mathbf{k}')
        T_{\rho,\rho'}(-\mathbf{k}, -\mathbf{k}')
        \\
        &\times
        \frac{1}{\beta}
        \sum_{ip_n}
        \mathcal{F}_{R, \sigma, \rho}^\dagger(\mathbf{k}, ip_n)
        \mathcal{F}_{L, \sigma', \rho'} (\mathbf{k}', ip_n-i\omega).
    \end{aligned}
\end{equation}
Here, $p_n = \pi(2n+1)/\beta$ with $n \in \mathbb{Z}$ are the fermion Matsubara frequencies, and $\mathcal{F}_{L, \sigma', \rho'}(\mathbf{k}, ip_n-i\omega)$ and $\mathcal{F}^\dagger_{R, \sigma, \rho}(\mathbf{k}, ip_n)$ are the Fourier transforms of the anomalous {Green's} functions,
$\mathcal{F}_{L, \sigma', \rho'}(\mathbf{k}, \tau) = \langle \mathcal{T}_\tau c_{L, -\mathbf{k}, \rho'}(\tau) c_{L, \mathbf{k}, \sigma'}(0) \rangle$ 
and $\mathcal{F}^\dagger_{R, \sigma, \rho}(\mathbf{k}, \tau) = \langle \mathcal{T}_\tau c^\dagger_{R, \mathbf{k}, \sigma}(\tau) c^\dagger_{R, -\mathbf{k}, \rho} (0) \rangle$, with $\mathcal{T}_\tau$ being the time-ordering operator.
It follows that the correlation function can be expressed as~\cite{Sigrist1991}
\begin{equation}
    \begin{aligned}
        \Phi(i\omega, \phi)
        =
        &-\frac{1}{\beta} \sum_{ip_n}
        \sum_{\mathbf{k}, \mathbf{k}'}
        \mathrm{Tr}
        [
        \mathcal{F}_L(\mathbf{k}, ip_n - i\omega)
        \\
        &\times
        T^\mathrm{T}(-\mathbf{k}, -\mathbf{k}')
        \mathcal{F}^{*}_R(\mathbf{k}, ip_n)
        T(\mathbf{k}, \mathbf{k}')
        ],
    \end{aligned}
    \label{correlation_fnc.general.appendix}
\end{equation}
where the trace is taken over any internal degrees of freedom (spin, orbital, \textit{etc.}).

Following Eq.~\eqref{BdG_hamiltonian_general},
we consider the following mean-field BdG Hamiltonian in $(\mathbf{c}_{\alpha, \mathbf{k}}, \mathbf{c}^\dagger_{\alpha, -\mathbf{k}})$ Nambu basis,
\begin{equation}
    \mathcal{H}_{\mathrm{BdG}, \alpha} (\mathbf{k})= 
    \left(
    \begin{array}{cc}
        \mathcal{H}_{\mathrm{kin}, \alpha}(\mathbf{k})  & \Delta_{\alpha}(\mathbf{k})e^{i\phi_{\alpha}}
        \\
        \Delta^\dagger_{\alpha}(\mathbf{k})e^{-i\phi_{\alpha}}
        & -\mathcal{H}^\mathrm{T}_{\mathrm{kin},\alpha}(-\mathbf{k})
    \end{array}
    \right),
\end{equation}
with $\alpha = L (R)$ corresponding to the left (right) side.
Here, $\mathcal{H}_{\mathrm{kin}, \alpha}(\mathbf{k})$ is the band Hamiltonian, $\Delta_\alpha(\mathbf{k})$ is the pairing matrix, and $\phi_\alpha$ is the overall $\mathrm{U}(1)$ superconducting phase, taken to be uniform.
The anomalous Green's functions can generally be expressed as~\cite{Luders1971}
\begin{widetext}
\begin{equation}
    \mathcal{F}_{\alpha}(\mathbf{k}, ip_n)
    =
    -\mathcal{G}_{\mathrm{kin},\alpha}(\mathbf{k}, ip_n)
    \Delta_\alpha(\mathbf{k})
    e^{i\phi_\alpha}
    \left[
    1 + \mathcal{G}^{\mathrm{T}}_{\mathrm{kin},\alpha}(-\mathbf{k}, -ip_n)
    \Delta^\dagger_\alpha(\mathbf{k})
    \mathcal{G}_{\mathrm{kin},\alpha}(\mathbf{k}, ip_n)
    \Delta_\alpha(\mathbf{k})
    \right]^{-1}
    \mathcal{G}_{\mathrm{kin},\alpha}^{\mathrm{T}}(-\mathbf{k}, -ip_n)
    \label{anomalous_GF.general}
\end{equation}
\end{widetext}
where $\mathcal{G}_{\mathrm{kin},\alpha}(\mathbf{k}, ip_n) = [ip_n - \mathcal{H}_\mathrm{kin}(\mathbf{k})]^{-1}$
is the single-particle Green's function of the band Hamiltonian in the absence of superconductivity.
We work in the weak-coupling regime and consider one of the two following scenarios, under which the anomalous Green's functions greatly simplify.

In the first scenario, the band Hamiltonian is spin-independent and thus the Green's functions $\mathcal{G}_{\mathrm{kin}, \alpha}$ are diagonal in spin space.
For simplicity, we consider the case where the Hamiltonian is in the spin-1/2 representation; though, generalization to systems with other internal degrees of freedom is straightforward.
Consider the pairing matrix given by
\begin{math}
    \Delta_\alpha(\mathbf{k}) = (d_{\alpha, 0} (\mathbf{k})\sigma_0 + \mathbf{d}_\alpha(\mathbf{k}) \cdot \boldsymbol{\sigma})i\sigma_y,
\end{math}
where $d_{\alpha,0}$ is the spin singlet pairing and $\mathbf{d}_\alpha$ is the $d$-vector corresponding to spin triplet pairing.
In this scenario,
the anomalous Green's functions are simplified as
\begin{widetext}
\begin{equation}
    \mathcal{F}_{\alpha}(\mathbf{k}, ip_n)
    =
    \frac{- \Big[f_{\alpha, 0}(\mathbf{k}, ip_n) + \mathbf{f}_{\alpha}(\mathbf{k}, ip_n)\cdot \boldsymbol{\sigma}\Big](i\sigma_y)}
    { \Big[p_n^2 + \xi_{\alpha, \mathbf{k}}^2 + \frac{1}{2} \mathrm{Tr}(\Delta_\alpha^\dagger\Delta_\alpha)\Big]^2
    -\Big[
    (\mathbf{d}^*_\alpha \cdot \mathbf{d}^*_\alpha)^2(d_{\alpha,0}^2)
    +
    2
    |\mathbf{d}_\alpha|^2 |d_{\alpha,0}|^2
    +
    (d^*_{\alpha, 0})^2(\mathbf{d}_\alpha \cdot \mathbf{d}_\alpha)
    +
    |\mathbf{d}_\alpha \times \mathbf{d}_\alpha^*|^2
    \Big]
    },
\end{equation}
where
\begin{subequations}
    \begin{align}
        f_{\alpha, 0} (\mathbf{k}, ip_n)
        &\equiv
        d_{\alpha, 0}\Big[p_n^2 + \xi_{\alpha, \mathbf{k}}^2 + \frac{1}{2}\mathrm{Tr}(\Delta_\alpha^\dagger\Delta_\alpha)\Big]
        -
        \Big[
        |\mathbf{d}|^2d_{\alpha, 0}
        +
        (\mathbf{d}_\alpha \cdot \mathbf{d}_\alpha)d_{\alpha, 0}^*
        \Big],
        \\
        \mathbf{f}_{\alpha} (\mathbf{k}, ip_n)
        &\equiv
        \Big[p_n^2 + \xi_{\alpha, \mathbf{k}}^2 + \frac{1}{2}\mathrm{Tr}(\Delta_\alpha^\dagger \Delta_\alpha) \Big]
        \mathbf{d}_\alpha
        -
        \Big[ d_{\alpha, 0}^2 \mathbf{d}_\alpha^*
        +
        |d_{\alpha, 0}|^2 \mathbf{d}_\alpha
        +
        \mathbf{d}_\alpha \times (\mathbf{d}_\alpha \times \mathbf{d}_\alpha^*)
        \Big]
    \end{align}
\end{subequations}
include corrections from nonunitary pairing orders.
Here,
$\xi_{\alpha, \mathbf{k}} = \xi_{\alpha, -\mathbf{k}}$ is the band energy dispersion at the left ($\alpha = L$) and right ($\alpha = R$) side of the junction. 
$d_{\alpha, 0}$ and $\mathbf{d}_\alpha$ are both functions of $\mathbf{k}$,  which we have dropped for convenience of notation.
When the pairing is unitary ($\Delta_\alpha^\dagger\Delta_\alpha \propto \sigma_0$), the anomalous Green's functions take a simpler form,
\begin{equation}
    \mathcal{F}_\alpha (\mathbf{k}, ip_n)
    = - \frac{e^{i \phi_\alpha} \Delta_\alpha(\mathbf{k})}{p_n^2 + E^2_{\alpha, \mathbf{k}}},
    \label{anomalous_GF}
\end{equation}
where $E_{\alpha, \mathbf{k}} = \sqrt{\xi_{\alpha, \mathbf{k}}^2 + \mathrm{Tr}[\Delta_\alpha^\dagger(\mathbf{k}) \Delta_\alpha(\mathbf{k})]/2}$
is the unperturbed dispersion of the BdG quasiparticle.

In the second scenario, there are spin-dependent 
terms in the band Hamiltonian.
Working in the weak-coupling regime, we transform to a band-diagonal representation and then project to the states at the Fermi surface 
(see Appendix~\ref{appendix:effective_pairing_channels}).
Then, the anomalous Green's functions take an analogous form,
\begin{align}
    \mathcal{F}_\alpha^{(bp)}(\mathbf{k}, ip_n)
    &=
    \nonumber
    -\mathcal{G}^{(bp)}_{\mathrm{kin},\alpha}(\mathbf{k}, ip_n)
    \Delta^{(bp)}_\alpha(\mathbf{k})
    e^{i\phi_\alpha}
    \left[
    1 + \mathcal{G}^{(b)\mathrm{T}}_{\mathrm{kin},\alpha}(-\mathbf{k}, -ip_n)
    \Delta^{(b)\dagger}_\alpha(\mathbf{k})
    \mathcal{G}^{(bp)}_{\mathrm{kin},\alpha}(\mathbf{k}, ip_n)
    \Delta^{(bp)}_\alpha(\mathbf{k})
    \right]^{-1}
    \mathcal{G}^{(b) \mathrm{T}}_{\mathrm{kin},\alpha}(-\mathbf{k}, -ip_n)
    \\
    & =
    - \frac{e^{i \phi_\alpha} \Delta^{(bp)}_\alpha(\mathbf{k})}{p_n^2 + E^{(bp)2}_{\alpha, \mathbf{k}}}.
\end{align}
Here, $\mathcal{G}^{(bp)}_\mathrm{kin, \alpha} (\mathbf{k}, ip_n) 
= P_+^{(b)}(ip_n - \Lambda(\mathbf{k}))^{-1}P_+^{(b)}$ is the Green's function for the band Hamiltonian evaluated in the band-diagonal representation and projected to the Fermi level, with $P_+^{(b)}$ being the projection operator and $\Lambda(\mathbf{k})$ the diagonal matrix of eigenvalues.
$\Delta_\alpha^{(bp)}(\mathbf{k})$ is the projected effective pairing matrix in the band-diagonal representation, as shown in Eq.~\eqref{band_projected_pairing_matrix.appendix} of Appendix~\ref{appendix:effective_pairing_channels}, and
$E^{(bp)}_{\alpha, \mathbf{k}} = \sqrt{\xi_{\mathbf{k}}^{(bp)2} + \mathrm{Tr}[\Delta^{(bp) \dagger}_{\alpha}(\mathbf{k}) \Delta^{(bp)}_{\alpha}(\mathbf{k})]}$ is the energy of the BdG quasiparticle after projection, with $\xi^{(bp)}_{\alpha, \mathbf{k}} = \xi^{(bp)}_{\alpha, -\mathbf{k}}$ being the band dispersion of the eigenstate at the Fermi level.
The above form 
is justified in the weak-coupling regime, as pairing only occurs between states near the Fermi surface.

Under the outlined scenarios,
the correlation function can simply be expressed as
\begin{equation}
    \Phi(i\omega, \phi)
    =
        - e^{-i\phi}
    \sum_{\mathbf{k}, \mathbf{k}'}
    w(E_{L, \mathbf{k}'}, E_{R, \mathbf{k}}; \beta, i\omega)
    \mathfrak{F}
    (\mathbf{k}, \mathbf{k}'),
\end{equation}
as shown in Eq.~\eqref{correlation_fnc.simplified}.
As described in Sec.~\ref{sec:JC}, the above correlation function can be considered in three distinct parts as follows: 

\textbf{(I)}~The first term, $e^{-i\phi}$, includes superconducting phase difference, $\phi = \phi_R - \phi_L$, and corresponds to the first-order $2\pi$-periodic Josephson current.
For higher-order contributions to the Josephson current, it is necessary to consider terms arising from multiple scatterings at the junction interface.

\textbf{(II)}~The second term is the function, $w(\mathbf{k}, \mathbf{k}'; \beta, i\omega)$,  which encodes the density of states and thermal weighting.
After performing the Matsubara frequency summation over fermion frequencies $ip_n$, the function is given by
\begin{equation}
    \begin{aligned}
        &w(E_{L, \mathbf{k}'}, E_{R, \mathbf{k}}; \beta, i\omega)
        =
                \frac{1}{4E_{R, \mathbf{k}} 
        E_{L, \mathbf{k}'}}
        \\
        &\times
        \Big[
        \Big(
        n_F(E_{L, \mathbf{k}'}) - n_F(E_{R, \mathbf{k}})
        \Big)
        \left(
        \frac{1}{E_{L, \mathbf{k}'} - E_{R, \mathbf{k}} - i\omega}
        +
        \frac{1}{E_{L, \mathbf{k}'} - E_{R, \mathbf{k}} + i\omega}
        \right)
        \\
        &\hspace{1em}+
        \Big(
        1
        -
        n_F(E_{L, \mathbf{k}'}) - n_F(E_{R, \mathbf{k}})
        \Big)
        \left(
        \frac{1}{E_{L, \mathbf{k}'} + E_{R, \mathbf{k}} + i\omega}
        +
        \frac{1}{E_{L, \mathbf{k}'} + E_{R, \mathbf{k}} - i\omega}
        \right)
        \Big],
    \end{aligned}
    \label{f_fnc.full}
\end{equation}
where 
$n_F$ is the Fermi-Dirac distribution.
The final expression for the retarded correlation function is obtained via analytic continuation, $i\omega \rightarrow  eV + i\delta$ as $\delta \rightarrow 0^+$.

\textbf{(III)}~The last and most essential term is the form factor, 
$
\mathfrak{F}(\mathbf{k}, \mathbf{k}')
=
\mathrm{Tr}
[
{\Delta_{L}(\mathbf{k}')}
{T^\mathrm{T}(-\mathbf{k},-\mathbf{k}')}
{\Delta_{R}^*(\mathbf{k})}
{T(\mathbf{k},\mathbf{k}')}
],
$
which encodes the symmetry of the order parameters in addition to the microscopic tunneling processes.
For a two-band basis, the pairing orders can be expressed as  $\Delta_{L(R)}(\mathbf{k}) = (d_{L(R), 0} (\mathbf{k}) + \mathbf{d}_{L(R)}(\mathbf{k})\cdot \boldsymbol{\sigma})i\sigma_y$, and the tunneling matrix as $T(\mathbf{k}, \mathbf{k}') = T_0(\mathbf{k}, \mathbf{k}') \sigma_0 + \mathbf{T}(\mathbf{k}, \mathbf{k}') \cdot \boldsymbol{\sigma}$.  
The form factor evaluates to
\begin{equation}
    \begin{aligned}
        \frac{1}{2} \mathfrak{F}(\mathbf{k}, \mathbf{k}')
        &=
        - d_{L,0}(\mathbf{k}')T_0(-\mathbf{k}, -\mathbf{k}')d^*_{R,0}(\mathbf{k}) T_0(\mathbf{k}, \mathbf{k}')
        +
        \Big[ \mathbf{d}_L(\mathbf{k}') \cdot \mathbf{T}(-\mathbf{k}, -\mathbf{k}')\Big] d^*_{R,0}(\mathbf{k}) T_0(\mathbf{k}, \mathbf{k}')
        \\
        &+
        d_{L,0}(\mathbf{k}') T_0(-\mathbf{k}, -\mathbf{k}') \Big[ \mathbf{d}_R^*(\mathbf{k}) \cdot \mathbf{T}(\mathbf{k}, \mathbf{k}') \Big]
        + d_{L,0}(\mathbf{k}') d^*_{R,0}(\mathbf{k})  \Big[ \mathbf{T}(-\mathbf{k}, -\mathbf{k}') \cdot \mathbf{T}(\mathbf{k}, \mathbf{k}')\Big]
        \\
        &- d_{L,0}(\mathbf{k}') T_0(\mathbf{k}, \mathbf{k}') \Big[\mathbf{T}(-\mathbf{k}, -\mathbf{k}') \cdot  \mathbf{d}_R^*(\mathbf{k})\Big]
        - T_0(-\mathbf{k}, -\mathbf{k}') d^*_{R,0}(\mathbf{k}) \Big[ \mathbf{d}_L(\mathbf{k}') \cdot \mathbf{T}(\mathbf{k}, \mathbf{k}')\Big]
        \\
        &
        + T_0(-\mathbf{k}, -\mathbf{k}')  T_0(\mathbf{k}, \mathbf{k}') \Big[ \mathbf{d}_L(\mathbf{k}') \cdot \mathbf{d}_R^*(\mathbf{k})\Big]
        {
        +id_{L,0}(\mathbf{k}') \mathbf{d}_R^*(\mathbf{k}) \cdot \Big[ \mathbf{T}(-\mathbf{k}, -\mathbf{k}') \times \mathbf{T}(\mathbf{k}, \mathbf{k}')\Big]
        }
                \\
        &
        {+ iT_0(-\mathbf{k}, -\mathbf{k}')  \mathbf{T}(\mathbf{k}, \mathbf{k}') \cdot \Big[ \mathbf{d}_L(\mathbf{k}') \times \mathbf{d}_R^*(\mathbf{k})  \Big]
                }
        + id^*_{R,0}(\mathbf{k}) \mathbf{d}_L(\mathbf{k}') \cdot \Big[\mathbf{T}(-\mathbf{k}, -\mathbf{k}') \times \mathbf{T}(\mathbf{k}, \mathbf{k}')\Big]
        \\
        &
        {+ iT_0(\mathbf{k}, \mathbf{k}') \mathbf{T}(-\mathbf{k}, -\mathbf{k}') \cdot \Big[ \mathbf{d}_L(\mathbf{k}') \times \mathbf{d}_R^*(\mathbf{k}) \Big]
                }
        +
        \Big[ \mathbf{d}_L(\mathbf{k}') \cdot\mathbf{d}^*_R(\mathbf{k}) \Big]
        \Big[ \mathbf{T}(-\mathbf{k}, -\mathbf{k}') \cdot \mathbf{T}(\mathbf{k}, \mathbf{k}') \Big]
        \\
        &-
        \Big[\mathbf{d}_L(\mathbf{k}') \cdot \mathbf{T}(-\mathbf{k}, -\mathbf{k}')\Big]
        \Big[\mathbf{d}_R^*(\mathbf{k}) \cdot \mathbf{T}(\mathbf{k}, \mathbf{k}')\Big]
        -
        \Big[\mathbf{d}_L(\mathbf{k}') \cdot \mathbf{T}(\mathbf{k}, \mathbf{k}')\Big]
        \Big[\mathbf{T}(-\mathbf{k}, -\mathbf{k}') \cdot \mathbf{d}_R^*(\mathbf{k})\Big],
    \end{aligned}
\end{equation}
which is independent of the choice of representation of the tunneling or pairing matrices.

\subsection{Effective spin-dependent tunneling in the presence of bulk spin-orbit coupling}
We derive the tunneling matrix in the band representation,
as introduced in Sec.~\ref{sec:JC} in the main text.
The effective spin-orbit interaction arises from the tunneling of band eigenstates, 
$
H_\mathrm{link} = \sum_{\mathbf{k}, s; \mathbf{k}', s'}
T_{s, s'}^{(b)}(\mathbf{k}, \mathbf{k}') \psi^\dagger_{R, \mathbf{k}, s} \psi_{L, \mathbf{k}', s'} + \mathrm{h.c.}
$
in Eq.~\eqref{link.band_basis}.
Here,
$\psi_{L(R), \mathbf{k}, s} = \sum_{\sigma=\uparrow, \downarrow} 
[U_{L(R)}^\dagger(\mathbf{k})]_{s, \sigma} 
c_{L(R), \mathbf{k}, \sigma}$
are the band eigenstates of the left (right) side in the absence of superconductivity, where $U_{L(R)} (\mathbf{k}) = u_{L(R),0} (\mathbf{k}) \sigma_0 + \mathbf{u}_{L(R)}(\mathbf{k}) \cdot \boldsymbol{\sigma}$ is the unitary transformation to the band-diagonal representation.
The tunneling matrix in the band-diagonal representation is expressed as
$T^{(b)}(\mathbf{k}, \mathbf{k}') = T^{(b)}_0(\mathbf{k}, \mathbf{k}') + \mathbf{T}^{(b)}(\mathbf{k}, \mathbf{k}') \cdot \boldsymbol{\sigma}$, 
where
\begin{subequations}
    \begin{equation}
        \begin{aligned}
            T^{(b)}_0(\mathbf{k}, \mathbf{k}') 
            &= 
            u^*_{R,0}(\mathbf{k})T_{0}(\mathbf{k}, \mathbf{k}')u_{L,0}(\mathbf{k}')
            + u^*_{R,0}(\mathbf{k}) \Big[ \mathbf{T}(\mathbf{k}, \mathbf{k}') \cdot \mathbf{u}_L(\mathbf{k}') \Big]
            + T_0(\mathbf{k}, \mathbf{k}') \Big[ \mathbf{u}^*_R(\mathbf{k}) \cdot \mathbf{u}_L(\mathbf{k}') \Big]
            \\
            &+ u_{L,0}(\mathbf{k}') \Big[ \mathbf{u}^*_R(\mathbf{k}) \cdot \mathbf{T}(\mathbf{k}, \mathbf{k}')\Big]
            + i \mathbf{u}^*_R(\mathbf{k}) \cdot \Big[\mathbf{T}(\mathbf{k}, \mathbf{k}') \times \mathbf{u}_L(\mathbf{k}')\Big],
        \end{aligned}
    \end{equation}
    \begin{equation}
        \begin{aligned}
            \mathbf{T}^{(b)}(\mathbf{k}, \mathbf{k}')
            &=
            u^*_{R,0}(\mathbf{k})T_0(\mathbf{k}, \mathbf{k}') \mathbf{u}_L(\mathbf{k}') 
            + u^*_{R,0}(\mathbf{k}) u_{L,0}(\mathbf{k}') \mathbf{T}(\mathbf{k}, \mathbf{k}') 
            + T_0(\mathbf{k}, \mathbf{k}') u_{L,0}(\mathbf{k}') \mathbf{u}^*_R(\mathbf{k})
            \\
            &+
            i \bigg\{
            u^*_{R,0}(\mathbf{k}) \Big[ \mathbf{T}(\mathbf{k}, \mathbf{k}') \times \mathbf{u}_L(\mathbf{k}') \Big]
            + T_0(\mathbf{k}, \mathbf{k}') \Big[ \mathbf{u}^*_R(\mathbf{k}) \times \mathbf{u}_L(\mathbf{k}') \Big]
            + u_{L,0}(\mathbf{k}') \Big[ \mathbf{u}^*_R(\mathbf{k}) \times \mathbf{T}(\mathbf{k}, \mathbf{k}') \Big]
            \bigg\}
            \\
            &- \Big[ \mathbf{u}^*_R(\mathbf{k}) \cdot \mathbf{u}_L(\mathbf{k}')\Big] 
            \mathbf{T}(\mathbf{k}, \mathbf{k}')
            + \Big[ \mathbf{T}(\mathbf{k}, \mathbf{k}') \cdot \mathbf{u}_L(\mathbf{k}')\Big] 
            \mathbf{u}^*_R(\mathbf{k})
            + \Big[ \mathbf{T}(\mathbf{k}, \mathbf{k}') \cdot \mathbf{u}^*_R(\mathbf{k})\Big] 
            \mathbf{u}_L(\mathbf{k}').
        \end{aligned}
    \end{equation}
\end{subequations}
In the weak-pairing regime, upon projecting the pairing order to the Fermi surfaces participating in Cooper pairing,
the above tunneling amplitudes account for the effective spin-dependent tunneling inherited from bulk interactions.
\end{widetext}

\section{Effective pairing in the presence of bulk spin-orbit coupling} 
\label{appendix:effective_pairing_channels}

We derive the effective monopole pairing order in the weak-coupling regime. 
Consider the general Hamiltonian of a monopole superconductor, given in Eq.~\eqref{MSC.general}:
\begin{equation}
    \mathcal{H}_\mathrm{BdG} (\mathbf{k})    =
    \left(
    \begin{array}{cc}
         \mathcal{H}_{\mathrm{kin},1}(\mathbf{k})         & 
         \Delta_\mathrm{inter}(\mathbf{k})
         \\
         \Delta_\mathrm{inter}^\dagger(\mathbf{k})
         & 
         -\mathcal{H}_{\mathrm{kin},2}^\mathrm{T} (-\mathbf{k})    
     \end{array}
    \right),
    \label{BdG_interFS.appendix}
\end{equation}
with $\Delta_\mathrm{inter}(\mathbf{k})$ describing the inter-Fermi surface pairing between Fermi surfaces $\mathrm{FS}_1$ and $\mathrm{FS}_2$.
Here, we treat the inter-Fermi surface pairing potential perturbatively.
Suppose that band Hamiltonians $\mathcal{H}_{\mathrm{kin}, 1}$ and $\mathcal{H}_{\mathrm{kin}, 2}$ are diagonalized by the unitary transformations
$U_{\mathrm{1}}(\mathbf{k})$ and $U_{\mathrm{2}}(\mathbf{k})$ respectively.
Without loss of generality, we define projection to the band eigenstates at the Fermi level in the band-diagonal representation by the idempotent matrix $P_+^{(b)} = \mathrm{diag}(1, 0, 0, \cdots)$.  
The inter-Fermi surface pairing matrix written in the band-diagonal representation is given by
\begin{equation}
    \Delta_{\mathrm{inter}}^{(b)}(\mathbf{k})
    = U^\dagger_1(\mathbf{k}) \Delta_\mathrm{inter}(\mathbf{k}) U_2^*(-\mathbf{k}).
\end{equation}
In the weak-coupling regime, we project to the helical states at the Fermi surface, so that the band projected pairing matrix is given by
\begin{equation}
    \Delta_{\mathrm{inter}}^{(bp)}(\mathbf{k})
    = P_+^{(b)} \Delta^{(b)} P_+^{(b)}.
    \end{equation}
Upon transforming back to the original representation in Eq.~\eqref{BdG_interFS.appendix}, the projected effective pairing order takes the following form:
\begin{align}
    \Delta^{(p)}_{\mathrm{inter}}(\mathbf{k})
    &= \nonumber
    U_1(\mathbf{k}) \Delta^{(bp)}_{\mathrm{inter}}(\mathbf{k}) U_2^\mathrm{T}(-\mathbf{k})
    \\
    &\equiv
    {P}_1(\mathbf{k}) \Delta_\mathrm{inter}(\mathbf{k}) {P}_2^\mathrm{T}(-\mathbf{k}),\label{band_projected_pairing_matrix.appendix}
\end{align}
where ${P}_{i=1,2}(\mathbf{k}) \equiv U_i(\mathbf{k}) P_+^{(b)} U_i^\dagger(\mathbf{k})$.
For example, consider a (pseudo)spin-$\uparrow,\downarrow$ basis and suppose that band eigenstates at the Fermi surface $\mathrm{FS}_{i=1,2}$ are given by $|\chi_{i, +}(\mathbf{k})\rangle = (u_i(\mathbf{k}), v_i(\mathbf{k}))^{\mathrm{T}}$.
It follows that the projection operator in the spin-$\uparrow, \downarrow$ representation is
\begin{equation}
    {P}_{i=1,2}(\mathbf{k}) = 
    \left(
    \begin{array}{cc}
         |u_i(\mathbf{k})|^2 & u_i(\mathbf{k}) v_i^*(\mathbf{k})
         \\
         u_i^*(\mathbf{k}) v_i(\mathbf{k}) & |v_i(\mathbf{k})|^2
    \end{array}
    \right).
\end{equation}

As an example, we show the effective spin triplet pairing channels in the $\Delta_\mathrm{MSC}^{(q_p=-1, l_z=0)}$ monopole superconducting order described in the main text in Sec.~\ref{sec:JJ_MSC}.
To recap, we consider a system with Fermi surfaces $\mathrm{FS}_1$ and $\mathrm{FS}_2$,  surrounding Weyl nodes $+\mathbf{K}_0$ and $-\mathbf{K}_0$ respectively, which are related by parity, $\mathcal{H}_{\mathrm{kin},1}(\mathbf{k}) = \sigma_z \mathcal{H}_{\mathrm{kin},2}(-\mathbf{k}) \sigma_z$.  
The respective unitary transformations which diagonalize the band Hamiltonians are given by
\begin{subequations}
    \begin{align}
            U_1(\mathbf{k})
            &=
            \sqrt{2\pi}
            \left(
            \begin{array}{cc}
                 \mathcal{Y}_{+{\frac{1}{2}}; \frac{1}{2}, 
                 - \frac{1}{2}}(\boldsymbol{\Omega}_{\tilde{k}})
                 & \mathcal{Y}_{{- \frac{1}{2}}; \frac{1}{2}, 
                 - \frac{1}{2}}(\boldsymbol{\Omega}_{\tilde{k}})
                 \\
                 - \mathcal{Y}_{+{\frac{1}{2}}; \frac{1}{2}, + \frac{1}{2}}(\boldsymbol{\Omega}_{\tilde{k}})
                 & - \mathcal{Y}_{{-\frac{1}{2}}; \frac{1}{2}, + \frac{1}{2}}(\boldsymbol{\Omega}_{\tilde{k}})
            \end{array}
            \right),
            \\
            U_2(-\mathbf{k})
            &=
            \sqrt{2\pi}
            \left(
            \begin{array}{cc}
                 \mathcal{Y}_{+{\frac{1}{2}}; \frac{1}{2}, 
                 - \frac{1}{2}}(\boldsymbol{\Omega}_{\tilde{k}})
                 & \mathcal{Y}_{{- \frac{1}{2}}; \frac{1}{2}, 
                 - \frac{1}{2}}(\boldsymbol{\Omega}_{\tilde{k}})
                 \\
                 \mathcal{Y}_{+{\frac{1}{2}}; \frac{1}{2}, + \frac{1}{2}}(\boldsymbol{\Omega}_{\tilde{k}})
                 & \mathcal{Y}_{{-\frac{1}{2}}; \frac{1}{2}, + \frac{1}{2}}(\boldsymbol{\Omega}_{\tilde{k}})
            \end{array}
            \right),
    \end{align}
\end{subequations}
where $\boldsymbol{\Omega}_{\tilde{\mathbf{k}}}$ is the spherical coordinate, with $\tilde{\mathbf{k}} = \mathbf{k} - \mathbf{K}_0$.  
The expressions for the half-integer monopole harmonics are given in Appendix~\ref{appendix:table_of_MHs}.
When there is inter-Fermi surface $s$-wave pairing $\Delta_\mathrm{inter}(\mathbf{k}) = \Delta_0 i \sigma_y$,  the projected pairing order in the spin-$\uparrow, \downarrow$ representation is given by
\begin{widetext}
\begin{align}
    \Delta^{(p)}_{\mathrm{inter}}(\mathbf{k}) 
    & = \nonumber
    8 \pi^2 \Delta_0
    \left(
    \begin{array}{cc}
         - |\mathcal{Y}_{+ \frac{1}{2}; \frac{1}{2}, -\frac{1}{2}}(\boldsymbol{\Omega}_{\tilde{k}})|^2
         \mathcal{Y}_{+ \frac{1}{2}; \frac{1}{2}, -\frac{1}{2}}(\boldsymbol{\Omega}_{\tilde{k}})
         \mathcal{Y}_{- \frac{1}{2}; \frac{1}{2}, -\frac{1}{2}}(\boldsymbol{\Omega}_{\tilde{k}})
         &
         |\mathcal{Y}_{+ \frac{1}{2}; \frac{1}{2}, -\frac{1}{2}}(\boldsymbol{\Omega}_{\tilde{k}})
         \mathcal{Y}_{+ \frac{1}{2}; \frac{1}{2}, +\frac{1}{2}}(\boldsymbol{\Omega}_{\tilde{k}})
         |^2
         \\
         - |\mathcal{Y}_{+ \frac{1}{2}; \frac{1}{2}, -\frac{1}{2}}(\boldsymbol{\Omega}_{\tilde{k}})
         \mathcal{Y}_{+ \frac{1}{2}; \frac{1}{2}, +\frac{1}{2}}(\boldsymbol{\Omega}_{\tilde{k}})
         |^2
         &
          - |\mathcal{Y}_{+ \frac{1}{2}; \frac{1}{2}, +\frac{1}{2}}(\boldsymbol{\Omega}_{\tilde{k}})|^2
         \mathcal{Y}_{+ \frac{1}{2}; \frac{1}{2}, +\frac{1}{2}}(\boldsymbol{\Omega}_{\tilde{k}})
         \mathcal{Y}_{- \frac{1}{2}; \frac{1}{2}, +\frac{1}{2}}(\boldsymbol{\Omega}_{\tilde{k}})
    \end{array}
    \right)
    \\
    &=
    \Delta_0
    \left(
    \begin{array}{cc}
         - \cos^2 \frac{\theta_{\tilde{k}}}{2}
         \sin \theta_{\tilde{k}} e^{-i \varphi_{\tilde{k}}}
         &
         \frac{1}{2} \sin^2 \theta_{\tilde{k}}
         \\
         - \frac{1}{2} \sin^2 \theta_{\tilde{k}}
         &
         - \sin^2 \frac{\theta_{\tilde{k}}}{2}
         \sin \theta_{\tilde{k}} e^{+i \varphi_{\tilde{k}}}
    \end{array}
    \right).
    \label{qp=-1.lz=0.projeted.spin_rep}
\end{align}
\end{widetext}
Here, due to the nontrivial spin texture from the Weyl spin orbit coupling in the band Hamiltonians, there are contributions from the $s_z = \pm 1$ spin triplet channels.
Specifically, there is nonvanishing $(p_x + ip_y){|\downarrow \downarrow\rangle}$ and $(p_x - ip_y){|\uparrow \uparrow\rangle}$
pairing for which the effective pairing amplitude varies over the Fermi surface.
Near the north pole, the $(p_x - ip_y){|\downarrow \downarrow \rangle}$ pairing is more heavily weighted and the $(p_x - ip_y){|\downarrow \downarrow \rangle}$ pairing is suppressed, while the opposite holds true near the south pole.
The spin singlet channel survives but likewise varies over the Fermi surface, reaching its maximum near the equator.
Moreover, the total angular momentum $j_z = l_z + s_z = 0$ is conserved.
This description is consistent with the effective pairing in the helical band basis, $\Delta_\mathrm{MSC}^{(q_p = -1, l_z=0)}$,  which globally transforms according to conserved angular momentum $l_{z, \mathrm{glob}} = 0$ but locally has features of a chiral $p$-wave superconductor, with $l_{z, \mathrm{loc}} = \pm 1$,  which arise from the induced spin triplet channels.

The effective pairing channels likewise can be seen in, for example, the tight-binding results discussed in Sec.~\ref{subsec:JG.tbm} for a Josephson junction along the $z$ direction between an $s$-wave superconductor and $\Delta_\mathrm{MSC}^{(q_p=-1, l_z=0)}$ monopole superconductor.
In the Josephson energy phase relation shown in Fig.~\ref{fig:WeylS.zjunction}(b), states with momentum $k_\parallel = k_{\parallel, 5}$ near $\tilde{k}_F$ have significant contributions to the first-order Josephson tunneling, analogous to a junction between two $s$-wave superconductors.
Similarly, the spin-singlet channel in the projected pairing order in Eq.~\eqref{qp=-1.lz=0.projeted.spin_rep} has maximal amplitude at the equator.
In contrast, for states with momentum $k_\parallel = k_{\parallel, 2}$ near the gap nodes, the first-order Josephson current is suppressed while the second-order $\pi$-periodic Josephson current is more significant.
This can be attributed to the effective spin triplet channels in Eq.~\eqref{qp=-1.lz=0.projeted.spin_rep}, which are more weighted relative to the spin singlet channel near the poles of the Weyl Fermi surface.
Similar to a junction between a spin singlet and spin triplet superconductor with spin-independent tunneling at the junction link, this leads to higher-order contributions to the Josephson current.

\section{Monopole harmonic functions}
\label{appendix:table_of_MHs}

Monopole harmonics are eigenfunctions of angular momentum operator in the presence of a magnetic monopole~\cite{Wu1976, Wu1977}.
Due to the magnetic monopole, the vector potential $\mathbf{A}$ must have singularities, corresponding to a Dirac string.
To avoid these singularities, one partitions the unit sphere into two regions covering the north and south poles.
For a magnetic monopole of strength $g$, the vector potential
is given by
    \begin{align}
        \mathbf{A}^{(\lambda)} (\theta, \varphi) &= \frac{{g}}{r} \left(\frac{- \cos \theta - \lambda}{\sin \theta} \right) \hat{\boldsymbol{\varphi}},
                        \label{vector_potential}
    \end{align}
    where the superscript $\lambda = \pm 1$ corresponds to a gauge where the vector potential is well-defined, with
the Dirac string located at $\theta = (\lambda - 1)\pi/2$.  
For a particle of electric charge $e$ moving in the presence of a magnetic monopole, let $q = {ge}/{(\hbar c)} = {n}/{2}$ for $n \in \mathbb{Z}$ denote the monopole charge.

Monopole harmonics satisfy
\begin{subequations}
    \begin{align}
        \tilde{L}^2 \mathcal{Y}_{q; l,l_z} (\theta, \varphi) &= \hbar^2 l (l+1) \mathcal{Y}_{q; l,l_z} (\theta, \varphi)
        \\
        \tilde{L}_z \mathcal{Y}_{q; l,l_z} (\theta, \varphi) &= \hbar l_z \mathcal{Y}_{q; l,l_z} (\theta, \varphi),
    \end{align}
\end{subequations}
where $q$ is the corresponding monopole charge, which sets the lower bound for the partial wave channels, $l \geq q$. 
Here, $\tilde{\mathbf{L}}= {\mathbf{r}} \times ({\mathbf{p}} + |e| \mathbf{A}^{(\lambda)}/\hbar c) - {\lambda} \hbar q \hat{\mathbf{r}}$ and $\tilde{L}_z = -i \hbar \partial_\varphi - {\lambda} \hbar q= L_z^{(\lambda)} - {\lambda} \hbar q$.
Due to the nontrivial monopole charge, functions defined in the $\lambda = +1$ and $\lambda = -1$ patch are related by a $\mathrm{U}(1)$ transformation,
\begin{equation}
    \mathcal{Y}_{q;l,l_z}^{(\lambda = +1)} (\theta, \varphi) = e^{i2q \varphi} \mathcal{Y}_{q;l,l_z}^{(\lambda = -1)} (\theta, \varphi),
\end{equation}
where the superscript corresponds to a gauge choice consistent with Eq.~\eqref{vector_potential}.
In the following, the monopole harmonics are normalized to the convention
$
\int_{0}^{-1} \mathrm{d}(\cos \theta) \int_0^{2\pi} \mathrm{d}\varphi |\mathcal{Y}_{q, l,l_z}(\theta, \varphi)|^2 = 1$.  

The general form of the monopole harmonic $\mathcal{Y}_{q; l,l_z}$ is given by~\cite{Wu1976}
\begin{equation}
    \mathcal{Y}^{(\lambda)}_{q; l,l_z}(\theta, \varphi)
    =
    M_{q, l,l_z} (1 - x)^{\frac{\alpha}{2}} 
    (1+x)^{\frac{\beta}{2}}
    P_n^{\alpha, \beta}(x)
    e^{i(l_z  + \lambda q) \varphi},
\end{equation}
where $\alpha = - q - l_z$, $\beta = q - l_z$, $n = j + l_z$,  $x = \cos \theta$, 
\begin{math}
    M_{q, l,l_z} = 2^{l_z}
    \sqrt{{(2l +1)}/{4\pi}}\sqrt{{(l-l_z)! (l + l_z)!}/{(l-q)!(l+q)!}},
\end{math}
and $P_n^{\alpha, \beta}$ are Jacobi polynomials,
\begin{equation}
    \begin{aligned}
        P_n^{\alpha, \beta} (x)
        =
        &\frac{(-1)^n}{2^n n!}
        (1-x)^{-\alpha} (1+x)^{-\beta} 
        \\
        &\times
        \frac{\mathrm{d}^n}{\mathrm{d}x^n}
        \left[
        (1-x)^{\alpha + n}
        (1+x)^{\beta +n}
        \right].
    \end{aligned}
\end{equation}
The local phase winding about the north or south pole, corresponding to $\lambda = +1$ and $\lambda = -1$ respectively, is given by 
\begin{equation}
    l_{z, \mathrm{loc}} = l_z + \lambda q,
\end{equation}
which differs from the global angular momentum $l_z = l_{z, \mathrm{glob}}$ by the monopole charge, $q$.
When the local phase winding is nonzero, there is always a point node at the corresponding pole.
Similarly, when the local phase winding is zero about the north pole (south pole), or equivalently $\alpha = 0$ ($\beta = 0$), the monopole harmonic functions are nonvanishing at the pole, corresponding to a locally gapped pairing in the context of monopole superconducting order.

Monopole harmonics of opposite charges are related by
$
    \mathcal{Y}_{q; l ,l_z} (\theta, \varphi) = (-1)^{q + l_z} \mathcal{Y}^*_{-q; l, -l_z} (\theta, \varphi).
$
The product of two monopole harmonics follows~\cite{Wu1977}
\begin{widetext}
\begin{eqnarray}
\mathcal{Y}_{q_1;l_1,l_{1,z}} \mathcal{Y}_{q_2;l_2,l_{2,z}}
        &=&\sum_{l_3}
        (-1)^{l_1 + l_2 + l_3 - q_3 - l_{3,z}}
        \sqrt{\frac{(2l_1 + 1)(2l_2 + 1)(2l_3+1)}{4\pi}} \nonumber
        \\
        &&\times
        \left(
        \begin{array}{ccc}
             l_1 & l_2 & l_3
             \\
             l_{1,z} & l_{2,z} & -l_{3,z}
        \end{array}
        \right)
        \left(
        \begin{array}{ccc}
             l_1 & l_2 & l_3
             \\
             q_1 & q_2 & -q_3
        \end{array}
        \right)
        \mathcal{Y}_{q_3; l_3, l_{3,z}},
\end{eqnarray}
\end{widetext}
where $q_3 = q_1 + q_2$ and $l_{3,z} = l_{1,z} + l_{2,z}$, and the round brackets are the Wigner $3j$ symbols.
In other words, the product of two monopole harmonics can be expressed as a superposition of monopole harmonics, all which belong in the same topological sector $q_3$ and have global angular momentum $l_{3,z}$, but may reside in different partial wave channels.

Below, we list the explicit forms for the monopole harmonics, $\mathcal{Y}_{q;j,l_z}(\theta, \varphi)$, used in the main text.
For $q= - {1}/{2}$, the monopole harmonics in the $l={1}/{2}$ partial wave channel are as follows:
\begin{equation*}
    \begin{array}{c|c}
         \mathbf{q= - \frac{1}{2}} & l=\frac{1}{2}
         \\
         \hline
         l_z = - \frac{1}{2}
         &
         \sqrt{\frac{1}{2\pi}} \sin \frac{\theta}{2} e^{-i (\lambda +1)\varphi/2}
         \\
         \hline
         l_z  = \frac{1}{2}
         &
         \sqrt{\frac{1}{2\pi}} \cos \frac{\theta}{2} e^{i (-\lambda +1)\varphi/2}
    \end{array}
\end{equation*}

For $q=-1$, the monopole harmonics in the $l=1,2$ partial wave channel take the form:
\begin{equation*}
\begin{array}{c|c}
     \textbf{q = -1}& l=1 
     \\
     \hline
     l_z =-1 
     &\sqrt{\frac{3}{4\pi}} \sin^2 \frac{\theta}{2} e^{-i(\lambda+1) \varphi}
     \\
     \hline
     l_z = 0
     & \sqrt{\frac{3}{8 \pi}} \sin \theta e^{-i\lambda \varphi}
     \\
     \hline
     l_z=1
     & \sqrt{\frac{3}{4\pi}} \cos^2 \frac{\theta}{2} e^{i(-\lambda+1) \varphi}
\end{array}
\end{equation*}
and
\begin{equation*}
\begin{array}{c|c}
     \textbf{q = -1}&  l=2
     \\
     \hline
     l_z=-2
     & \sqrt{\frac{5}{4\pi}} \sin^2 \frac{\theta}{2} \sin \theta e^{-i(\lambda+2) \varphi}
     \\
     \hline
     l_z =-1 
     & \sqrt{\frac{5}{4\pi}}  \sin^2 \frac{\theta}{2} (1 + 2 \cos \theta) e^{-i(\lambda+1) \varphi}
     \\
     \hline
     l_z = 0
     & \sqrt{\frac{15}{8 \pi}} \cos \theta \sin \theta e^{-i\lambda \varphi}
     \\
     \hline
     l_z=1
     & \sqrt{\frac{5}{4\pi}}  \cos^2 \frac{\theta}{2} (-1 + 2 \cos \theta) e^{i(-\lambda+1) \varphi}
     \\
     \hline
     l_z = 2
     & \sqrt{\frac{5}{4\pi}} \cos^2 \frac{\theta}{2} \sin \theta e^{i(-\lambda+2) \varphi}
\end{array}
\end{equation*}

For $q=-3$, the monopole harmonics in the $l=3$ partial wave channel are:
\begin{equation*}
\begin{array}{c|c}
     \textbf{q = -3}& l=3
     \\
     \hline
     l_z=-3
     & \sqrt{\frac{7}{4\pi}} \sin^6 \frac{\theta}{2} e^{-i(3\lambda+3) \varphi}
     \\
     \hline
     l_z =-2
     & \sqrt{\frac{21}{8\pi}} \sin^4 \frac{\theta}{2}  \sin \theta e^{-i(3\lambda+2) \varphi}
     \\
     \hline
     l_z = -1
     & \sqrt{\frac{105}{64\pi}} \sin^2 \frac{\theta}{2} \sin^2 \theta e^{-i(3\lambda+1) \varphi} 
     \\
     \hline
     l_z = 0
     & \sqrt{\frac{35}{64\pi}} \sin^3 \frac{\theta}{2} e^{-i3\lambda \varphi}
     \\
     \hline
     l_z=1
     & \sqrt{\frac{105}{64\pi}} \cos^2 \frac{\theta}{2} \sin^2 \theta e^{i(-3\lambda+1) \varphi} 
     \\
     \hline
     l_z = 2
     & \sqrt{\frac{21}{8\pi}} \cos^4 \frac{\theta}{2}  \sin \theta e^{i(-3\lambda+2) \varphi}
     \\
     \hline
     l_z = 3
     &
     \sqrt{\frac{7}{4\pi}} \cos^6 \frac{\theta}{2} e^{i(-3\lambda+3) \varphi}
\end{array}
\end{equation*}

\section{Tight-binding model of Josephson junction}
\label{appendix:TBM_JJ}

We describe the tight-binding model used to model a Josephson junction, as used in Sec.~\ref{subsec:JG.tbm}, \ref{sec:JL}, and \ref{sec:distinguish_from_chiral_SCs}.
For the following analysis, we consider a junction along the $x$ direction without loss of generality.
We take periodic boundary conditions in the $y$ and $z$ directions, with transverse momenta $k_y$ and $k_z$ being conserved across the junction barrier. 
For given conserved momenta $k_y$ and $k_z$, we treat the junction as a pseudo-$1$D junction in the $x$ direction with $2N_x$ sites, where the barrier is located between $n_x = N_x$ and $n_x = N_x+1$.
We consider the Hamiltonian of a Josephson junction,
\begin{math}
    H_\mathrm{JJ}(k_y, k_z) = H_{\mathrm{BdG},L}(k_y, k_z) + H_{\mathrm{link}}(k_y, k_z) + H_{\mathrm{BdG},R}(k_y, k_z)
\end{math}
in Eq.~\eqref{BdG_JJ}.
The bulk superconductors at the left (right) side are given by
\begin{widetext}
\begin{equation}
    \begin{aligned}
        H_{\mathrm{BdG}, L(R)}(k_y, k_z)
        & =
                \sum_{n_x, n_x'} 
        \sum_{\sigma, \sigma' = \uparrow, \downarrow}
        c^\dagger_{k_y, k_z; \sigma, n_x}
        \left[\mathcal{H}_{\mathrm{kin}, L(R)}(k_y, k_z)\right]_{\sigma, n_x; \sigma', n_x'}
        c_{k_y, k_z; \sigma', n_x'}
        \\
        &+
                \sum_{n_x, n_x'} 
        \sum_{\sigma, \sigma' = \uparrow, \downarrow}
        \left(
        c^\dagger_{k_y, k_z; \sigma, n_x}
        \left[e^{i\varphi_{L(R)}} \Delta_{L(R)}(k_y, k_z)\right]_{\sigma, n_x; \sigma', n_x'}
        c^\dagger_{-k_y, -k_z; \sigma', n_x'}
        + \mathrm{h.c.}
        \right)
    \end{aligned}
    \label{JJ.BdG_hams}
\end{equation}
where  sum over $n_x$ runs from $1$ to $N_x$ for the left superconductor ($N_x + 1$ to $2N_x$ for the right superconductor).
Here, $\mathcal{H}_{\mathrm{kin}, L(R)}(k_y, k_z)$ is the kinetic Hamiltonian, $\Delta_{L(R)}(k_y, k_z)$ is the superconducting pairing order parameter, and $\phi_{L(R)}$ is the $\mathrm{U}(1)$ superconducting phase, taken to be uniform in the respective superconductor.
$c_{k_y, k_z; \sigma, n_x}$ is the annihilation operator for an electron at site $n_x$ with spin $\sigma$ and conserved momentum $k_y \hat{y} + k_z \hat{z}$.

The tight-binding analogue of the tunneling Hamiltonian for the Josephson junction in Eq.~\eqref{tunneling_hamiltonian.spin_basis} is given by
\begin{equation}
    \begin{aligned}
    H_\mathrm{link} (k_y, k_z) = 
        \sum_{\sigma, \sigma' = \uparrow, \downarrow}
         c^\dagger_{k_y, k_z; \sigma, N_x + 1}
         \left[ \hat{t}(k_y, k_z)\right]_{\sigma, \sigma'}
        c_{k_y, k_z; \sigma', N_x} + \mathrm{h.c.},
    \end{aligned}
    \label{Hlink.appendix}
\end{equation}
where $\hat{t}(k_y, k_z) = t_0(k_y, k_z)\sigma_0 + \mathbf{t}(k_y, k_z) \cdot \boldsymbol{\sigma}$ describes spin-independent and spin-dependent hopping across the junction barrier.
Here, we consider only one site at the barrier for simplicity.  
Including more sites will increase the amount of in-gap Andreev states but not affect the overall qualitative features of the Josephson current, \textit{e.g.} its periodicity.
\end{widetext}

\section{Tight-binding models of monopole superconductors}
\label{appendix:TBM}

In this section, we outline the tight-binding model of the monopole superconductor described in Eq.~\eqref{MSC.general} used for numerical calculations of monopole superconductors.
The general Hamiltonian of a monopole superconductor takes the following form
\begin{equation}
    \begin{aligned}
        H_\text{MSC}
        = &\sum_{\mathbf{n}, \sigma; \mathbf{n}', \sigma'} 
        c^{\dagger}_{\mathbf{n}, \sigma} [\mathcal{H}_\mathrm{kin}]_{\mathbf{n}, \sigma; \mathbf{n}', \sigma'}
        c_{\mathbf{n}', \sigma'}
        \\
        &+  
        \sum_{\mathbf{n}, \sigma; \mathbf{n}', \sigma'} 
        \left(
        c^{\dagger}_{\mathbf{n}, \sigma} [\Delta_\mathrm{inter}]_{\mathbf{n}, \sigma; \mathbf{n}', \sigma'}
        c^\dagger_{\mathbf{n}', \sigma'}
        + \mathrm{h.c.}
        \right)
    \end{aligned}
    \label{q=1.msc_ham}
\end{equation}
where $\mathcal{H}_\mathrm{kin}$ and $\Delta_{\mathrm{inter}}$ are respectively the kinetic and inter-Fermi-surface pairing kernels of the BdG Hamiltonian. 
Here, $c_{\mathbf{n}, \sigma}$ annihilates a single-particle state at lattice site $\mathbf{n}$ and with spin $\sigma = \uparrow, \downarrow$. 
We take the following convention for the tensor product basis, with the two-dimensional spin basis  indexed by $\sigma$ nested under lattice site basis indexed by $\mathbf{n}$. 

\subsection{Tight-binding model of \texorpdfstring{$q_p=-1$}{qp=-1} monopole superconductor}
\label{appendix:subsec.TBM.qp=-1}

For the completeness of the manuscript, we reproduce the tight-binding model of the monopole superconductor with pair monopole charge $q_p = -1$~\cite{Sun2019} which corresponds to the continuum model in Eq.~\eqref{qp=-1.MSC.hamiltonian}.
The Fourier transform of the tight-binding model of the magnetic doped Weyl semimetal band Hamiltonian on a cubic lattice is given by
\begin{equation}
    \begin{aligned}
        \mathcal{H}_\mathrm{kin}^{\mathrm{WSM}}(\mathbf{k}) = 
        &t_z(\cos K_{0,z} - \cos k_z)\sigma_z
        \\
        &+ m(\cos k_x + \cos k_y - 2)\sigma_z 
        \\
        &+ t_x \sin k_x \sigma_x
        + t_y \sin k_y \sigma_y
        -\mu \sigma_0,    
    \end{aligned}
    \label{tbm_qp=1.band_part.fourier_transform}
\end{equation}
where
$\mu$ is the chemical potential and
$t_{i=x,y,z}$ 
are the hopping amplitudes along the $i^\text{th}$ direction (we take $t_x=t_y=t_z \sin K_{0,z}= v_F$ to compare to Eq.~\eqref{continuum_models.c=1}).
At appropriate doping, the system gives rise to two disjoint Fermi surfaces surrounding Weyl nodes at $\pm \mathbf{K}_0 = (0,0,\pm K_{0,z})^\mathrm{T}$. 
The parameter $m$ is chosen satisfying $m/t_z\le -2$ so that the sign of the $\sigma_z$-term in momentum space only changes between the two Weyl nodes at $\pm \mathbf{K}_0$.

Nonvanishing elements of the tight-binding model in the real-space representation are as follows:
\begin{equation}
    \begin{aligned}
        [\mathcal{H}_\mathrm{kin}^{\mathrm{WSM}}]_{\mathbf{n}; \mathbf{n}}
        &=
        (t_z \cos K_{0,z} - 2m)\sigma_z - \mu \sigma_0,
        \\
        [\mathcal{H}_\mathrm{kin}^{\mathrm{WSM}}]_{\mathbf{n}; \mathbf{n}+\boldsymbol{\delta}_x}
        &=[\mathcal{H}_\mathrm{kin}^{\mathrm{WSM}}]_{\mathbf{n}+\boldsymbol{\delta}_x; \mathbf{n}}^\dagger
        =
        \frac{1}{2} (m\sigma_z -it_x \sigma_x),
        \\
        [\mathcal{H}_\mathrm{kin}^{\mathrm{WSM}}]_{\mathbf{n}; \mathbf{n}+\boldsymbol{\delta}_y}
        &=
        [\mathcal{H}_\mathrm{kin}^{\mathrm{WSM}}]_{\mathbf{n}+\boldsymbol{\delta}_y; \mathbf{n}}^\dagger
        =
        \frac{1}{2}(m \sigma_z  -it_y \sigma_y),
        \\
        [\mathcal{H}_\mathrm{kin}^{\mathrm{WSM}}]_{\mathbf{n}; \mathbf{n}+\boldsymbol{\delta}_z}
        &=
        [\mathcal{H}_\mathrm{kin}^{\mathrm{WSM}}]_{\mathbf{n}+\boldsymbol{\delta}_z; \mathbf{n}}^\dagger
        =
        -\frac{1}{2} t_z \sigma_z.
    \end{aligned}
\end{equation}
Here, $\boldsymbol{\delta}_x=(1,0,0)^\mathrm{T}$, $\boldsymbol{\delta}_y=(0,1,0)^\mathrm{T}$, and $\boldsymbol{\delta}_z=(0,0,1)^\mathrm{T}$ are lattice unit vectors along the crystalline axes. 
The $s$-wave pairing in Eq.~\eqref{q=1.msc_ham} is defined locally on-site as
\begin{equation}
    [\Delta_\mathrm{inter}]_{\mathbf{n}; \mathbf{n}} = 
    i\Delta_0 \sigma_y,
                                    \label{TBM.inter_FS.swave_pairing}
\end{equation}
where $\Delta_0$ is the pairing amplitude.

\begin{figure}[tb]
    \centering
    \includegraphics[width=.95\linewidth]{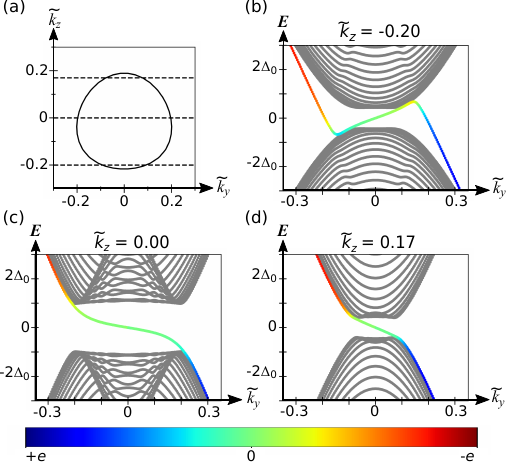}
    \caption{
    (a)~Fermi surface $\mathrm{FS}_1$ of the tight-binding model $\mathcal{H}_\mathrm{kin}^{\mathrm{WSM}}$ in Appendix~\ref{appendix:subsec.TBM.qp=-1} (black, solid) at $\tilde{k}_x = 0$. 
    Three  dashed lines correspond to three values of relative momenta, $\tilde{k}_z = -0.20$, $0.00$, and $0.17$.
    (b)-(d)~The bulk (gray) and surface  (colored) BdG spectra of the $\Delta_{\mathrm{MSC}}^{(q_p=-1, l_z=0)}$ monopole superconductor with boundary at $x= 0$ 
    and at the three values of $\tilde{k}_z$ shown in (a).  
    The color scheme 
    indicates the charge of the surface BdG quasiparticles states, with particle-like (hole-like) states corresponding to red (blue).
    We use parameters $t_x = t_y = t_z \sin K_0 = -0.5$, $K_0 = 1.0$, $\mu = -0.1$, $m = 1.0$, and $\Delta_0 = 0.02$.
    }
    \label{fig:surface_modes}
\end{figure}

Under open boundary conditions, the system exhibits helical surface states. 
Consider the partial Fourier-transform of the above tight-binding model, for given conserved $\tilde{k}_z$, the $\Delta_\mathrm{MSC}^{(q_p=-1, l_z = 0)}$ monopole pairing order can be regarded as an effective two-dimensional chiral pairing order characterized by local angular momentum $l_{z, \mathrm{loc}} = \pm 1$.
Away from $\mathrm{FS}_1$, the chiral surface modes result from the Weyl Fermi arcs; however, between the gap nodes at the poles of $\mathrm{FS}_1$ and $\mathrm{FS}_2$,
a single surface mode appears within the gap.
The single surface mode appearing between the gap nodes is a result of the effective local chiral $p$-wave pairing, as described in Sec.~\ref{sec:JL}.
As $\tilde{k}_z$ varies from the south to north pole,
corresponding to Fig.~\ref{fig:surface_modes}~(b-d),
though the topology of the pairing order, which is given by pair monopole charge $q_p = -1$, does not change, 
the sign of the in-gap states' group velocity changes.
The change in the group velocity is due to the shift in the local angular momentum, in agreement to the effective pairing order in Eq.~\eqref{q=-1.lz=0.effective_pairing}.

\begin{figure}[tb]
    \centering
    \includegraphics[width=.95\linewidth]{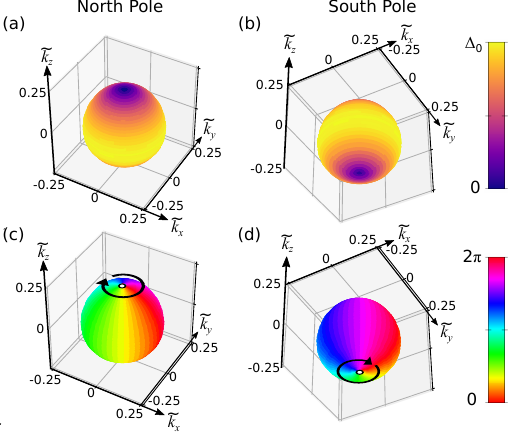}
    \caption{
    The magnitude [(a) and (b)] and phase winding [(c) and (d)] of the monopole pairing order $\Delta_\mathrm{MSC}^{(q_p=-1, l_z=0)}$ 
    based on the tight-binding model in Appendix~\ref{appendix:subsec.TBM.qp=-1}.
    The left and right columns respectively correspond to the north and south pole views of the pairing order at FS$_1$.
    (c) and (d) are plotted under the gauge that the pairing phase is well-defined at the north and south pole respectively.
    Here, we use the same parameters as those for Fig.~\ref{fig:surface_modes}.
    }
    \label{fig:q=1.projected_pairing}
\end{figure}

The bulk monopole pairing order 
of the tight-binding model is shown in Fig.~\ref{fig:q=1.projected_pairing}.
The 
pairing between two helical Fermi surfaces $\mathrm{FS}_1$ and $\mathrm{FS}_2$, as introduced in Sec.~\ref{subsec:MSC.microscopic}, is defined as $\Delta_{\mathrm{MSC}}^{(q_p)}(\tilde{\mathbf{k}}) = \langle \chi_{1, +}(\tilde{\mathbf{k}}) | \Delta_\mathrm{inter}(\mathbf{k})| \chi_{2, +}^*(-\tilde{\mathbf{k}})\rangle$ in Eq.~\eqref{MSC.effective_gap_function.weak_coupling}.  
Here, $|\chi_{1(2), +}(\tilde{\mathbf{k}})\rangle$ is the helical eigenstate of $\mathrm{FS}_{1(2)}$.
The magnitude of the effective pairing is shown in Fig.~\ref{fig:q=1.projected_pairing}(a) and \ref{fig:q=1.projected_pairing}(b), where there are two gap nodes pinned to the north and south pole of $\mathrm{FS}_1$.
The phase winding, chosen in a gauge where the pairing is analytic at the north and south pole, is shown in Fig.~\ref{fig:q=1.projected_pairing}(c) and (d) respectively.
Near the south pole, consistent with the group velocity of the surface modes in Fig.~\ref{fig:surface_modes}(b), the monopole pairing order behaves locally as a $p_x + ip_y$ chiral pairing order.
In contrast, near the north pole, the effective pairing order behaves locally as a $p_x - ip_y$ chiral pairing order.
The shift in local angular momentum $l_{z, \mathrm{loc}} = \pm 1$ from the global angular momentum, $l_{z, \mathrm{glob}} = 0$,  is indicative of the nontrivial pair monopole charge, $q_p = -1$.
For the tight-binding model, the $\Delta_{\mathrm{MSC}}^{(q_p=-1, l_z = 0)}$ monopole superconducting pairing order can be expressed as a superposition of monopole harmonics, all which have pair monopole charge $q_p = -1$ and global angular momentum $l_z = 2$, but reside in different partial wave channels ($l \geq |q_p|$) due to the lattice symmetry.

\subsection{Tight-binding model of \texorpdfstring{$q_p=3$}{qp=3} monopole superconductor}
\label{appendix:TBM.qp=3}

We construct a tight-binding model for the monopole superconducting order having pair monopole charge $q_p=3$,  as introduced in Sec.~\ref{subsec:qp=3}.
Here, the band Hamiltonian describes a magnetic doped triple Weyl semimetal.
As $C_6$ symmetry is needed to stabilize the triple Weyl nodes, the tight-binding model is defined on a three-dimensional hexagonal lattice with $D_{6h}$ symmetry, which consists of $AA$ stacked two-dimensional triangular lattices.

We consider the following tight-binding mean-field BCS Hamiltonian,
\begin{eqnarray}
        && H_\mathrm{MSC}^{(q_p = 3, l_z = 0)} 
        = \sum_{\mathbf{k}} \sum_{\sigma, \sigma' = \uparrow, \downarrow} c^{\dagger}_{\mathbf{k}, \sigma} \left[\mathcal{H}_\mathrm{kin}^{\mathrm{TWSM}}
        (\mathbf{k})\right]_{\sigma,\sigma'}c_{\mathbf{k}, \sigma'} \nonumber \\
        &&+
        \sum_{\mathbf{k}} \sum_{\sigma, \sigma' = \uparrow, \downarrow} \Big( c^{\dagger}_{\mathbf{k}, \sigma} \big[\Delta_\mathrm{inter}(\mathbf{k})\big]_{\sigma,\sigma'}c^{\dagger}_{-\mathbf{k}, \sigma'}
        + \mathrm{h.c.}
        \Big),
        \label{q=3.msc_ham}
\end{eqnarray}
with $\Delta_\mathrm{inter}(\mathbf{k})$
being the inter-Fermi surface 
pairing.
Here, the Fourier transform of the band Hamiltonian kernel for the triple Weyl semimetal is
\begin{equation}
    \mathcal{H}_\mathrm{kin}^{\mathrm{TWSM}}     (\mathbf{k})
    = \mathbf{g}    (\mathbf{k}) \cdot \boldsymbol{\sigma} - \mu,
    \label{C=3.kin}
\end{equation}
where
$g_x(\mathbf{k}) 
= 2t_{1} ( \sin k_1 - \sin k_2 + \sin k_3)$, 
$g_y(\mathbf{k}) 
= 
({-2 t_{2} }/{3\sqrt{3}})
[\sin(k_1 + k_2) - \sin(k_2 + k_3) 
+ \sin(k_3-k_1)]$, 
and
$g_z(\mathbf{k}) 
= {2t_z(\cos K_{0,z} - \cos k_z)} + {2m (3 - \cos k_1 - \cos k_2 - \cos k_3 )}$.  
The above system hosts two triple Weyl nodes pinned along the $k_z$ axis at $\pm \mathbf{K}_0 = \left(0, 0, \pm K_{0,z}\right)$ with opposite winding numbers $\mp 3$. In the low-energy regime, the above tight-binding model of the band Hamiltonian reduces to that of the continuum model  given in Eq.~\eqref{C=3.kin.continuum}.
Nearest and next-nearest neighbor hopping amplitudes are denoted by $t_1$ and $t_2$ respectively which, for simplicity, we choose to be $t_1 = t_2 = 8t_z \sin K_{0,z}$ to compare to the continuum model.
Here, $k_i \equiv \boldsymbol{\delta}_i \cdot \mathbf{k}$, where $\boldsymbol{\delta}_1 = (1,0,0)$, $\boldsymbol{\delta}_2 = (\frac{1}{2}, \frac{\sqrt{3}}{2}, 0)$, and $\boldsymbol{\delta}_3 = (-\frac{1}{2}, \frac{\sqrt{3}}{2}, 0)$ are the lattice vectors connecting neighboring sites in the $ab$-plane, and the lattice constant has been set to unity.
At appropriate doping, the above system realizes two disjoint Fermi surfaces, $\mathrm{FS}_1$ and $\mathrm{FS}_2$, situated about triple Weyl nodes at $\pm \mathbf{K}_0$ and related by parity, 
$\mathcal{H}_{\mathrm{kin}, 1}^{(\nu= -3)}(
\mathbf{k}
) 
= \sigma_z \mathcal{H}_{\mathrm{kin}, 2}^{(\nu= +3)}(
-\mathbf{k}
)\sigma_z$.

The nonzero matrix elements,  which feature nearest and next-nearest neighbor hopping, read
\begin{equation}
    \begin{aligned}
        [\mathcal{H}_\mathrm{kin}^{\mathrm{TWSM}}]_{\mathbf{n}; \mathbf{n}}
        &=
        (2t_z \cos K_{0,z} + 6m)\sigma_z - \mu \sigma_0,
        \\
        [\mathcal{H}_\mathrm{kin}^{\mathrm{TWSM}}]_{\mathbf{n}; \mathbf{n}+\boldsymbol{\delta}_1}
        &=
        [\mathcal{H}_\mathrm{kin}^{\mathrm{TWSM}}]_{\mathbf{n}+\boldsymbol{\delta}_1; \mathbf{n}}^\dagger
        =
        -m \sigma_z -i t_1 \sigma_x
        \\
        [\mathcal{H}_\mathrm{kin}^{\mathrm{TWSM}}]_{\mathbf{n}; \mathbf{n}+\boldsymbol{\delta}_2}
        &=
        [\mathcal{H}_\mathrm{kin}^{\mathrm{TWSM}}]_{\mathbf{n}+\boldsymbol{\delta}_2; \mathbf{n}}^\dagger
        =
        -m \sigma_z + i t_1 \sigma_x
        \\
        [\mathcal{H}_\mathrm{kin}^{\mathrm{TWSM}}]_{\mathbf{n}; \mathbf{n}+\boldsymbol{\delta}_3}
        &=
        [\mathcal{H}_\mathrm{kin}^{\mathrm{TWSM}}]_{\mathbf{n}+\boldsymbol{\delta}_3; \mathbf{n}}^\dagger
        =
        -m \sigma_z -i t_1 \sigma_x
        \\
        [\mathcal{H}_\mathrm{kin}^{\mathrm{TWSM}}]_{\mathbf{n}; \mathbf{n}+\boldsymbol{\delta}_1 + \boldsymbol{\delta}_2}
        &=
        [\mathcal{H}_\mathrm{kin}^{\mathrm{TWSM}}]_{\mathbf{n}+\boldsymbol{\delta}_1 + \boldsymbol{\delta}_2; \mathbf{n}}^\dagger
        =
        - \frac{it_2 \nu}{3\sqrt{3}}  \sigma_y
        \\
        [\mathcal{H}_\mathrm{kin}^{\mathrm{TWSM}}]_{\mathbf{n}; \mathbf{n}+\boldsymbol{\delta}_2 + \boldsymbol{\delta}_3}
        &=
        [\mathcal{H}_\mathrm{kin}^{\mathrm{TWSM}}]_{\mathbf{n}+\boldsymbol{\delta}_2 + \boldsymbol{\delta}_3; \mathbf{n}}^\dagger
        =
        \frac{it_2 \nu}{3\sqrt{3}}  \sigma_y
        \\
        [\mathcal{H}_\mathrm{kin}^{\mathrm{TWSM}}]_{\mathbf{n}; \mathbf{n}+\boldsymbol{\delta}_3 - \boldsymbol{\delta}_1}
        &=
        [\mathcal{H}_\mathrm{kin}^{\mathrm{TWSM}}]_{\mathbf{n}+\boldsymbol{\delta}_3 - \boldsymbol{\delta}_1; \mathbf{n}}^\dagger
        =
        - \frac{it_2 \nu}{3\sqrt{3}}  \sigma_y
        \\
        [\mathcal{H}_\mathrm{kin}^{\mathrm{WSM}}]_{\mathbf{n}; \mathbf{n}+\boldsymbol{\delta}_z}
        &=
        [\mathcal{H}_\mathrm{kin}^{\mathrm{WSM}}]_{\mathbf{n}+\boldsymbol{\delta}_z; \mathbf{n}}^\dagger
        =
        - t_z \sigma_z.
    \end{aligned}
\end{equation}
with $\boldsymbol{\delta}_{i=1,2,3}$ and 
$\mathbf{\delta}_z = (0,0,1)$ being the lattice vectors connecting nearest neighbors.
When there is inter-Fermi surface $s$-wave pairing, as in Eq.~\eqref{TBM.inter_FS.swave_pairing}, the system gives rise to the monopole superconducting order $\Delta_\mathrm{MSC}^{(q_p=3, l_z=0)}$ confined to the topological sector $q_p=3$ and with global angular momentum $l_{z, \mathrm{glob}} = 0$, as described in Sec.~\ref{subsec:qp=3} in the main text.

\begin{figure}
    \centering
    \includegraphics[width=0.95\linewidth]{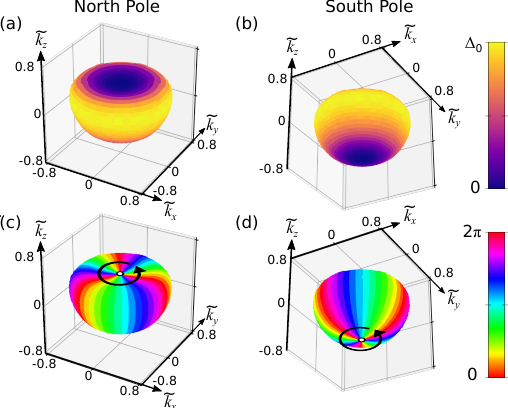}
    \caption{
    Magnitude [(a) and (b)] and phase winding [(c) and (d)] of the $\Delta_\mathrm{MSC}^{(q_p=3, l_z=0)}$ monopole superconducting pairing order based on the tight-binding model in Appendix~\ref{appendix:TBM.qp=3}.
    The left and right columns respectively correspond to the north and south pole views of the pairing order at $\mathrm{FS}_1$.
    (c) and (d) are computed in a gauge where the pairing phase is well-defined at the north and south pole respectively.
    Parameters are chosen as $\mu = 0.1$, $K_0 = \pi/3$, $t_{xy} = t_z \sin K_0 /8 = -1$, and $m = 0.05$.
            }
    \label{fig:q=3.effective_pairing}
\end{figure}

We further examine the effective monopole pairing order $\Delta_\mathrm{MSC}^{(q_p=3, l_z=0)}$ in Fig.~\ref{fig:q=3.effective_pairing}.
Here, we have computed the band-projected pairing gap function in Eq.~\eqref{MSC.effective_gap_function.weak_coupling}, using the tight-binding model given in Eq.~\eqref{q=3.msc_ham} and projecting onto $\mathrm{FS}_{1}$ of the band Hamiltonian in Eq.~\eqref{C=3.kin}.
As depicted in Fig.~\ref{fig:q=3.effective_pairing}(a) and \ref{fig:q=3.effective_pairing}(b), the magnitude of the gap function reaches its maximum near the equator and vanishes at the poles, consistent with the effective pairing gap function in Eq.~\eqref{q=3.eff_pairing}, calculated for the related continuum model.
There are two gap nodes located at the north and south pole, which each display a local effective chiral-$f$ wave symmetry but with opposite chirality.
Figures~\ref{fig:q=3.effective_pairing}(a) and \ref{fig:q=3.effective_pairing}(c) are computed in the ``north pole gauge'' where the Dirac string pierces the Fermi surface at $\theta_{\tilde{k}} = \pi$.
In this gauge, the $\Delta_\mathrm{MSC}^{(q_p=3, l_z=0)}$ shows a local $f+if$ pairing, as evident by the winding in Fig.~\ref{fig:q=3.effective_pairing}(c).
In contrast, Figs.~\ref{fig:q=3.effective_pairing}(b) and \ref{fig:q=3.effective_pairing}(d) are computed in the ``south pole gauge,'' with the Dirac string piercing through $\theta_{\tilde{k}}=0$.
In this gauge, the $\Delta_\mathrm{MSC}^{(q_p=3, l_z=0)}$ is described by a local $f-if$ pairing, as shown in Fig.~\ref{fig:q=3.effective_pairing}(d).
The $\mathrm{U}(1)$ topological obstruction due to the nonvanishing pair monopole charge $q_p=3$ leads to nonvanishing net vorticity, given by twice the pair monopole charge $2 q_p = 6$.
The global angular momentum, which is a conserved quantity, is given by $l_{z, \mathrm{glob}}=0$.
Moreover, the system with $\Delta_\mathrm{MSC}^{(q_p=3, l_z=0)}$ pairing order exhibits three chiral surface modes in the $ab$ plane under open boundary conditions.
Although the local angular momentum $l_{z, \mathrm{loc}}$ of the pairing order changes as $k_z$ is varied between the gap nodes, as shown in the projection of the pairing gap function
in Fig.~\ref{fig:q=3.effective_pairing}, the leading term in the surface state dispersion is of order $\mathcal{O}(k^2)$ and consequently overshadows terms of order $\mathcal{O}(k^3)$ which encode the shift in local angular momentum.
Nonetheless, the three surface modes are analogous to those that would arise in a system with chiral $f$-wave pairing.

\section{Josephson current for a \texorpdfstring{$z$}{z} direction junction between a \texorpdfstring{$\Delta_\mathrm{MSC}^{(q_p=-1, l_z = 2)}$}{(qp=-1, lz=2)} monopole superconductor and a \texorpdfstring{$d_{x^2-y^2} + id_{xy}$}{d+id}-wave superconductor}
\label{appendix:dwave}

\begin{figure}[htb]
    \centering
    \includegraphics[width=.95\linewidth]{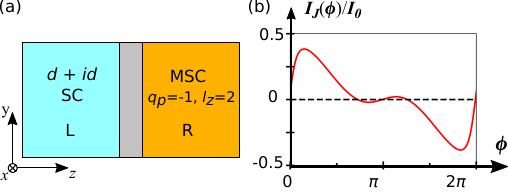}
    \caption{
        (a)~Josephson junction along the $z$ direction between a $\Delta_{\mathrm{MSC}}^{(q_p-1, l_z=2)}$ monopole superconductor and a $d_{x^2-y^2} + id_{xy}$-wave superconductor.
                (b)~Josephson current phase relation between the $\Delta_{\mathrm{MSC}}^{(q_p-1, l_z=2)}$ monopole superconductor and $d_{x^2-y^2} + id_{xy}$-wave superconductor (red, solid), which shows nonvanishing first-order coupling.
    In contrast, the first-order Josephson current phase relation for a junction with an $s$-wave SC (black, dashed), is zero.
    }
    \label{fig:WeylProxP.junction}
\end{figure}

We demonstrate the nonvanishing first-order Josephson critical current for a junction between a $\Delta_\mathrm{MSC}^{(q_p=-1, l_z = 2)}$ monopole superconductor and a $d_{x^2-y^2} + id_{xy}$-wave superconductor, as detailed in Sec.~\ref{subsec:different_lz}.
We consider a junction along the $z$ direction, shown schematically in Fig.~\ref{fig:WeylProxP.junction}(a) and compute the Josephson current using a tight-binding model of the Josephson junction with spin-independent hopping at the junction link, analogous to that described in Sec.~\ref{subsec:JG.tbm}.
The Josephson current phase relation is shown in red in Fig.~\ref{fig:WeylProxP.junction}(b).
For the same symmetry reasons as discussed in Sec.~\ref{subsec:different_lz}, there is nonvanishing first-order Josephson current between a $\Delta_{\mathrm{MSC}}^{(q_p=-1, l_z=2)}$ monopole superconductor and $d_{x^2-y^2}+id_{xy}$ superconductor, \textit{i.e.} because both pairing orders transform under rotation $R_z$ according to global angular momentum $l_z = 2$, there is nonvanishing first-order Josephson coupling.
The magnitude of the Josephson current is largely suppressed due to the gap nodes along the $k_z$ axis for both the chiral $d$-wave and monopole pairing orders.
Nonetheless, the $2\pi$-periodic first-order Josephson current persists, indicative of the global angular momentum of the monopole superconducting pairing order, $l_z = 2$.
This is in contrast to, for example, a junction with a junction between an $\Delta_{\mathrm{MSC}}^{(q_p=-1, l_z=2)}$ monopole superconductor and $s$-wave superconductor, for which the total Josephson current would cancel upon summing the contributions from states with different conserved transverse momentum $k_\parallel$.

%

\end{document}